\documentclass[superscriptaddress,11pt]{article}
\pdfoutput=1 
\usepackage{authblk}
\usepackage{array}
\usepackage{type1cm}
\usepackage{lettrine}
\usepackage{graphicx}
\usepackage{geometry}
\usepackage{psfrag}
\usepackage{amsmath}
\usepackage{amsfonts}
\usepackage{appendix}
\usepackage[margin=10pt,font=footnotesize,labelfont=sc,format=hang,labelformat=parens]%
{caption}
\usepackage{subcaption}
\usepackage{amssymb}%
\providecommand{\U}[1]{\protect\rule{.1in}{.1in}}
\numberwithin{equation}{section}
\def\be{\begin{equation}}
\def\ee{\end{equation}}
\def\ba{\begin{eqnarray}}
\def\ea{\end{eqnarray}}
\def\bi{\begin{itemize}}
\def\ei{\end{itemize}}

\def\bra{\langle}
\def\ket{\rangle}

\begin{document}

\title{On quantum propagation in Smolin's weak coupling limit of 4d Euclidean Gravity}

\author{Madhavan Varadarajan}
\affil{Raman Research Institute\\Bangalore-560 080, India}

\maketitle

\begin{abstract}

Two desireable properties of a  quantum dynamics for Loop Quantum Gravity (LQG) are that its generators provide
an anomaly free representation of the classical constraint algebra and that 
physical states which lie in the kernel of these generators encode propagation.  
A physical state in LQG is expected to be a sum over  graphical $SU(2)$ spin network states. By propagation we mean that a quantum perturbation at one vertex of a
spin network state propagates to vertices which are  `many links away' thus yielding a new spin network state which is related to the 
old one by this propagation. A physical state encodes propagation if its spin network summands are related by propagation.
Here we study propagation in an LQG quantization of Smolin's weak coupling limit of Euclidean Gravity based on graphical $U(1)^3$ `charge' network states.
Building on our earlier work on anomaly free quantum constraint actions for this system, we analyse the extent to which physical states encode 
propagation. In particular, we show that 
a slight modification of the constraint actions constructed in our previous work
leads to physical states which encode robust propagation. Under appropriate conditions, this propagation  merges, seperates  and  entangles vertices of 
charge network states. The  `electric' diffeomorphism constraints introduced in prevous work play a key role in our considerations. The main import of our work is that there are choices of quantum constraint constructions 
through LQG methods which are consistent with vigorous propagation thus 
providing a counterpoint to Smolin's early observations on the difficulties of  propagation in the context of LQG type operator constructions. 
Whether the choices considered in this work are physically appropriate is an open question
worthy of further study.

\end{abstract}


\section{\label{sec1}Introduction}

Loop Quantum Gravity \cite{aajurekreview,ttbook,mebook} is an attempt at non-perturbative canonical quantization of General Relativity based on a classical Hamiltonian description in terms of triads (or, equivalently $SU(2)$ electric fields) and conjugate connections.
While the $SU(2)$ rotation constraint and the 3d diffeomorphism constraints can be satisfactorily represented and solved in quantum theory, the construction of the Hamiltonian constraint operator involves an infinity of choices.
In order to identify the correct choice of the Hamiltonian constraint, these choices should be confronted with physical criteria. Two such critera are that the constraint action be compatible with an anomaly free representation 
of the classical constraint algebra and that the constraint action should be consistent with the propagation of quantum perturbations.
Whereas the `anomaly free' criterion ensures spacetime covariance, the `propagation' criterion is motivated by the existence of classical solutions to General Relativity which describe propagating degrees of freedom.

A clear statement  of the propagation criterion in the context of LQG states which live on graphs was first provided by Smolin \cite{leeprop} as follows. Given a spin network state, the action of a constraint deforms the state
in the vicinity of one of its vertices to yield a `perturbed' spin net state. If the quantum dynamics is  such that this perturbation moves to distant (in terms of graph connectivity) vertices of the original spin net state,
the quantum dynamics will be said to encode propagation and the initially perturbed state  and the final state can be said to be related by propagation.
Further, Smolin envisaged putative propagation as being generated by successive actions of the Hamiltonian constraint and concluded that 
LQG techniques were unable to generate propagation when viewed in this way.  
\footnote{\label{fnlee}While this represents the broad lesson drawn by Smolin, it is a drastic  oversimplification of Smolin's analysis. 
The reader is urged to consult Reference \cite{leeprop} for a detailed account of this analysis. 
It is also pertinent to note here
that Smolin's analysis ruled out propagation in the part of the joint kernel of the Hamiltonian and 
diffeomorphism constraints, 
erroneously claimed to
be the full kernel, in a preprint version of \cite{qsd2}. In collaboration with Thiemann \cite{ttmv}, we are currently engaged
in an analysis of propagation for states in the full kernel not considered by Smolin. We return to a discussion of this 
analysis towards the end of this paper. The reader unwilling to be left in suspense may directly consult
the last paragraph of section \ref{sec6}.
}

However, the study of Parameterised Field Theory (PFT) \cite{proppft}  implies that rather than visualising propagation as being generated by 
successive actions of constraint operators, propagation should be seen as a property of physical states as follows. Physical states lie in the joint kernel  of the quantum constraints and
are constructed as sums of infinitely many spin net states. 
Given one such spin net summand, if there exists another summand which is related to the first by propagation then we shall say that 
the physical state encodes propagation.
Note that while this notion of propagation is not necessarily generated by the action of the constraints, it is nevertheless crucially dependent on structural properties of the constraints 
The reason is as follows. The structural properties of the constraint operators  determine the 
structural properties of 
any  physical state in their joint kernel. More in detail since every physical state is a sum over spin network summands, how these summands relate to each other is determined by the structure of the constraints. 
As mentioned above, propagation is encoded in these relations and it is in this indirect way that the structure of the constraint operators dictate if propagation ensues or not. It is in this sense that 
the propagation criterion restricts the choice of the Hamiltonian constraint.

Next, consider the anomaly free criterion.
While LQG does provide an anomaly free representation of a very non-trivial subalgebra of the constraint algebra, namely 
that of the spatial diffeomorphism constraints \cite{alm2t,habitat1,mediff},  
the implementation of a {\em non-trivial}
anomaly  free commutator corresponding to the Poisson bracket between a pair of Hamiltonian constraints is still an open problem \cite{habitat1,habitat2,mebook}. 
A key identity discovered in \cite{p1},  implies that  this Poisson bracket is the same as that of a sum of Poisson brackets between  pairs
of {\em diffeomorphism}  constraints smeared with {\em electric field dependent} shifts which we shall call electric shifts. Such diffeomorphism constraints are called 
{\em electric diffeomorphism constraints}.  Hence an implementation of non-trivial anomaly free commutators between Hamiltonian constraints is equivalent to the 
imposition of this identity in quantum theory. This requires the construction not only of the Hamiltonian constraint operator but also these electric diffeomorphism operators in such a way
that the commutator between a pair of Hamiltonian constraints equals the 
appropriate sum of commutators of electric diffeomorphism constraints.
This requirement is extremely non-trivial.
However, precisely because of this fact, the anomaly free criterion is expected to prove
extremely restrictive for the choice of Hamiltonian constraint operator.  

Given the non-triviality of the two criteria and the involved nature of full blown LQG, it is useful to first develop intuition for structural properties of the Hamiltonian constraint which are compatible with these criteria in simpler toy models.
In this regard 
Smolin's weak coupling limit of Euclidean Gravity\cite{leeu13} 
offers an ideal testing ground. 
Since this system may be  obtained simply be replacing the $SU(2)$ electric and connection fields of Euclidean gravity
by their $U(1)^3$ counterparts, we shall refer to this model as the $U(1)^3$ model. 
Its constraint algebra is isomorphic to that of Euclidean gravity and hence displays structure functions.
An LQG quantization of this sytem \cite{p1,p2,jureklin,p3} leads to a representation space for  holonomy- flux operators spanned by $U(1)^3$ spin network states
which we call `charge' network states. These states  are labelled by  graphs whose edges are colored by representations of $U(1)^3$.
Our recent work \cite{p3} concerns the imposition of the {\em anomaly free criterion}, as articulated above, in the context of this  model.
Here  we confront the ideas of \cite{p3}  with the {\em propagation criterion} as applied to physical states.
In doing so we shall not be concerned with the enormous amount of technical detail entailed in the constructions of \cite{p3}. Rather, we shall abstract what we regard as the key features of those constructions and 
base our analysis of propagation on these features and minimal generalizations and modifications thereof.  Whether this broad treatment can then be endowed with the level of technical detail in \cite{p3} so as to demonstrate
compliance with anomaly freedom is an open question. 

We now turn to an account of the key features of the work in \cite{p3}.
The  Hamiltonian constraint operator constructed in that work acts non-trivially on a charge network state only at those of its vertices which have valence greater than 3 provided these vertices are 
non-degenerate in a precise sense.
\footnote{The Hamiltonian and electric diffeomorphism constraint actions of \cite{p3} are specified on charge net states with only a single vertex whereas the discussion of propagation necessarily involves multivertex chargenet states. 
However, since these actions at different vertices are independent of each other, the actions of \cite{p3} automatically define actions on multivertex  charge nets through a sum over actions on individual vertices.
}
The Hamiltonian constraint action on any such  $N$ ($N>3$) valent  vertex of a `parent' charge net results in a sum over deformed `child' charge nets. 
The graph underlying a child  charge net is obtained by deforming the parental graph in the vicinity of its vertex along some  parental edge at that vertex. 
Roughly speaking, this deformation along  a parental  edge corresponds to a `singular diffeomorphism' wherein the 
the remaining $N-1$ edges are 
pulled `almost' along this edge so as to form a cone with axis along this edge.  
The resulting `child graph' now has the original $N$ valent parent vertex 
and
an 
$N$ valent  child vertex, the child vertex
structure being conical in the manner described. In addition the charges in the vicinity of these two vertices are altered, these alterations arisng from flipping the signs of certain parental charges due to which
the original parent vertex is now no longer non-degenerate in the child.
Thus, the constraint acts through a combination of `singular diffeomorphisms'
and `charge flips'.  For obvious reasons, and for future reference, we shall refer to the Hamiltonian and electric diffeomorphism constraint actions as $N\rightarrow N$ actions.

Reference \cite{p3} constructs  a space  of `anomaly free' states which support the action of this Hamiltonian constraint operator so as to yield an anomaly free constraint algebra 
in the detailed sense articulated in that work and sketched briefly above.
Each such state is a specific linear combination of certain charge net  states.  
Thus each such state is specified by a `Ket Set' of charge net states and a set of coefficients, one for each element of the Ket Set
associated with the anomaly free state. The sum over all the elements of the Ket Set with these coefficients then yields the anomaly free  state so specified, 
on which constraint commutators are anomaly free in the sense described above.
In particular the Hamiltonian constraint commutators are shown to equal the appropriate sum of electric diffeomorphism constraint commutators
in accordance with the identity discovered in \cite{p1}. The electric diffeomorphism constraint operators
can be constructed very naturally in a manner similar to the Hamiltonian constraint. The  operators so constructed  move the original parent vertex by exactly the same 
`singular diffeomorphisms' as employed by the Hamiltonian 
constraint, but with {\em no} charge flips, with the singular nature of the `singular diffeomorphisms
arising from the distributional nature of the quantum electric shift.

The constructions of \cite{p3} also imply that 
in the special case that  the coefficients are chosen to be unity for all elements of the Ket Set, it turns out that  the anomaly free state is killed by the Hamiltonian and diffeomorphism constraints 
as well as the electric diffeomorphism constraints. Such a state is then a 
physical state which supports a trivial anomaly free realization of the commutators. Nontrivial anomaly free commutators arise only if the coefficients are chosen in a specific non-trivial way and the resulting state
is then an {\em off shell} state.  From \cite{p3}, such an  anomaly free state can be thought of as an off-shell deformation of the physical state obtained with unit coefficients. 
The Ket Set associated with a physical state and its off shell deformations in \cite{p3} satisfy the following properties.
The first property is that the Ket Set
is closed with respect to deformations generated not only by the action of the Hamiltonian constraint but also by the electric diffeomorphism constraints. Thus, in the parent-child language used
above, this property says that if a certain charge net is in the Ket Set then so are all its deformed children produced by the action of the Hamiltonian and electric diffeomorphism constraints.
The second property is more subtle and can be phrased succintly in the parent- child language as follows. If a chargenet is in the Ket Set then so are all its possible parents. Here  a `possible parent' $p$
of a charge net $c$
refers to any charge net  which when acted upon by the Hamiltonian or electric diffeomorphism constraints gives rise to deformed children, one of which is the charge net $c$. 
In addition to these properties, the Ket Set is also closed with respect to semianalytic diffeomorphisms; amongst other things, 
this is necessary for an anomaly free representation of the commutators involving the  (usual) diffeomorphism constraints smeared with $c$-number shifts.
We summarise these properties in the form of the statement (a) below:
\\

\noindent (a) If the Ket Set has a certain charge net then \\
(a.1): All possible children generated by the action of the Hamiltonian and electric diffeomorphism  constraints on this charge net are also in the Ket Set.\\
(a.2): All possible {\em parents} of this charge net (i.e. all charge nets which when acted upon by these constraints generate children one of which is the charge net in question) are also in the Ket Set.
\\
(a.3) The  Ket Set is closed with respect to the action of semianalytic diffeomorphisms.
\\

In terms of Ket Sets subject to property (a),
\footnote{Reference \cite{p3} only constructs Ket Sets each of whose elements have single non-degenerate vertex. Here we implictly assume that the considerations of \cite{p3}
can be generalized to multivertex Ket Sets. We shall discuss this further in section \ref{sec5}.}
our discussion of propagation may be re-stated as follows.
Smolin's visualisation of propagation  is based on the repeated action of constraint operators. Such actions concern property (a1)  but not (a2). The key new element to be confronted when 
we analyse physical states are the summands which owe their existence due to property (a2).

The fact that the sum, with unit coefficients, over  elements of the Ket Set subject to Property (a) defines an {\em anomaly free, physical} state, is a direct 
consequence of a particular structural property of the Hamiltonian and electric diffeomorphism constraint approximants employed in \cite{p3}.
We shall describe this property in section \ref{sec2.8}. We note here, that
if we drop the anomaly free requirement on physical states, there is no reason to consider the electric diffeomorphism constraints. 
Then, as will be apparent in section \ref{sec2.8}, for Hamiltonian constraints with this particular structural property, physical states may be constructed as sums over elements of Ket Sets
with a weaker version of  property (a) wherein any mention of the electric diffeomorphism constraint is removed so that all children and `possible' parents are 
only with reference to the  Hamiltonian  constraint. However such states will not, in general, support anomaly free commutators since we have no control
on the `right hand side' of the key identity of \cite{p1}. 
If we now construct electric diffeomorphism constraint operators with the structural property described
in section \ref{sec2.8}, then physical states which support anomaly free commutators may be naturally constructed as sums over  elements of  Ket Sets with unit coefficients, these Ket Sets
being subject to property (a) in which both  the Hamiltonian and electric diffeomorphism constraints play a role. Such physical states are then killed by both the 
Hamiltonian and electric diffeomorphism constraints and thereby provide a consistently trivial implementation of the anomaly free requirement.
To summarise: for Hamiltonian and electric diffeomorphism constraint actions  with the structural property described in section \ref{sec2.8}, anomaly free physical
states can be constructed as sums, with unit coefficients, over elements of Ket Sets subject to property (a) above.

In this work, it will prove necessary to slightly modify the constraint approximants of \cite{p3} so as to engender propagation. The modified actions also have the  special 
structural property of section \ref{sec2.8} . Hence the Ket Sets we consider in this work will all be subject to Property (a) and will define anomaly free physical states.
Whether  propagation ensues or not for a particular choice of 
such actions is then dependent 
entirely on the properties of the `possible' parents in property (a2).
\footnote{
The structural property of constraint approximants alluded to above was 
first discovered in the context of PFT \cite{proppft}. While we did not phrase that  analysis
explicitly in terms of Ket Sets, it is straightforward to check that such a rephrasing is 
immediate and that the key lesson of that analysis  is the role played by the properties of `possible parents' of (a2) in enabling propagation.}
Hence our strategy in this work is to analyse whether the Ket Sets relevant to choices of constraint actions with the structure described in section \ref{sec2.8}
have possible parents which facilitate propagation. If they do, it follows that the physical states obtained as sums over elements of such Ket Sets encode propagation. 
In what follows we shall often use a more direct language and simply say that such Ket Sets encode propagation.
As stated above, while  physical states constructed as sums over elements of  Ket Sets are guaranteed to be anomaly free with respect to the 
associated constraint actions of the type discussed in section \ref{sec2.8} , whether off shell deformations of these physical states can be constructed which support {\em non-trivial}
anomaly free commutator brackets is an open question which we leave for future work.

We are now  in a position to discuss the work done in this paper.
As a warm up exercise, we start with an  exploration of propagation of specific perturbations between the vertices of a simple 
2 vertex state. This exercise serves to illustrate the notion of propagation (or lack thereof) in the language of Ket Sets
and `possible parents' articulated above.
The simple 2 vertex state $|s_{A,B,N}\ket$ that we study
consists of a pair of vertices $A,B$ joined by $N$ edges with charges subject to a genericity condition.
We create a `perturbation' of this state in the vicinity of the vertex $A$ by the action of the Hamiltonian constraint.
We show that this perturbation cannot `propagate' to vertex $B$ and be `absorbed' there in the context of the constraint actions
constructed in \cite{p3}. In the language of Ket Sets, we 
show that the minimal Ket Set which satisfies Property (a) with respect to the 
constraint actions of \cite{p3} and which contains $|s_{A,B,N}\ket$
{\em does not} encode propagation of this specific type of perturbation.
%
%
%
Equivalently,  the  physical state annihilated by the constraint actions of \cite{p3} and  obtained by summing over the elements of this Ket Set   {\em  does not} encode propagation
of this specific perturbation. 
Next, we study physical states subject to a further physically reasonable condition. This condition implies that physical states satisfy additional operator equations
which are also of the form discussed in section \ref{sec2.8}. A natural class of anomaly free physical states which are annhilated by the constraints of \cite{p3} and 
satisfy the new condition may then be constructed as sums over elements of Ket Sets which satisfy, in addition to property (a),  the property of closure with respect to
children and `possible parents' appropriate to the new condition. Once again, we study the minimal Ket Set which contains the 2 vertex state $|s_{A,B,N}\ket$.
In this case, we study the perturbation created at the vertex $A$ by the action of operators involved in the specification of this additional condition.
We show that this perturbation can propagate to vertex $B$ and be absorbed there. In the language of Ket Sets, this minimal Ket Set
 {\em does}
encode propagation of this specific perturbation from $A$ to $B$ by virtue of the richer class of children and possible parents whose existence is traced to the new condition.
Thus,  there is a natural class of physical states which satisfy an additional physical condition and which do encode propagation between vertices of 
a generic 2 vertex states of the type $|s_{A,B,N}\ket$. We note here that the new condition is closely related to the combination of constraints appearing in Reference \cite{aanewpersp} 
(see vi), pg 85, Chapter 6 of this reference). This concludes our study of propagation of specific types of perturbations of this simple 2 vertex state.

Next, we consider generic multivertex states with more than 2 vertices. We show that the `$N\rightarrow N$' constraint actions of \cite{p3} cannot generate propagation of any perturbations
between pairs of vertices
of different valence. We argue that at best, even for states subject to the additional condition mentioned above, only a certain `1d' propagation may be possible for special 
multivertex states.
%
Therefore, in order to engender a more vigorous, 3d and long range propagation, we slightly modify the `$N\rightarrow N$' constraint actions of \cite{p3}. 
The key features of the modified constraint actions can be seen to arise from valid quantization choices and differ from those of 
\cite{p1,p2,p3} in that they {\em change} the valence of the vertex on which they act. It then turns out that  Ket Sets satisfying condition (a) with respect to this modified action have a significantly
richer structure which 
 encodes vigorous propagation. The key property of the modified action 
is that it changes the valence of both the original parent vertex as well as the child vertex in the deformed children charge nets relative to \cite{p3}. This is achieved by visualising the 
`singular diffeomorphism' deformation involved in the action of the Hamiltonian and electric diffeomorphism constraints
slightly differently from that described above as follows.

We imagine the generalised diffeomorphisms to act by pulling all but 3 of the remaining edges {\em exactly} along the 
$I$th edge with the remaining 3 edges being pulled `almost' along the $I$th edge in a conical manner. This results in a child graph in which the original parent vertex valence drops to $N-3$ and the 
child vertex has a valence of 4. 
With the incorporation of this modified action into that of the Hamiltonian and electric diffeomorphism constraints, 
it turns out that the Ket Sets subject to condition (a)  do encode propagation. We shall refer to this modification of the  constraint action  as  an `$N\rightarrow 4$ modification. 
As we shall see, the  deformations generated by the electric diffeomorphism constraints play 
a  crucial role in this encoding of propagation.
Since the only reason the electric diffeomorphism constraints appear in our considerations is to ensure compliance with the anomaly free criterion, this suggests   
that  the two criteria of anomaly freedom and
propagation work in unision. 

The layout of the paper is as follows.
Section \ref{sec2} starts with  a brief review of earlier material in \cite{p1,p2,p3,proppft} which is of direct relevance to our work here. 
Specifically, in sections \ref{sec2.1}- {\ref{sec2.5}  we  review  the constraint actions of \cite{p3}.  In the interests of pedagogy we suppress certain important details
in our treatment; these details are collected and described in section \ref{sec2.7} and may be skipped by readers unfamiliar with \cite{p3}.
In section \ref{sec2.8} we review the  structural property of constraint approximants alluded to above which is  connected with property (a).
Section \ref{sec3} studies propapagation in the context of the $N\rightarrow N$ actions of \cite{p3}.
In section \ref{sec3.1} we study propagation of specific perturbations in a simple 2 vertex state as discussed above. 
In section \ref{sec3.2}
we show  that the constraint actions of \cite{p3} cannot engender propagation between  pairs of vertices of different valence
in multivertex states and argue that, at best, a `1d propagation' may be possible for a very restrictive class of multivertex states.
In section \ref{sec4} we describe the $N\rightarrow 4$ modification of the constraint action and show that the Ket Set compatible with this modified action encodes vigorous propagation.
As mentioned above, a comprehensive proof that the constraint action considered in section \ref{sec4}
has a non-trivial  anomaly free implementation would be at least as involved as the considerations of \cite{p3} 
and is out of the 
scope of this work. In section \ref{sec5.0} we discuss the new challenges to be confronted relative to Reference \cite{p3} in the construction of such a putative proof
as well as certain technicalities related to our treatment of propagation hitherto.
In section \ref{sec5.1} we discuss an important  consequence of the $N\rightarrow 4$ action, namely the phenomenon of 
vertex mergers.
%
Section \ref{sec6} contains a discussion of our results and of open issues.

Our work here may be considered as a continuation of that in the series of papers \cite{p1,p2,p3}.
While a detailed understanding
of the considerations of this work, especially that of sections \ref{sec2.7}  and \ref{sec5.0}, 
requires  familiarity with these works, an understanding of the broad features of this work requires familiarity only with the reasonably self contained
expositon of sections \ref{sec2.1}- \ref{sec2.5} and section \ref{sec2.8}  of this work. Readers not familiar with \cite{p1,p2,p3} may skip
sections \ref{sec2.7} and \ref{sec5.0} on a first reading.
Further, the reader interested mainly in the long range 3d propagation results may skip section \ref{sec3.1} entirely.

\section{\label{sec2} Brief Review of Relevant Material from References \cite{p3,proppft}}
\subsection{\label{sec2.1}Elements of the classical theory}
The phase space variables $(A_a^i, E^a_i, i=1,2,3)$ are a triplet of  $U(1)$  connections and conjugate density weight one electric fields
on the Cauchy slice $\Sigma$ with canonical Poisson brackets\\  $\{A_a^i (x), E^b_j(y)\}= \delta^b_a\delta_j^i \delta^3(x,y)$.
The Gauss law, diffeomorphism, and Hamiltonian constraints of the theory are:
\begin{align}
G[\Lambda]  &  =\int\mathrm{d}^{3}x~\Lambda^{i}\partial_{a}E_{i}^{a}\\
D[\vec{N}]  &  =\int\mathrm{d}^{3}x~N^{a}\left(  E_{i}^{b}F_{ab}^{i}-A_{a}%
^{i}\partial_{b}E_{i}^{b}\right) \label{defclassd}\\
H[N]  &  =\tfrac{1}{2}\int\mathrm{d}^{3}x~{N}q^{-1/3}\epsilon^{ijk}E_{i}^{a}E_{j}%
^{b}F_{ab}^{k}, \label{defclassh}
\end{align}
with $F_{ab}^{i}:=\partial_{a}{A}_{b}^{i}-\partial_{b}{A}_{a}^{i}\;\;\;$,
 $qq^{ab}:= \sum_i E^a_iE^b_i\;\;\;$, $q = \det q_{ab}$.
 A key identity \cite{p1} holds on the Gauss Law constraint surface:
\be 
\{H[N],H[M]\}  = (-3) \sum_{i=1}^3\{D[{\vec N}_i],D[{\vec M}_i]\} 
\label{key}
\ee
where the Electric Shifts $N^a_i$ are defined as:
\be
N^a_i = NE^a_i q^{-1/3}
\label{defes}
\ee
and the Electric Diffeomorphism Constraints $D({\vec N}_i)$ by 
\be
D[\vec{N}_i]   =\int\mathrm{d}^{3}x~N_i^{a}  E_{j}^{b}F_{ab}^{j}
\ee

\subsection{\label{sec2.2}Quantum Kinematics}
A charge network label $c$ is the collection $(\alpha, {\vec {q_I}}, I=1,,N )$ where $\alpha$ is an oriented graph with $N$ edges, the 
$I$th edge $e_I$ colored with a triplet of $U(1)$ charges $(q^1_I,q^2_I, q^3_I)\equiv {\vec q_I}$ such that the net outgoing charge at every vertex vanishes.
 The gauge invariant holonomy associated with $c$ is $h_c$,
\be
h_{c}:=\prod_{I=1}^{N}\; 
\mathrm{e}^{\mathrm{i}\kappa\gamma q_I^{j}\int_{e_{I}}A_{a}%
^{j}\mathrm{d}x^{a}}, 
\ee
where $\kappa$ is a constant with dimensions $ML^{-1}$, $\gamma$ is a dimensionless Immirzi parameter. Henceforth we use  units such
that $\kappa \gamma =1$.
The Hilbert space is spanned by charge network states $\vert c\ket$ which are eigen states of the electric field operator.
The eigen value of the electric shift operator ${\hat N^a_i}(x)$ (see (\ref{defes})) is non-zero only at vertices of the charge net state and requires
a regulating coordinate patch at each of these  vertices for its evaluation:
\be
{\hat N^a_i}(v) |c\ket=  N^a_i (v) |c\ket:= \sum_{I_v} N^{a}_{I_v i} |c\ket, \;\;\; N^{a}_{I_v i} := \frac{3}{4\pi} N(x(v)) \nu_v^{-2/3}q^i_{I_v}{\hat e}^a_{I_v} .
\label{qsev}
\ee
Here $v$ is a vertex of $c$, $I_v$ refers to the $I_v$th edge at $v$, and ${\hat e}^a_{I_v}$  to the unit $I_v$th edge tangent vector, unit with respect to the coordinates $\{x\}$ 
at $v$ and $N(x(v))$ denotes the 
evaluation of the density weighted lapse $N$ at $v$ in this coordinate system. $\nu^{-2/3}_v$ is proportional to the eigen value of the ${\hat q}^{-1/3}$ operator, this eigen value
being (possibly) non-trivial only for vertices of valence greater  than 3. 
We refer to the eigenvalue $N^a_i (v)=\sum_{I_v} N^{a}_{I_v i}$ as the {\em quantum shift}. We emphasise that for each vertex of valence $N>3$ we need a choice of
regulating coordinates to evaluate this quantum shift. 

\subsection{\label{sec2.3}Discrete Hamiltonian Constraint from P1}
The action of the discrete approximant to the Hamiltonian constraint operator of \cite{p3} is motivated as follows.
A charge net state can be thought of heuristically as a wave function of
the connection  which is itself a holonomy. Accordingly we use the following notation for the this wave function:
\be
c(A)= h_c(A)= \prod_{I=1}^{N}\; 
\mathrm{e}^{\mathrm{i}\kappa\gamma q_I^{j}\int_{e_{I}}A_{a}%
^{j}\mathrm{d}x^{a}}= 
\exp\left(  \int\mathrm{d}^{3}x~c_{i}^{a}A_{a}^{i}\right)
\ee
where we have defined:
\begin{equation}
c_{}^{ai}(x): =c_{}^{ai}(x;\{e_{I}\},\{q_{I}\})=\sum_{I=1}^{M}\mathrm{i}%
q_{I}^{i}\int\mathrm{d}t_{I}~\delta^{(3)}(e_{I}(t_{I}),x)\dot
{e}_{I}^{a}(t_{I}). \label{chrgcoord}%
\end{equation}
Holonomy operators  act by multiplication and the electric field operator by functional differentiation on charge net wave functions.
Using the identity 
$N_{i}^{a}F_{ab}^{k}=\pounds _{\vec{N}_{i}}A_{b}^{k}- \partial_{b}(N_{i}%
^{c}A_{c}^{i})$, 
the classical  Hamiltonian constraint can be written on the Gauss Law constraint surface as:
\begin{eqnarray}
H[N]&=& \tfrac{1}{2}\int_{\Sigma}\mathrm{d}^{3}x~\epsilon^{ijk}N^a_iF_{ab}^{k}%
E_{j}^{b}  + \tfrac{1}{2}\int_{\Sigma}\mathrm{d}^{3}x~N^a_iF_{ab}^{i}%
E_{i}^{b}  \nonumber\\
&=& \tfrac{1}{2}\int_{\Sigma}\mathrm{d}^{3}x~\left(
-\epsilon^{ijk}(\pounds _{\vec{N}_{j}}A_{b}^{k})E_{i}^{b}+%
{\textstyle\sum\nolimits_{i}}
(\pounds _{\vec{N}_{i}}A_{b}^{i})E_{i}^{b}\right) 
\label{hamconst1}
\end{eqnarray}
where we have added the classically vanishing second term  on the right hand side of the first line.
The action of the corresponding operator on the state $c(A)$ is obtained
by replacing the electric shift by the action of its operator correspondent (\ref{qsev}) which is, in turn, replaced by its 
eigen value $N^a_i (v)= \sum_{I_v} N^{a}_{I_v i}$ to yield:
\be
\hat{H}[N]c(A) 
=\sum_{I_v}
-\frac{\hbar}{2\mathrm{i}}c(A)\int
_{\triangle_{\delta(v)}}\mathrm{d}^{3}x~A_{a}^{i}\left(  \epsilon
^{ijk}\pounds _{\vec{N}_{j}^{I_{v}}}c_{k}^{a}+\pounds _{\vec{N}_{i}^{I_{v}}%
}c_{i}^{a}\right)
\ee
where  for the purposes of our heuristics we have replaced the quantum shift $N^a_i$,  which is strictly speaking non zero only at the point $x=v$ on the Cauchy slice $\Sigma$,
by  some regulated version thereof which 
is of small compact support $\Delta_{\delta}(v)$ of coordinate size $\delta^3$ about $v$ (in the coordinates we used to define the quantum shift).
Next, we approximate  the Lie derivative with respect to the quantum shift in $\Delta_{\delta}(v)$
by the difference of the {\em pushforward} action of a  small diffeomorphism and the identity as follows:
\begin{equation}
(\pounds _{\vec{N}_{i}^{I}}c_{j}^{a})A_{a}^{k}=-\frac{3}{4\pi}N(x(v))\nu
_{v}^{-2/3}
\frac{\varphi(q_{I_{v}}^{i}\vec{{\hat{e}}}_{I},\delta)^{\ast}%
c_{j}^{a}A_{a}^{k}-c_{j}^{a}A_{a}^{k}}{\delta}+O(\delta). \label{liee}%
\end{equation}
where we imagine extending the unit coordinate edge tangents $\vec{{\hat{e}}}_{I}$ to  $\Delta_{\delta}(v)$ in some smooth compactly supported way  and define
$\varphi(q_{I_{v}}^{i}\vec{{\hat{e}}}_{I},\delta)$ to be the finite diffeomorphism corresponding to translation by an affine amount $q_{I_{v}}^{i}\delta$  
along this edge tangent vector field.
Using (\ref{liee}), we obtain:
\begin{equation}
\hat{H}[N]c(A)=\frac{1}{\delta}\frac{\hbar}{2\mathrm{i}%
}c(A)\frac{3}{4\pi}\sum_v N(x(v))\nu_{v}^{-2/3}\sum_{I_v, i}\int_{\Sigma}%
\mathrm{d}^{3}x~\left[  \cdots\right]  _{\delta}^{I_{v},i}+O(\delta), 
\label{qoutsidep1}
\end{equation}
\begin{align}
\left[  \cdots\right]  _{\delta}^{I_{v},1}  &  =\left[  (\varphi c_{2}%
^{a})A_{a}^{3}-c_{2}^{a}A_{a}^{3}\right]  +\left[  (\varphi\bar{c}_{3}%
^{a})A_{a}^{2}-\bar{c}_{3}^{a}A_{a}^{2}\right]  +\left[  (\varphi c_{1}%
^{a})A_{a}^{1}-c_{1}^{a}A_{a}^{1}\right] \nonumber\\
\left[  \cdots\right]  _{\delta}^{I_{v},2}  &  =\left[  (\varphi c_{3}%
^{a})A_{a}^{1}-c_{3}^{a}A_{a}^{1}\right]  +\left[  (\varphi\bar{c}_{1}%
^{a})A_{a}^{3}-\bar{c}_{1}^{a}A_{a}^{3}\right]  +\left[  (\varphi c_{2}%
^{a})A_{a}^{2}-c_{2}^{a}A_{a}^{2}\right] \nonumber\\
\left[  \cdots\right]  _{\delta}^{I_{v},3}  &  =\left[  (\varphi c_{1}%
^{a})A_{a}^{2}-c_{1}^{a}A_{a}^{2}\right]  +\left[  (\varphi\bar{c}_{2}%
^{a})A_{a}^{1}-\bar{c}_{2}^{a}A_{a}^{1}\right]  +\left[  (\varphi c_{3}%
^{a})A_{a}^{3}-c_{3}^{a}A_{a}^{3}\right]  , \label{postlien3}%
\end{align}
where we have written ${\bar c}^a_i\equiv -c^a_i$ and where we have suppressed the  labels $I_v, i $ to set  $\varphi c_{j}^{a}\equiv\varphi(\vec{  q_{I_{v}}^{i} {\hat{e}}}_{I_{v}},\delta)^{\ast
}c_{j}^{a}$.
The   integral in (\ref{qoutsidep1}) is of order $\delta$ and we approximate it by its exponential minus the identity to get our final expression:
\be
\hat{H}[N]c(A) =\frac{\hbar}{2\mathrm{i}}c(A)\frac{3}{4\pi}\sum_v N(x(v))\nu
_{v}^{-2/3}\sum_{I_{v}}\sum_{i}\frac{\mathrm{e}^{\int_{\Sigma
}\left[  \cdots\right]  _{\delta}^{I_{v},i}}-1}{\delta}+O(\delta).\label{p1cnf1}
\ee
For each fixed $(I_v,i)$ the  exponential term is a  product of edge holonomies corresponding to the chargenet labels specified through  (\ref{postlien3}). This product
may be written as 
\be
h^{-1}_{c_{(i, flip})}h_{c_{(i, flip, I_v, \delta)}},
\label{holflipprod1}
\ee
where 
$c_{(i, flip, I_v, \delta)}$ is the deformation of $c_{(i, flip)}$ by $\varphi(q_{I_{v}}^{i}\vec{{\hat{e}}}_{I},\delta)$ and $c_{i, flip}$ has the same graph as $c$ but `flipped' charges. To see what these
charges are,  fix $i=1$ and some edge $I_v$ corresponding to the 
the first line of (\ref{postlien3}).  In $c_{(1,flip)}$, the  
connection $A_a^3$ corresponding to the 3rd copy of $U(1)$ is multiplied by the charge net $c_2^a$ corresponding to the second copy of $U(1)$. This implies that in
the holonomy $h_{c_{(1,flip)}}$ the charge label  in the 3rd copy of $U(1)$ for any edge is exactly the charge label in the second copy of $U(1)^3$ of the same edge in $c$ i.e. in obvious notation
$q^3|_{c_{(1,flip)}} = q^2|_{c}$ where we have suppressed the edge label. A similar analysis for all the remaining terms in (\ref{postlien3}) indicates that the  charges  $^{(i)}\!q^{j}, j=1,2,3$  on any edge of  $c_{(i,flip)}$
are given by the following `$i$- flipping' of the charges on the same edge of $c$.
\begin{equation}
\left.  ^{(i)}\!q^{j}\right.  =\delta^{ij}q^{j}-%
{\textstyle\sum\nolimits_{k}}
\epsilon^{ijk}q^{k} \label{defchrgeflip}%
\end{equation}
The exact nature of the deformed chargenet $c_{(i, flip, I_v, \delta)}$  depends on the definition of the deformation. Since the deformation is of compact support around $v$, 
the combination
$h^{-1}_{c_{(i, flip)}}h_{c_{(i, flip, I_v, \delta})}$ is the  identity except for a small region around $v$. 
From (\ref{p1cnf1}), this term multiplies $c(A)$. 
We call the resulting chargenet
as $c^{}_{(i,I_v,1,\delta )}$
\footnote{The `$1$' in the subscript refers to the `positive' $i$-flip (\ref{defchrgeflip}) as distinct from a `negative' $i$-flip which we shall 
encounter in (\ref{-defchrgeflip}) below.}
so that in terms of holonomies we have that:
\be
h_{ c^{}_{(i,I_v,1,\delta )}   } (A) = h^{-1}_{c_{(i, flip})}(A) h_{c_{(i, flip, I_v, \delta)}} (A) h_c (A)
\label{holflipprod2}
\ee

 Our final expression 
 for the discrete approximant to  the Hamiltonian constraint then reads:
\be
\hat{H}[N]_{\delta}c(A) =\frac{\hbar}{2\mathrm{i}}\frac{3}{4\pi}\sum_v N(x(v))\nu_{v}^{-2/3}\sum_{I_{v}}\sum_{i}
\frac{c_{(i,I_v,1,\delta)}- c}{\delta}
\label{hamfinalp1}
\ee
A similar analysis for the action of the electric diffeomorphism constraint yields the following counterpart of (\ref{p1cnf1}):
\be
\hat{D}[{\vec N}_i ]c(A) =\frac{\hbar}{\mathrm{i}}c(A)\frac{3}{4\pi}\sum_v N(x(v))\nu
_{v}^{-2/3}\sum_{I_{v}}\frac{\mathrm{e}^{\int_{\Sigma}
\varphi(q_{I_{v}}^{i}\vec{{\hat{e}}}_{I},\delta)^{\ast}%
c_{j}^{a}A_{a}^{j}-c_{j}^{a}A_{a}^{j}
}-1}{\delta}
+O(\delta).\label{p1dn1}
\ee
which then yields the final result
\be
\hat{D}_{\delta}[\vec{N}_{i}]c   =\frac{\hbar}{\mathrm{i}}\frac{3}{4\pi}%
\sum_v N(x(v))\nu_{v}^{-2/3}\sum_{I_{v}}\frac{1}{\delta
}(c^{}_{(I_{v},i,0, \delta)}-c)
\label{dnfinalp1}
\ee
where $c^{}_{(I_{v},i,0, \delta )}$ is obtained from $c$ only by a singular deformation without any charge flipping so that 
\begin{equation}
(c^{}_{({I_{v},i,0, \delta})})_{i}^{a}(x):=\varphi(q_{I_{v}}^{i}\vec{{\hat{e}}}_{I_v}%
,\delta)^{\ast}c_{i}^{a}(x).
\label{defd1cdef}
\end{equation}

It remains to specify the deformation $\varphi(q_{I_{v}}^{i}\vec{{\hat{e}}}_{I_v}, \delta)$. 
We do so in the next section. As we shall see, this deformation is visualised as an abrupt pulling of the vertex structure 
along the $I_v$th edge. Due to its `abruptness' we refer to this deformation as a {\em singular diffeomorphism}.
In this language, equations (\ref{hamfinalp1}), (\ref{dnfinalp1}) imply that whereas the action of the Hamiltonian constraint is a combination of charge flips
and singular diffeomorphisms, the action of the electric diffeomorphism constraints is exactly that of singular diffeomorphisms without any charge flips.

\subsection{\label{sec2.4}Linear Vertices, Upward and Downward Cones and Negative Charge flips} 

The following discussion implies that the detailed specification of the deformation $\varphi(q_{I_{v}}^{i}\vec{{\hat{e}}}_{I_v}, \delta)$ is
only needed for a special class of chargenet vertices which are called {\em linear} vertices. In this regard,
recall from section \ref{sec1}  that any state of interest is associated with a corresponding Ket Set and is built out of  linear combinations of charge net states
in this Ket Set. 
The action of the Hamiltonian and electric diffeomorphism constraints on any such state is then determined by their action on elements of the Ket Set.
Charge net elements of these  Ket Sets  are characterised by a certain  {\em linearity} property \cite{p3}. In order to define this linearity property recall 
from section \ref{sec2.3} that  the action of these constraints  on an element $c$ of the Ket Set requires the evaluation of  the quantum shift (\ref{qsev}) at  vertices of $c$
which have valence greater than 3 which, in turn, requires the choice of a coordinate patch around each such vertex.
Let $v$ be any $N$ valent ($N>3$) vertex of an element $c$
of a Ket Set. Then the following linearity property holds: there exists a small enough neighbourhood of $v$ such that the edges of $c$ at $v$ in this neighbourhood 
are {\em straight lines} with respect to the coordinate patch at $v$.  Such vertices are called {\em linear} with respect to the coordinate patches associated with them
and these coordinate patches are referred to as linear coordinate patches. Thus, for our purposes, it suffices to specify the deformations 
$\varphi(q_{I_{v}}^{i}\vec{{\hat{e}}}_{I_v}, \delta)$ for linear vertices.

Accordingly consider any such vertex of $c$. 
\footnote{The deformations described in this section are appropriate for  linear vertices sibject to a further restriction, namely that at such vertices 
no triple of edge tangents  is  linearly dependent.
In the language of \cite{p1,p2,p3} and of the next section, such vertices are called `GR' vertices.}
From the discussion above this deformation must distort the graph underling $c$ in the vicinity of its vertex $v$ in such a way 
that its vertex
is displaced by a coordinate distance $\delta$ along the  $I_v$th edge direction to leading order in $\delta$. Due to the vanishing of the  quantum shift everywhere 
except at $v$, 
this regulated deformation is visualised to
abruptly pull the vertex structure at $v$ along the $I_v$th edge.
Due to the `abrupt' pulling the original edges develop kinks signalling the point from which they are suddenly pulled. 
Since these kinks are points at which the edge tangents differ we call them $C^0$ kinks.
The final picture of the distortion is one in which the  displaced vertex 
lies along  the $I_v$th edge and is connected to the 
kinks on the remaining edges by edges which point  `almost' exactly opposite to the $I_v$th one.  The structure in the vicinity of the displaced vertex 
is  exactly that of a `downward' cone formed by these edges with axis along the $I_v$th one. For small enough $\delta$ the linear nature of the vertex
provides the necessary linear structure to define this conical deformation, with the cone getting stiffer  as the regulating parameter  $\delta$ decreases.

The downward conical structure is appropriate for vertex  displacement by 
$\varphi(q_{I_{v}}^{i}\vec{{\hat{e}}}_{I_v}, \delta)$ 
along the {\em outgoing} `upward' direction   $\vec{{\hat{e}}}_{I_v}$     along the $Iv$th edge which, in turn, 
is appropriate for positive $q_{I_{v}}^{i}$. For negative $q_{I_{v}}^{i}$, the displacement is downward along an {\em extension} of the $I_v$th edge
past $v$, with the remaining $N-1$ edges forming an {\em upward} cone around the cone axis along the $I_v$th edge.

Note also that we can equally well replace equation (\ref{liee}) through the judicuous placement of negative signs by:
\begin{equation}
(\pounds _{\vec{N}_{i}^{I}}c_{j}^{a})A_{a}^{k}=\frac{3}{4\pi}N(x(v))\nu
_{v}^{-2/3}\frac{\varphi(q_{I_{v}}^{i}(-\vec{{\hat{e}}}_{I}),\delta)^{\ast}%
c_{j}^{a}A_{a}^{k}-c_{j}^{a}A_{a}^{k}}{\delta}+O(\delta). \label{-liee}%
\end{equation}
This would then result in {\em upward} conical deformations for $q_{I_{v}}^{i} >0$
and downward ones for $q_{I_{v}}^{i}<0$.

A similar use of negative signs in equation (\ref{hamconst1}) offers a different starting point for our heuristics and leads to 
`negative' charge flips for the deformed charge nets generated by the Hamiltonian constraint approximant:
\begin{equation}
\left.  ^{(-i)}\!q^{j}\right.  =\delta^{ij}q^{j} +%
{\textstyle\sum\nolimits_{k}}
\epsilon^{ijk}q^{k} \label{-defchrgeflip}%
\end{equation}
We denote the negative $i$-flipped child of the parent $c$ by 
\be
c_{(i, I_v, -1,\delta)},
\label{ci-1}
\ee
 the `$-1$' denoting the negative flip (\ref{-defchrgeflip}).

To summarise: Using the parent-child language of section \ref{sec1}, we have that (a) legitimate approximants to the Hamiltonian and electric diffeomorphism constraints 
generate both upward and downward conically deformed children irrespective of the sign of the edge charge labels and (b) legitimate approximants to the 
Hamiltonian constraint generate both  positive and negative flipped charge net children.
While our notation for deformed children will not reflect the choice of upward and downward deformation (which we shall specify explicitly
as and when required), our notation will reflect the choice of positive or negative charge flip as follows. We have already denoted
a `positive flipped child  by $c^{}_{(i,I_v,1,\delta )}$ with `1' signifying a positive $i$-flip. We shall denote a negative flipped child by
$c^{}_{(i,I_v,-1,\delta )}$ with the `$-1$' signifying a negative flip. As in section \ref{sec2.1} we shall denote the holonomies 
associated with these states by $h_{c^{}_{(i,I_v,1,\delta )} }, h_{c^{}_{(i,I_v,-1,\delta )}}$.

Next, we note  that 
it turns out \cite{p3} that, by virtue of the linearity of the conical deformation in the vicinity of the displaced vertex,  the displaced vertices
are also linear. In addition, it also turns out that the constraint approximants (\ref{hamfinalp1}), (\ref{dnfinalp1})  preserve  a second set of properties called GR and CGR
properties of the 
vertex structure. We describe these properties in the next section.

\subsection{\label{sec2.5}GR and CGR vertices}
A linear GR vertex is defined as a linear vertex  which has valence greater than 3 and at which 
no triple of edge tangents is linearly dependent. 
A linear vertex $v$ of a charge net $c$  will be said to be linear CGR if:\\
(i) The union of 2 of the edges at $v$ form a single straight line so that $v$ splits this  straight line into 2 parts\\
(ii) The set of remaining edges  together with any one of the two edges in (i) constitute a GR vertex in the following sense. Consider, at $v$, the set of out going edge tangents
to each of the remaining edges together with the outgoing edge tangent to one of the two edges in (i). Then any triple of elements of this set is linearly dependent.\\
In (i) and (ii) above, the notion of straight line is with respect to the linear regulating coordinate system associated with the linear vertex $v$.
The edges in (i) are called conducting edges, the line in (i) is called the conducting line and the remaining edges are called  non-conducting edges.

It is straightforward to see that the conical deformations generated by the Hamiltonian and electric diffeomorphism constraint approximants
on parental vertices which are GR  result in displaced vertices in the children
which are either GR or CGR. While the  displaced vertices in the children generated by the electric diffeomorphism constraints preserve the GR  nature of the parental vertex, 
those generated  by the Hamiltonian constraint, depending on the charge labellings and the
edge along which the deformation is generated,  could be GR or CGR.

If the parental vertex is CGR, the conical deformations are, in general,  constructed slightly differently from those encountered in section \ref{sec2.4}.
However in the specific case that the deformation at a CGR vertex  is along a conducting  edge, this deformation 
is identical to that described in section \ref{sec2.4}, in that the non-conducting edges lie on a cone with axis along the conducting line.
In the main body of this work
\footnote{A minor exception is the state depicted in Fig \ref{1dgraph}; however we do not discuss its deformations in any detail.}
we shall only encounter parental vertices which are GR or CGR, and in the latter case, will only encounter 
 deformations along conducting edges.  We depict these deformations in Figures \ref{gr} and \ref{cgrk=i}.
More in detail, 
Figure  \ref{gr}  depicts downward conical  deformations of a GR vertex. 
For simplicity of depiction we have chosen the 
valence to be $N=4$. Figure \ref{cgrk=i} shows a deformation along the set of collinear `conducting'  edges at a CGR vertex. 

\begin{figure}
  \begin{subfigure}[h]{0.3\textwidth}
    \includegraphics[width=\textwidth]{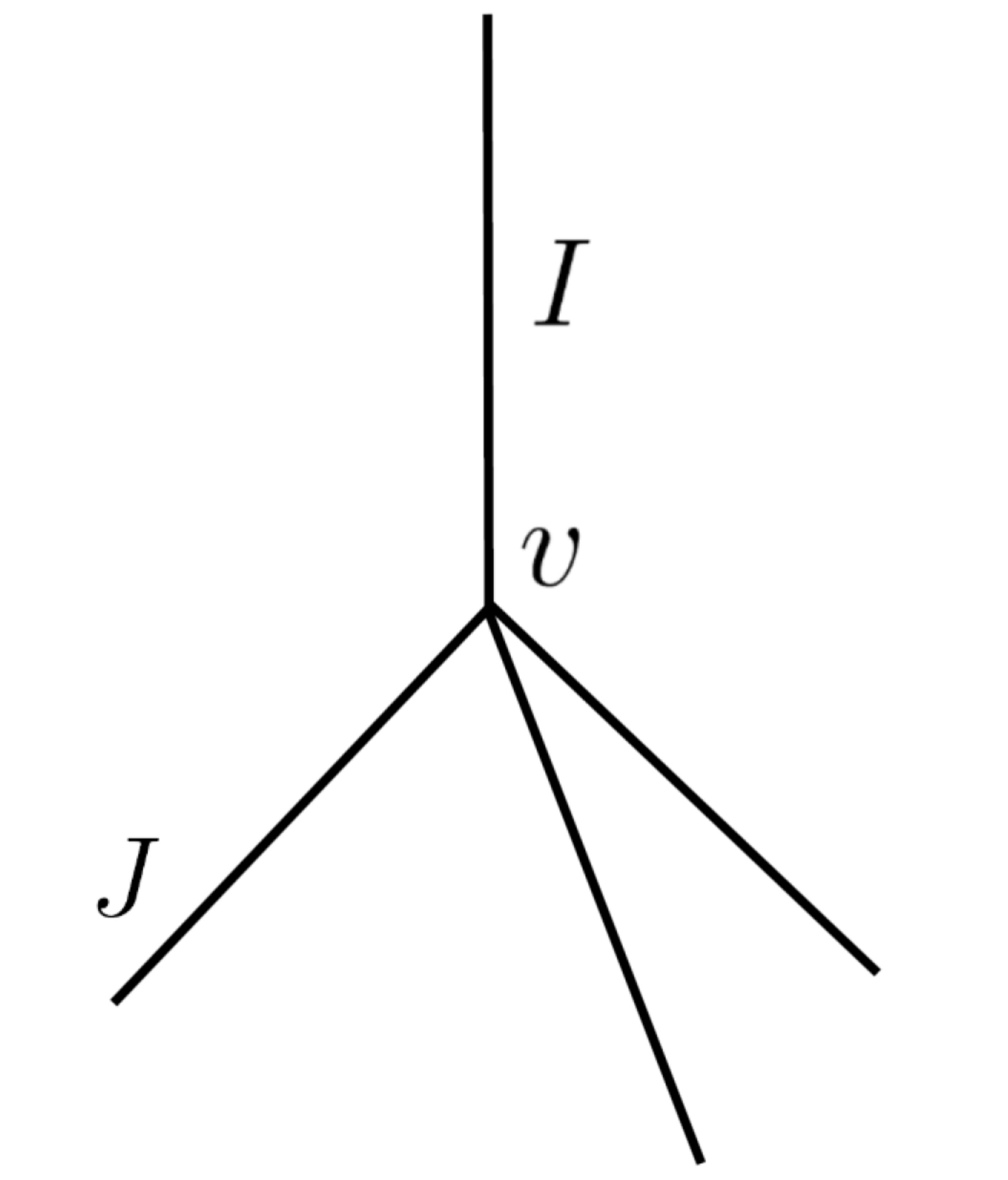}
    \caption{}
 \label{gra}
  \end{subfigure}
  \begin{subfigure}[h]{0.3\textwidth}
    \includegraphics[width=\textwidth]{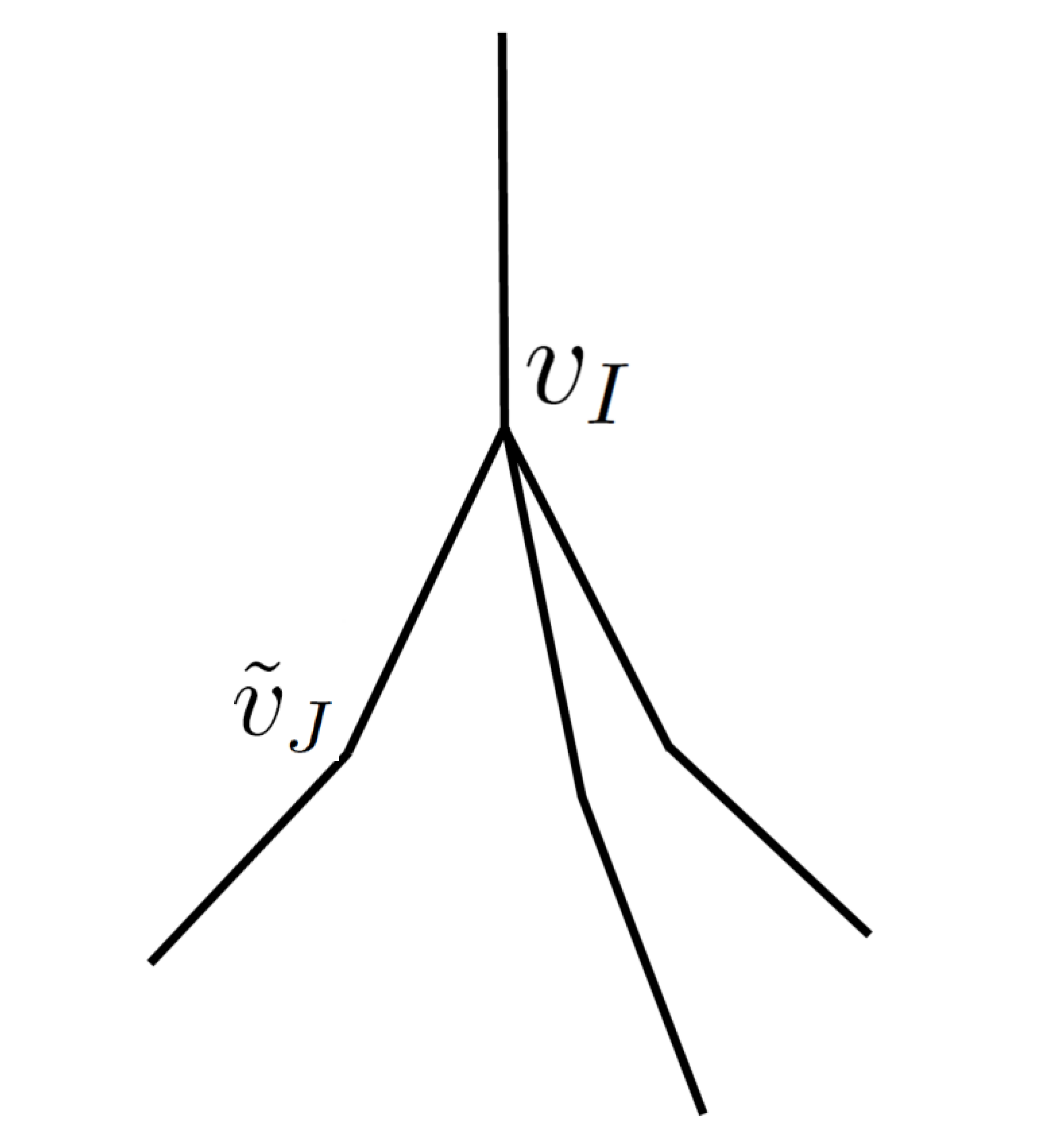}
    \caption{}
   \label{grb}
  \end{subfigure}
\begin{subfigure}[h]{0.3\textwidth}
    \includegraphics[width=\textwidth]{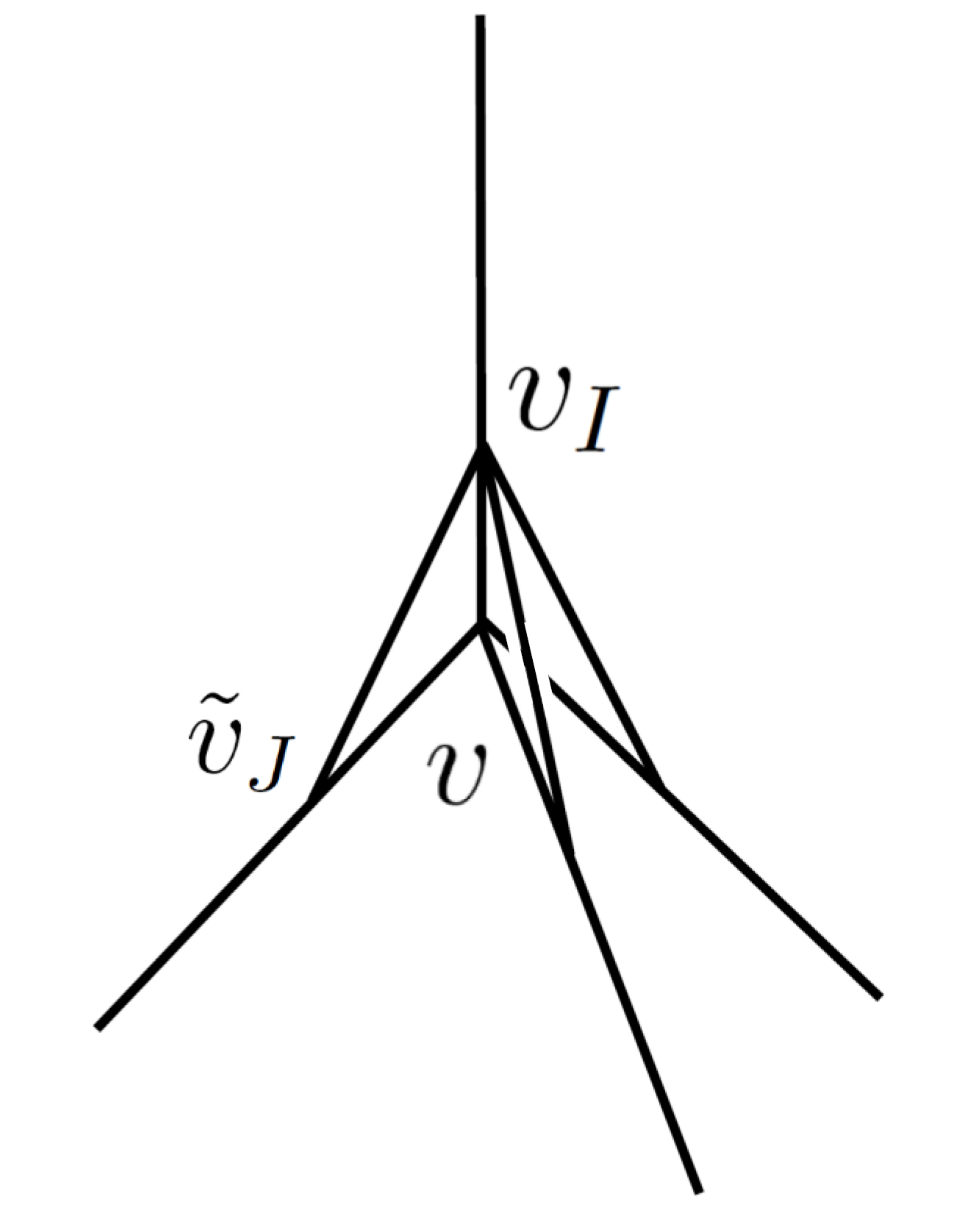}
    \caption{}
   \label{grc}
  \end{subfigure}
  \caption{ Fig \ref{gra} shows an undeformed GR vertex $v$ of a chargenet $c$  with its $I$th and $J$th edges as labelled. The vertex is deformed along its $I$th edge in Fig \ref{grb} wherein the displaced
vertex $v_I$ and the $C^0$ kink, ${\tilde v}_J$ on the $J$th edge are labelled. Fig \ref{grc} shows the result of a Hamiltonian type deformation 
obtained by multiplying the chargenet holonomies obtained by 
coloring  the edges of Fig \ref{grb}  by 
flipped images of charges  on their counterparts in $c$ ,  Fig \ref{gra}  by negative of these flipped charges and Fig \ref{gra} by the charges on $c$. If the edges of Fig \ref{grb} are colored by the charges on 
their counterparts in $c$ then one obtains an electric diffemorphism deformation.
}%
\label{gr}%
\end{figure}

\begin{figure}
  \begin{subfigure}[h]{0.3\textwidth}
    \includegraphics[width=\textwidth]{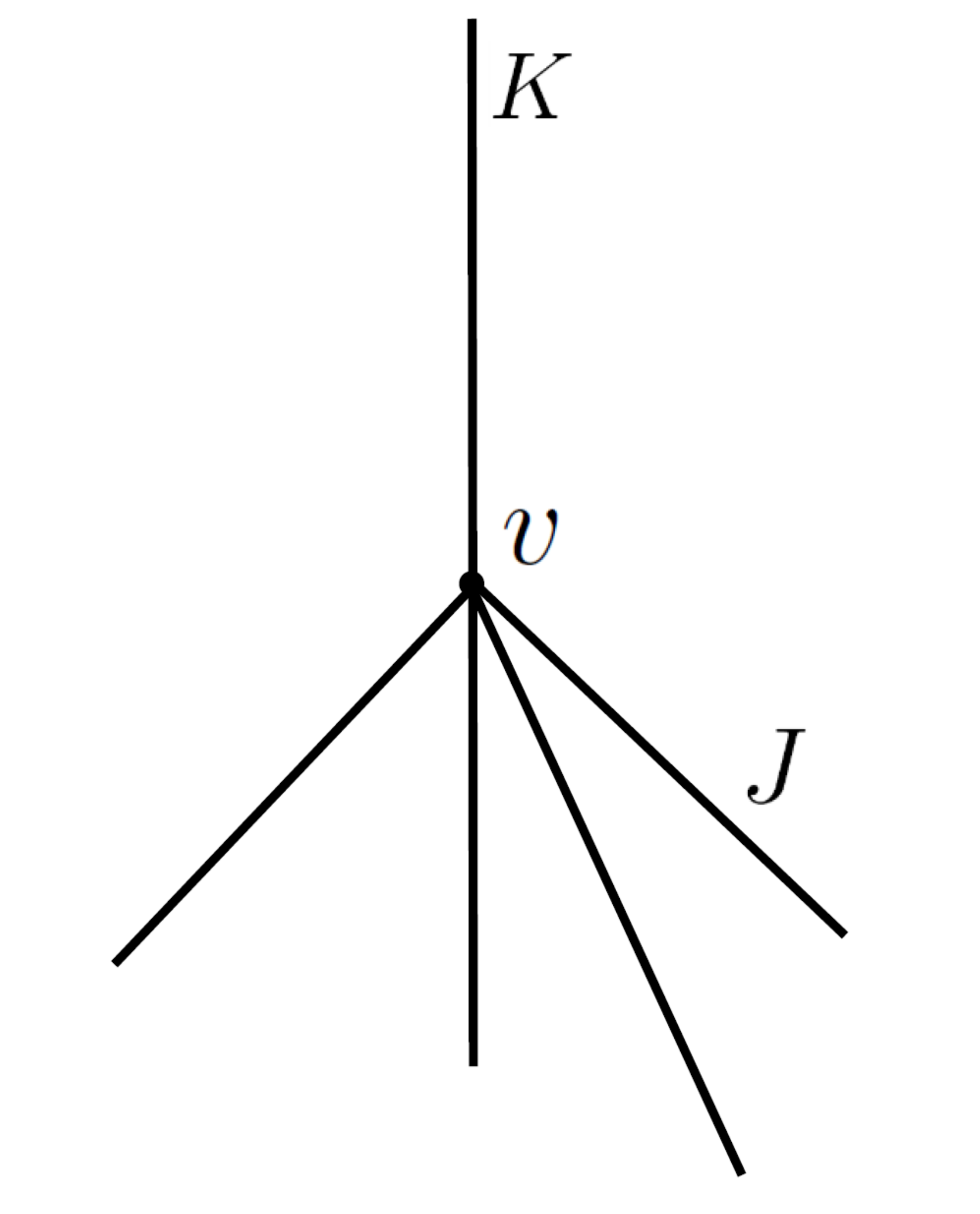}
    \caption{}
    \label{cgrk=ia}
  \end{subfigure}
  \begin{subfigure}[h]{0.3\textwidth}
    \includegraphics[width=\textwidth]{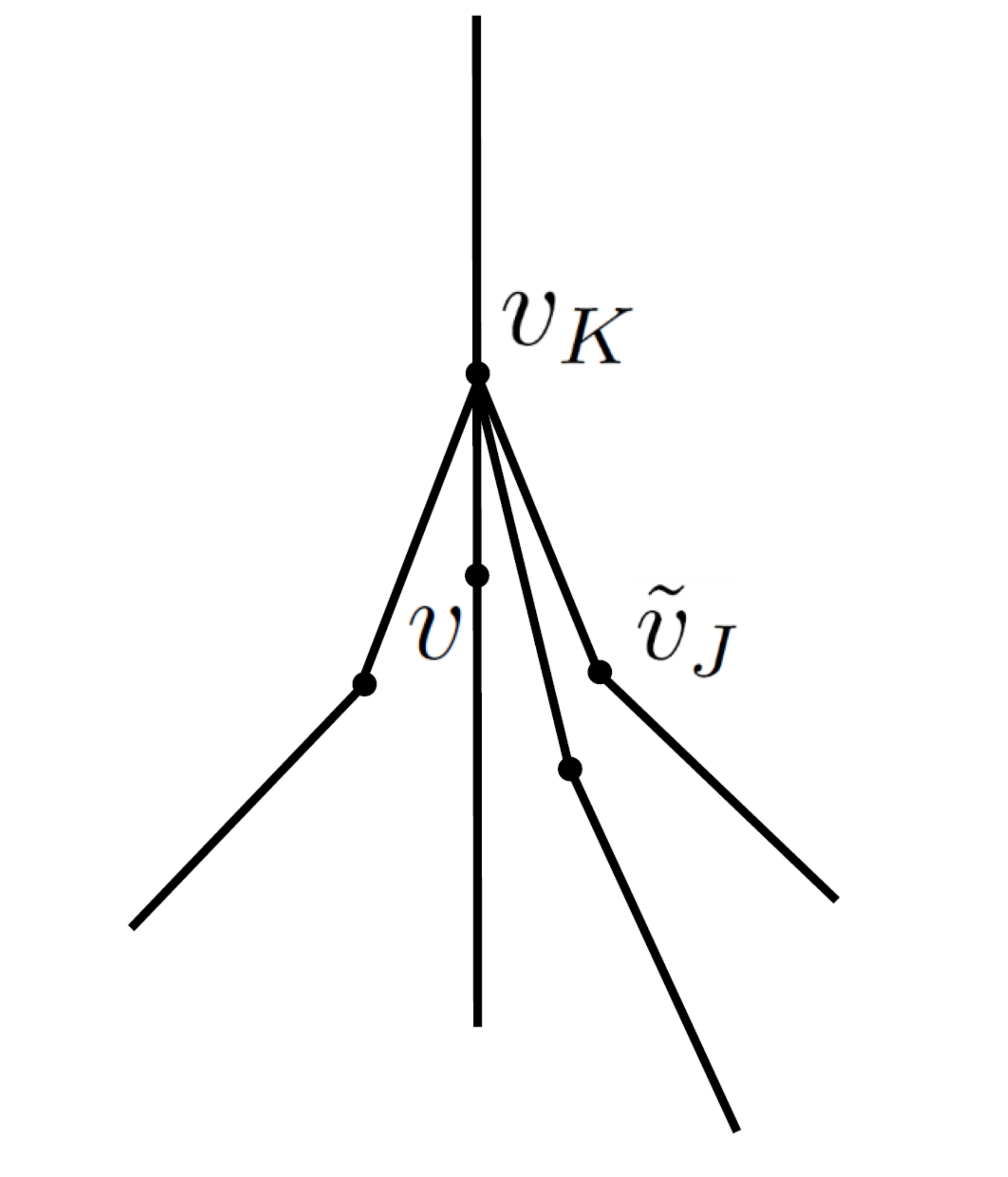}
    \caption{}
    \label{cgrk=ib}
  \end{subfigure}
\begin{subfigure}[h]{0.3\textwidth}
    \includegraphics[width=\textwidth]{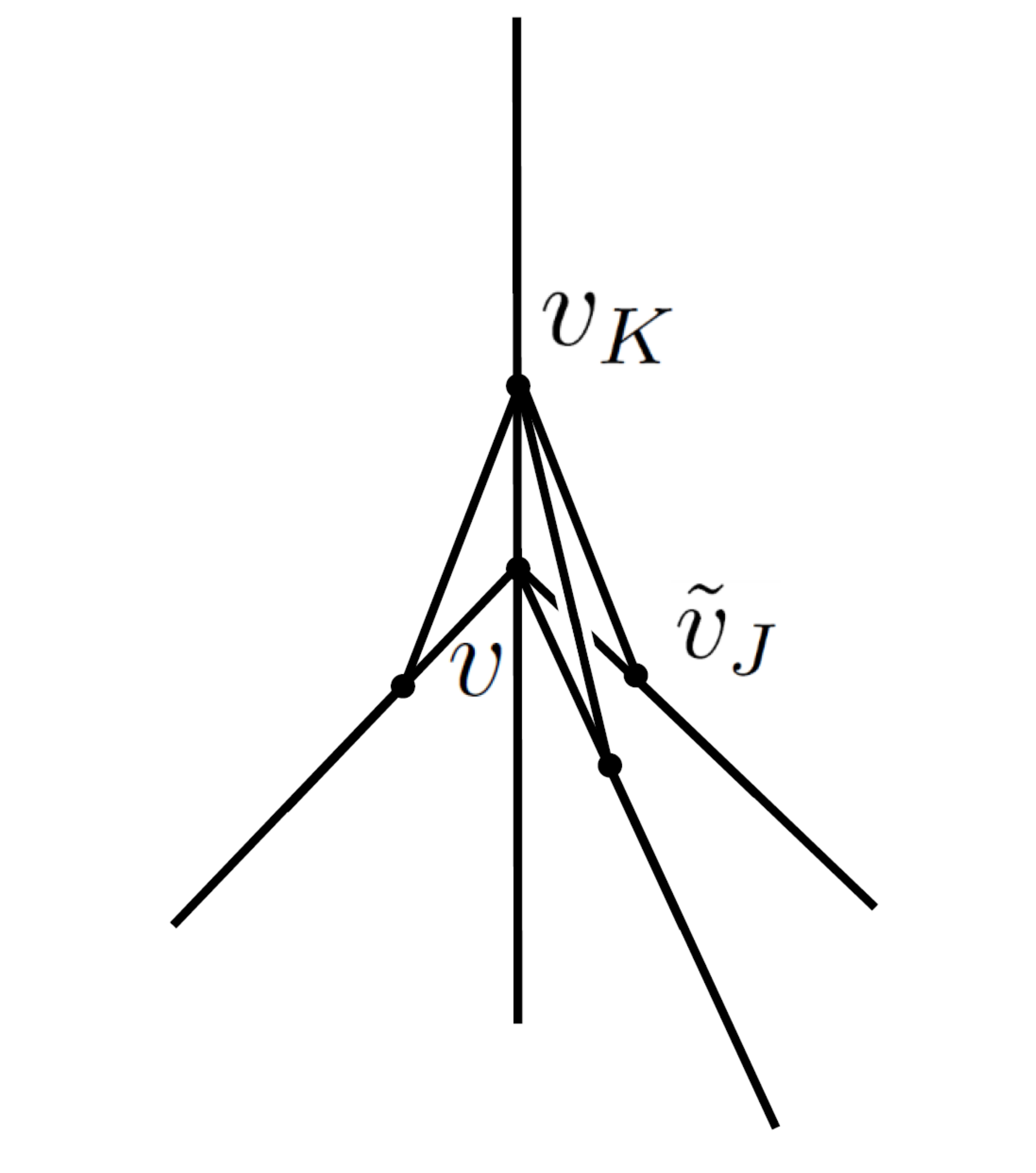}
    \caption{}
    \label{cgrk=ic}
  \end{subfigure}
  \caption{ In  Fig \ref{cgrk=ib}  the vertex  structure of Fig \ref{cgrk=ia} is deformed
along its $K$th edge  and the displaced
vertex $v_K$ and the $C^0$ kink ${\tilde v}_J$ on the $J$th edge are as labelled. Fig \ref{cgrk=ic} shows the result of a Hamiltonian type deformation  obtained by multiplying the 3 chargenet holonomies obtained by 
coloring  the edges of Fig \ref{cgrk=ib} by the  flipped images of the charges  on their counterparts in $c$ ,  the edges of Fig \ref{cgrk=ia} by the negative of these  flipped charges and the edges of Fig \ref{cgrk=ia} 
by the charges on $c$. If the edges of Fig \ref{cgrk=ib} are colored by the charges on 
their counterparts in $c$ then one obtains an electric diffemorphism deformation.
}%
\label{cgrk=i}%
\end{figure}

%

In our review hitherto we have skipped certain technicalities and, more importantly,  {\em extrapolated} some of the results and structures  of \cite{p3} in a manner
plausible to us. We comment on these matters in the next section.

\subsection{\label{sec2.7}Technical caveats to our hitherto broad exposition}

\noindent (1) {\em Interventions, upward directions and $C^1,C^2$ kinks:} 
The singular diffeomorphisms which are responsible for the deformations underlying 
equations (\ref{hamfinalp1}), (\ref{dnfinalp1}) are of the form  $\varphi(q_{I_{v}}^{i}\vec{{\hat{e}}}_{I_v}, \delta)$. Here  $\varphi(q_{I_{v}}^{i}\vec{{\hat{e}}}_{I_v}, \delta)$
approximates the Lie derivative
with respect to the quantum shift through equation (\ref{liee}). 
As sketched above one may choose to approximate the Lie derivative of the quantum shift  through equation (\ref{-liee}) in which case 
(\ref{hamfinalp1}), (\ref{dnfinalp1})  would be modifed by appropriate negative signs. 
In both cases the `upward' direction for the cone was defined to be along the outward point edge tangent $\vec{{\hat{e}}}_{I_v}$.
We note here that this is {\em not} how we proceeded
in \cite{p3}. 


There we assigned an upward direction to each edge at the vertex $v$ of $c$ based on the graph topology of $c$. 
This direction was either outward or inward pointing. 
In the latter case,   we first multiplied the parental charge net by judiciously chosen
small loop holonomies which we called {\em interventions}. The loops were chosen so that (a) these interventions  were classically equal to unity to  $O(\delta^2)$ and (b) the action of the
intervening holonomies modified the parental vertex structure so as to replace 
the edges which were assigned inward pointing directions  by edges whose outward pointing directions at $v$ coincided with the assigned upward directions (c) the interventions
resulted in the conversion of CGR vertices to GR ones. 
The resulting modified parent state was then acted upon by (\ref{hamfinalp1}), (\ref{dnfinalp1}) and then multiplied by the inverse of the intervening holonomies.
\footnote{As discussed in \cite{p3} the displaced vertex in the resulting child is either CGR or GR.}
As a result the deformations of the parental state were dictated by the assigned upward directions rather than the outward pointing parental tangents and no extra negative
signs were introduced. In addition certain ``$C^1$ and $C^2$'' kinks were placed on the parental edge (and/or its extension) along which the child vertex was displaced so 
as to serve as markers for the choice of upward direction \cite{p3}. The demonstration
 of anomaly free action in \cite{p3} was based on this complicated choice of approximant and off-shell and physical states were constructed from  Ket Sets
 satisfying property (a) with respect to these choices of approximants.

As we shall discuss further in section \ref{sec5}, we believe that it is possible to repeat the demonstration of anomaly free action by interpreting the choice of upward direction
as a regulating choice rather than as being fixed once and for all by the graph topology as in \cite{p3}. 
In other words, we may specify the choice of 
upward directions at the parental vertex being acted upon by a product of constraint operators as inward or outward for each edge freely. 
We shall {\em assume} that with this freedom of  choice, we will still be able to provide a demonstration of anomaly free constraint action along lines similar to 
that in \cite{p3} albeit without the introduction of the $C^1$ and $C^2$ kinks referred to above. The Ket Set satisfying property (a) appropriate to 
the incorporation of this freedom of choice then contains conically deformed children for cone axes which may be chosen along or opposite to the outward pointing 
parental edges at any vertex $v$ of the parent  {\em independent} of the sign of the parental edge charges $q_{I_{v}}^{i}$.
\footnote{Thus, the Ket Sets considered here differ from those of \cite{p3} in that (a) there is no placement  of $C^1,C^2$ kinks in children (b) children which arise from 
{\em both} directions of conical deformations of parental vertices 
irrespective of the signs of parental edge charges are in the Ket Set rather than  children which arise only from uniquely prescribed  choices of  these directions as in \cite{p3}; 
in this sense the Ket Sets here are slightly larger than those of \cite{p3}.}


\noindent (2) {\em Multivertex states:}
The detailed demonstration of anomaly freedom in \cite{p3} is in the context of Ket Sets with elements which have {\em only 
a single vertex} where the Hamiltonian and electric diffeomorphism constraints act non-trivially. However the notion of propagation between vertices
can only be formulated for {\em multivertex}  charge nets. Note that the action of the constraints as derived in section \ref{sec2.1}
at one vertex is {\em independent} of the action at a distinct vertex. Hence the action of the constraints derived in \cite{p3}
can be easily generalised to multivertex charge nets and the sum over $v$ in (\ref{hamfinalp1}), (\ref{dnfinalp1}) 
constitutes exactly this generalization. It is then necessary to also generalise the detailed demonstration of anomaly freedom in  \cite{p3} to the  case of Ket Sets
satisfying property (a)
whose elements have multiple vertices on which  constraint approximants act non-trivially in accordance with this generalization.
While such a demonstration is outside the scope of this work, its existence  does seem plausible to us and we shall {\em assume} this existence for the 
considerations in this paper. We comment on this matter further in section \ref{sec5}.




\subsection{\label{sec2.8} A key  structural property of constraint actions of interest}


The structural property of constraint approximants ${\hat C}_{\delta}$ connected with property (a) of section \ref{sec1} and alluded to in that section is that 
any such approximant takes the following form:
\be
{\hat C}_{\delta} |s\ket  =\sum_{v}\sum_{  {\rm deformation}, v}a_{{\rm deformation},v}\frac{{\hat O}_{{\rm deformation},v} -{\bf 1}}{\delta} |s\ket .
\label{prop0}
\ee
Here $|s\ket$ is (the appropriate counterpart of) a spin net state. The operator ${\hat O}_{{\rm deformation},v}$ is a kinematically well defined operator
which deforms the vertex structure  of the `parent' $|s\ket$  in a $\delta$ coordinate sized vicinity of   its vertex $v$ in a specific way and yields a deformed `child' spin net
${\hat O}_{{\rm deformation},v}|s\ket$, and $a_{{\rm deformation},v}$ is  a non-zero complex coefficient. The sums are over different deformations at each vertex and 
then over all vertices. 

We now show that this form implies that the state obtained as the sum, with unit coefficients,  over all elements of any Ket Set which satisfies property (a) is an anomaly free physical state.
More precisely, since the sum is kinematically non-normalizable, it is more appropriate to define the state as a sum over {\em bra} correspondents of elements of the 
Ket Set. Such a state $\Psi$ lies in the algebraic dual space of complex linear mappings on the finite span of (the appropriate analog of) spin network states.
The constraints operators act through dual action on such a state. 
We show below that such a state is an anomaly free physical state with respect to the dual action of the constraint operators of the form (\ref{prop0}).

The constraint  approximants act by dual action on such a state as follows:
\be
\Psi ({\hat C}_{\delta} |s\ket ) =  \Psi (\sum_{v}
\sum_{   {\rm deformation}, v } a_{   {\rm deformation},v  }
\frac{   {\hat O}_{  {\rm deformation},v } -{\bf 1}   }{\delta} |s\ket)  
\label{prop1}
\ee
and their continuum limit action is defined as
\be
\lim_{\delta \rightarrow  0} \Psi ({\hat C}_{\delta} |s\ket ) = \lim_{\delta \rightarrow 0} \Psi (\sum_{v}
\sum_{   {\rm deformation}, v } a_{   {\rm deformation},v  }
\frac{   {\hat O}_{  {\rm deformation},v } -{\bf 1}   }{\delta} |s\ket)  
\label{prop2}
\ee
We show that the contribution of each term in the sum vanishes i.e. we show that 
\be
\Psi (|s\ket) = \Psi ({\hat O}_{{\rm deformation},v}|s\ket)  \forall |s\ket
\label{prop3}
\ee
First let $|s\ket$ lie in the complement of the Ket Set. Then the left hand side (lhs) of (\ref{prop3}) vanishes. The right hand side (rhs) involves the action of $\Psi$
on a child of $|s\ket$. 
This vanishes by virtue of property (a2) of the Ket Set, for if it did not vanish, that would imply the existence of a possible parent $|s\ket$ of the child 
${\hat O}_{{\rm deformation},v}|s\ket$  such that this possible parent is {\em not} in the Ket Set even though its child is.
Next let $|s\ket$ be in the Ket Set. Then the lhs is  equal to $1$ because $\Psi$ is a superposition of (bra correspondents of) elements of the Ket Set with {\em unit} coefficients.
The rhs is then also equal to $1$ by virtue of property (a1). Thus $\Psi$ is in the kernel of the electric diffeomorphism and Hamiltonian constraint operators.
Finally, $\Psi$ is diffeomorphism invariant by virtue of property (a3). 
Since the diffeomorphism invariant  state $\Psi$ is killed by the Hamiltonian constraint, $\Psi$ is a physical state. Since it is also killed by the electric diffeomorphism
constraint, constraint commutators consistently trivialise and the state is also anomaly free.
This completes the proof.

To summarise: The constraint approximants considered  in this work will all have the structure (\ref{prop0}).
For any Ket Set  which satisfies  property (a) with respect to these constraint approximants, the state obtained by summing over elements of this Ket Set with 
unit coefficients
is a physical state i.e. it is a diffeomorphism invariant state annihilated by these constraint approximants, and, hence, by their continuum limits.
Such a state also supports  trivial anomaly free constraint commutators. 
As discussed in section \ref{sec1} (see also \cite{proppft}), whether a physical state based on a specific Ket Set supports propagation
depends crucially on the nature of the `possible parents' of property (a2). 
In contrast to the single vertex Ket Sets considered in \cite{p3}, the Ket Sets
considered in this work are based on multi-vertex kets because the very notion of propagation as that between vertices is defined only for the multivertex case.

As  mentioned in section \ref{sec1}, whether off-shell deformations of these multivertex physical states can be 
constructed in a manner similar to the single vertex case of \cite{p3} 
so as to  support non-trivial anomaly free commutators is a question which  is outside the scope of the work in this paper. 
We return to this point in section \ref{sec5}.

\section{\label{sec3} Insufficient Propagation}
In  section \ref{sec3.1}, in order to illustrate the various structures involved in our discussion of propagation in section \ref{sec1}, we 
study these structures in the context of a simple example, namely  
that of a  2 vertex charge network state, each vertex having the same valence. 
In section \ref{sec3.1a} we consider a perturbation created by the action of a single Hamiltonian constraint on this state.
We show that the minimal Ket Set, consistent with the constraint actions of \cite{p3}, which contains this state does {\em not} encode propagation of this
perturbation.  In section \ref{sec3.1b} we enlarge this Ket Set by requiring that the physical state it defines be subject to additional 
conditions. We consider a specific perturbation of the  simple 2 vertex state created by the action of  an operator associated with these additional conditons. 
We show that the enlarged Ket Set {\em does} encode propagation of this perturbation from one vertex of the state to the other.
Besides their pedagogic value in illustrating our articulation of propagation in terms of Ket Sets, the considerations of section \ref{sec3.1b}
display an intriguing  connection with the existence of a certain elegant combination of constraints in Reference \cite{aanewpersp} (see Footnote \ref{fnaa} in this regard).


In section \ref{sec3.2} we investigate propagation between vertices of different  valence.
Specifically, we show that 
the $N\rightarrow N$ constraint actions of \cite{p3} are {\em inconsistent} with propagation between vertices of different valence of a multivertex state, and that, at best these actions
may engender `1d' propagation between vertices of special multivertex states.
The arguments in section \ref{sec3.2} are simple and robust and the reader mainly interested in the motivation for the $N\rightarrow 4$ modification
may skip the slightly more involved considerations of section \ref{sec3.1}.

\subsection{\label{sec3.1} Propagation in a simple 2 vertex state}
Consider the simple case of a $U(1)^3$ gauge invariant parent charge net state $p$ with 2 $N$-valent vertices connected by $N$ edges  as depicted in Fig \ref{2N1a}.
Let the set of $U(1)^3$ edge charges be ${\vec q}_J, J=1,..,N$ with ${\vec q}_J= (q_J^1,q_J^2,q_I^3)\in U(1)^3$.
Consider the `generic' case where $p$ has no symmetries and none of its charge components vanishes so that:
\be
q^i_J  \neq 0 \;\;i=1,2,3 \;\;\; J=1,..,N.
\label{generic}
\ee
 We are interested in establishing propagation or the lack thereof of specific perturbations
between vertices $A$ and $B$ in the context of  $N\rightarrow N$ constraint actions.

\subsubsection{\label{sec3.1a} No propagation}
Here we consider  the smallest Ket Set subject to property (a) which contains $p$.
A single $N\rightarrow N$ action of the Hamiltonian constraint on this parent state yields various children in this Ket Set. We focus on the child 
$c$ obtained by deforming the vertex structure at $A$  along its $I$th edge as shown in Fig \ref{2N1b}.  The deformation renders the original
parent vertex at $A$ degenerate so that it has vanishing volume eigen value and creates the new non-degenerate vertex $v_I$.
The vertex $v_I$ is connected in $c$ to the $C^0$ kinks $\{{\tilde v}_{J\neq I}\}$ on the parental edges $\{e_{J\neq I}\}$ 
by the deformed counterparts of the latter as show in Figure \ref{2N1b}.

\begin{figure}
  \begin{subfigure}[h]{0.3\textwidth}
    \includegraphics[width=\textwidth]{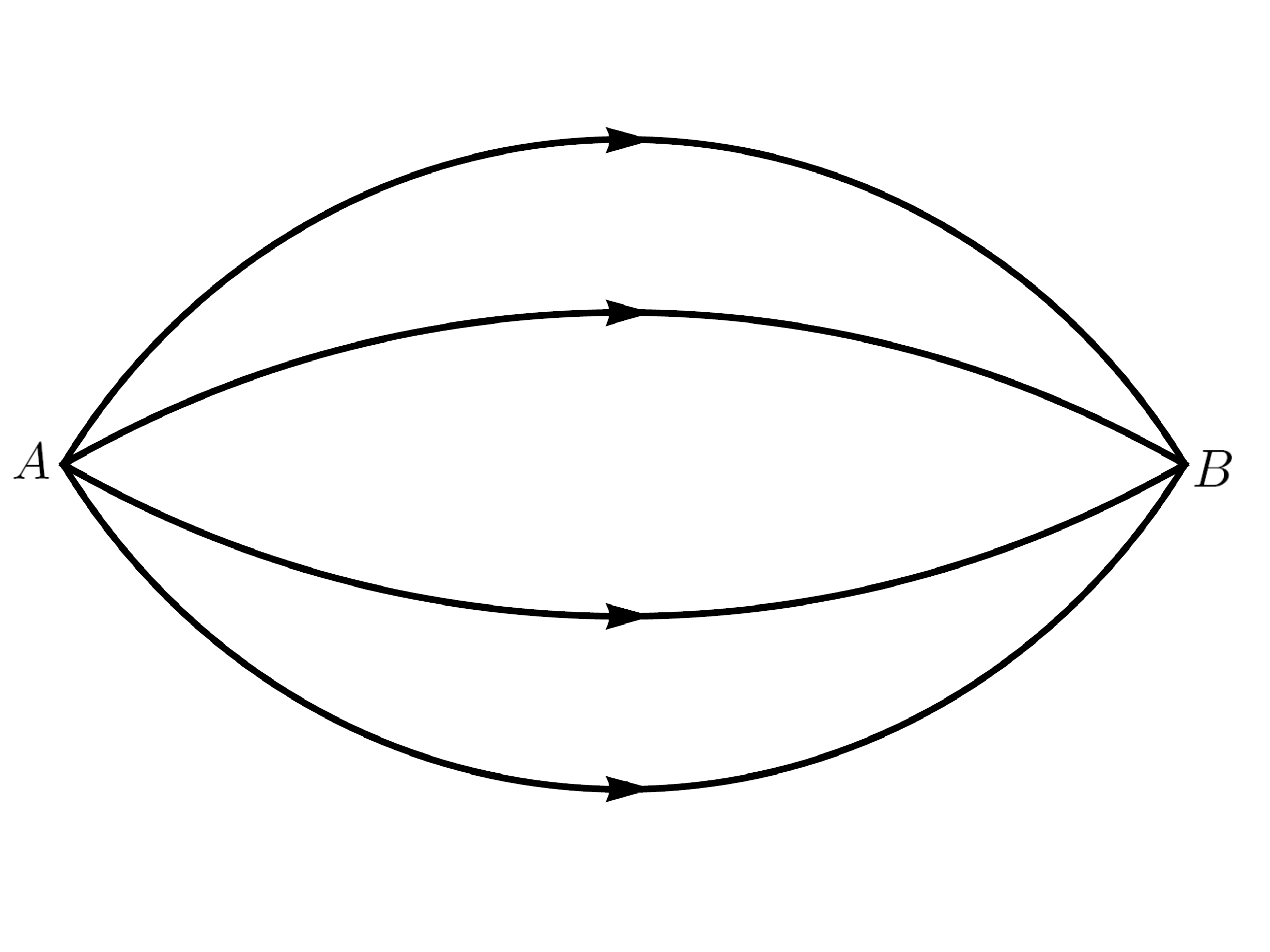}
    \caption{}
 \label{2N1a}
  \end{subfigure}
  \begin{subfigure}[h]{0.3\textwidth}
  \hspace{15mm}
  \includegraphics[width=\textwidth]{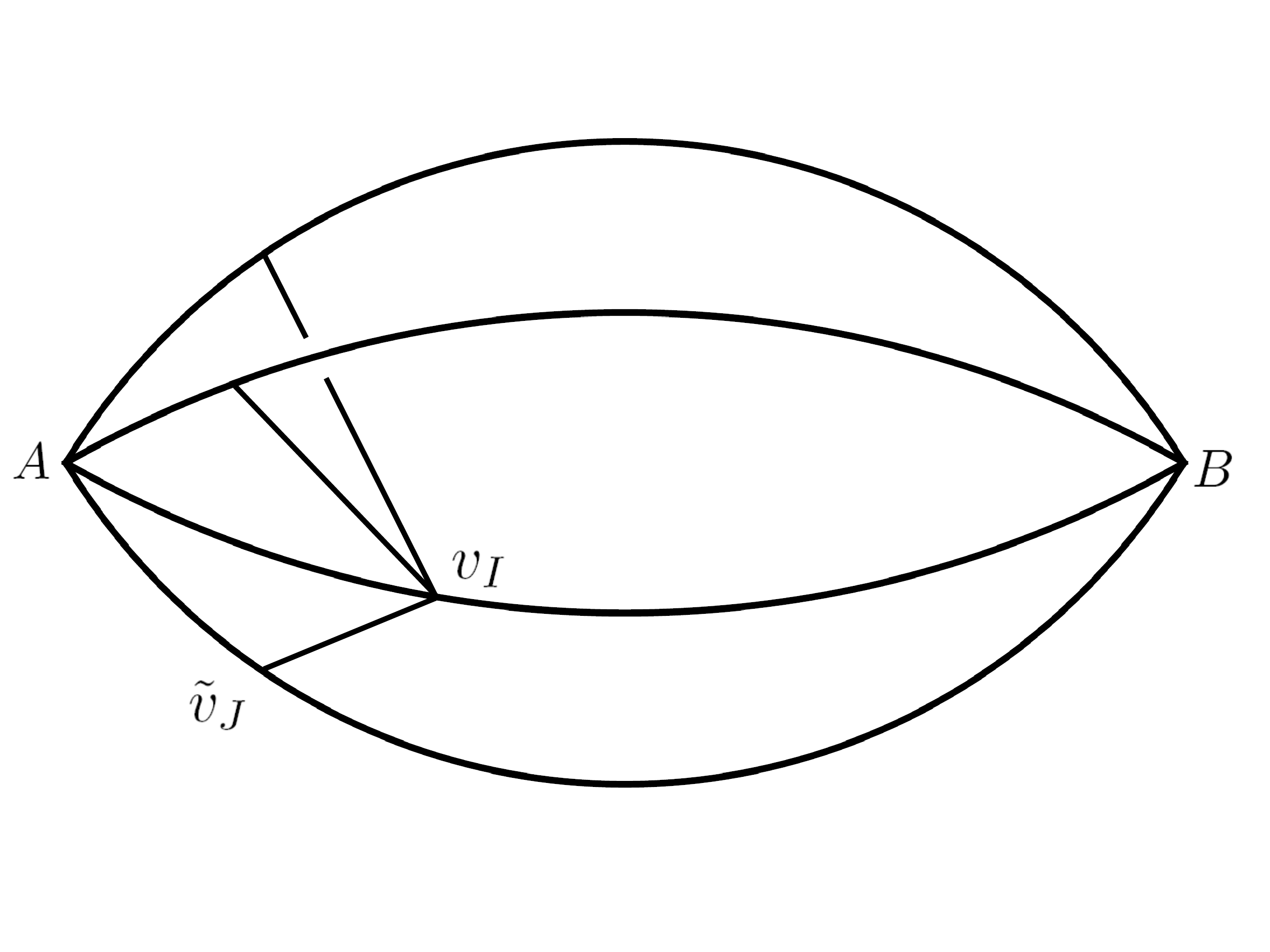}
    \caption{}
   \label{2N1b}
  \end{subfigure}
\caption{ Fig \ref{2N1a} shows the simple 2 vertex chargenet of interest. Fig \ref{2N1b} shows the result of a Hamiltonian constraint deformation
of  the chargenet of Fig \ref{2N1a} along its $I$th edge with the displaced vertex $v_I$ and the $C^0$ kink ${\tilde v}_{J}$ as labelled. 
}%
\label{f2N}%
\end{figure}

We are interested in the existence of other possible' parents of $c$ in the Ket Set. 
By a possible parent we mean a chargenet $p^{\prime}$ whose deformation by at least one
$N\rightarrow N$ constraint action yields $c$ upto diffeomorphisms. 
By `other' we that $p^{\prime}$ is  not diffeomorphic to $p$.
Thus, we are interested in the existence of  $p^{\prime}$ not diffeomorphic to $p$
such that $c$ is generated from $p^{\prime}$ by any combination of at least one Hamiltonian or  electric diffeomorphism constraint deformation,
ordinary diffeomorphisms and, possibly, further Hamiltonian/electric diffeomorphism deformations.

Since $c$ contains only one set of {\em trivalent}  kinks, it can only be generated (upto diffeomorphisms) by a {\em single Hamiltonian  constraint}
action on a state $p^{\prime}$ with 2 $N$ valent vertices and no such kinks so that 
$c= \phi^{\prime}\; {\widehat {\rm Ham}}\; \phi \; p^{\prime}$ where ${\widehat {\rm Ham}}$ refers to a Hamiltonian constraint deformation. Clearly by redefining $p^{\prime}$
appropriately we may set $\phi$ equal to the identity with no loss of generality. Let the vertices of $p^{\prime}$ be $A^{\prime}, B^{\prime}$ and let 
${\widehat {\rm Ham}}$ act at $A^{\prime}$ to yield $c^{\prime}$ so that  we have that $\phi^{\prime}\; c^{\prime}= c$.

Next, denoting the flipped charges on the deformed edges in $c$ by the subscript `$flip$' we have the following:\\
(a)By virtue of the genericity condition (\ref{generic}) on the charge labels of $p$, it follows straightforwardly that 
 the charges $q^k_J- (q_{flip})^k_{J}$ on the segments between $A$ and ${\tilde v}_J$ are non vanishing, thus implying that these segments are present in $c$.
\\
(b) the parental graph can be immediately reconstructed from that of $c$ simply by removing the deformed edges in $c$ from $v_I$ to each ${\tilde v}_J$.\\
(c) the parental vertex whose deformation yields $c$ can be identified uniquely as $A$ by virtue of $A$ being degenerate in $c$.\\
(d) the parental edge charges $q^k_{J\neq I}$ in $p$ can be uniquely identified with the charges $q^k_{J\neq I}$ on the segments in $c$ from 
${\tilde v}_J$ to $B$ and the parental edge charges on the $I$th edge in $p$ can be idnetified with those on the edge from $v_I$ to $B$ in $c$.\\
From (a)-(d) , $p$ can be uniquely reconstructed from $c$. 

Since $\phi^{\prime}$ preserves kink structure, vertex degeneracy, and colorings, it immediately follows that $\phi^{\prime} p^{\prime}=p$. 
Hence this example illustrates the lack of propagation of this particular perturbation. 

\subsubsection{\label{sec3.1b} Propagation from an additional condition}

As discussed in sections \ref{sec1} and \ref{sec2.8}, anomaly free physical states are annhilated by the diffeomorphism,  Hamiltonian and electric diffeomorphism 
constraints. Here, we demand that these states be further annihilated by certain operator implementations of  the linear combinations
\be
H_{\pm}(N) =(\pm H(N))  + \frac{1}{2}\sum_{i=1}^3 D({\vec N}_i) \label{hpm}
\ee
of the Hamiltonian and electric diffeomorphism constraints. 
\footnote{\label{fnaa}As mentioned earlier these combinations are reminiscent of the 
elegant combinations of the diffeomorphism and Hamiltonian constraints for Lorentzian gravity constructed in \cite{aanewpersp}. 
These combinations in that work obtain an elegant form when expressed in terms of 
spinors. The trace part of the combination yields the Hamiltonian constraint and the trace free part yields 
electric diffeomorphism
constraints smeared with an additional electric field. 
For details see (see vi), pg 85, Chapter 6 of \cite{aanewpersp}). }
If the operators ${\hat H}_{\pm} (N)$ are regulated simply as sums of the
regulated versions (\ref{hamfinalp1}) and (\ref{dnfinalp1}) of the individual Hamiltonian and electric diffeomorphism constraints, this
condition is already satisfied by anomaly free physical states by virtue of their being annihilated by the individual constraints.
Here we do not regulate ${\hat H}_{\pm} (N)$ in this trivial way. Instead we proceed as follows.

From equations (\ref{p1cnf1}) and (\ref{p1dn1}), it immediately follows that
\ba
\hat{H}_{+}[N]c(A) &=&\frac{\hbar}{2\mathrm{i}}c(A)\frac{3}{4\pi}\sum_v N(x(v))\nu
_{v}^{-2/3} 
%
 \sum_{I_{v}}\sum_{i} \frac{\mathrm{e}^{\int_{\Sigma
}\left[  \cdots\right]  _{\delta}^{I_{v},i} +
\varphi(q_{I_{v}}^{i}\vec{{\hat{e}}}_{I},\delta)^{\ast}%
c_{j}^{a}A_{a}^{j}-c_{j}^{a}A_{a}^{j}
}
-1}{\delta}+O(\delta)\;\;\;\;\;\;\; \label{hp1}\\
&=&\frac{\hbar}{2\mathrm{i}}\frac{3}{4\pi}\sum_v N(x(v))\nu
_{v}^{-2/3} 
%
 \sum_{I_{v}}\sum_{i}  \frac{\mathrm{e}^{\int_{\Sigma
}\left[  \cdots\right]  _{\delta}^{I_{v},i} +
\varphi(q_{I_{v}}^{i}\vec{{\hat{e}}}_{I},\delta)^{\ast}%
c_{j}^{a}A_{a}^{j}}
-c(A)}{\delta}+O(\delta) \label{hp2}\\
&=&\frac{\hbar}{2\mathrm{i}}\frac{3}{4\pi}\sum_v N(x(v))\nu
_{v}^{-2/3} 
%
 \sum_{I_{v}}\sum_{i}  \frac{
c_{(i,I_v,+,\delta)}(A)
-c(A)}{\delta}+O(\delta) \label{hp3}
\ea
where in the second line we used that $c(A)= \exp \int_{\Sigma} c^{a}_i A_a^i$.
and in third we defined the deformed state $c_{(i,I_v,+,\delta)}(A)$ as
\be
c_{(i,I_v,+,\delta)}(A):=
\mathrm{e}^{\int_{\Sigma
}\left[  \cdots\right]  _{\delta}^{I_{v},i} +
\varphi(q_{I_{v}}^{i}\vec{{\hat{e}}}_{I},\delta)^{\ast}%
c_{j}^{a}A_{a}^{j}} .
\label{hp4}
\ee
In the notation of equations 
(\ref{hamfinalp1}), (\ref{dnfinalp1}), we have that:
\be
\hat{H}_+[N]_{\delta}c(A) =\frac{\hbar}{2\mathrm{i}}\frac{3}{4\pi}\sum_v N(x(v))\nu_{v}^{-2/3}\sum_{I_{v}}\sum_{i}
\frac{c_{(i,I_v,+,\delta)}- c}{\delta}
\label{hpfinal}
\ee
It is straightforward to check that in the notation developed in the beginning of section \ref{sec2.1}, the holonomy 
underlying the deformed state 
$c_{(i,I_v,+,\delta)}$ is obtained as the product of the holonomy corresponding to an $i$-flipped child generated by the Hamiltonian constraint
and the holonomy corresponding to an electric diffeomorphism child as follows:
\be
h_{c_{(i,I_v,+,\delta)}}(A) = 
(h^{-1}_{c_{i,flip}}(A) h_{c_{i,flip,I_v,\delta}}(A))h_{c_{(i,I_v,0,\delta)}}(A)
=(h_{c_{(i,I_v,1,\delta)}}(A) h^{-1}_c(A))
h_{c_{(i,I_v,0,\delta)}}(A).
\label{holprod+}
\ee
Here, from  (\ref{holflipprod1}) the term in brackets in the first equality corresponds to 
the $\mathrm{e}^{\int_{\Sigma
}\left[  \cdots\right]  _{\delta}^{I_{v},i}}$ contribution to equation (\ref{hp4}), and the second equality follows from (\ref{holflipprod2}).

It is also straightforward to check that if the charge flip (\ref{defchrgeflip})
underlying the term $\left[  \cdots\right]  _{\delta}^{I_{v},i}$ in 
(\ref{p1cnf1})
is replaced by the negative charge flip (\ref{-defchrgeflip}), then the line of argumentation which leads to (\ref{hp1})-(\ref{hp3})
yields the following regulated action of ${\hat H}_-(N)$:
\be
\hat{H}_-[N]_{\delta}c(A) =\frac{\hbar}{2\mathrm{i}}\frac{3}{4\pi}\sum_v N(x(v))\nu_{v}^{-2/3}\sum_{I_{v}}\sum_{i}
\frac{c_{(i,I_v,-,\delta)}- c}{\delta}.
\label{hmfinal}
\ee
The holonomy underlying the deformed state 
$c_{(i,I_v,-,\delta)}$ is given by the product:
\be
h_{c_{(i,I_v,-,\delta)}}(A) = (h_{c_{(i,I_v,-1,\delta)}}(A)h^{-1}_c (A)) 
h_{c_{(i,I_v,0,\delta)}}(A),
\label{holprod-}
\ee
where  we have used the notation $c_{(i,I_v,-1,\delta)}$  as in (\ref{ci-1}).
The discussion of section \ref{sec2.4}  may then be repeated in the context of 
the deformed children generated by ${\hat H}_{\pm}(N)$. It follows that legitimate approximants to these
operators can be constructed so as to generate both upward and downward conically deformed children irrespective
of the sign of the edge charge labels. 
Figure \ref{fighpm} depicts a downward conical  deformation of a parental GR vertex by these operators.
The holonomy underlying the deformed child is  obtained as the product of holonomies based on the graphs  depicted in 
the figure. The graphs involved are the same irrespective of whether the child is generated by ${\hat H}_+(N)$ or ${\hat H}_-(N)$;
however the colorings in the two cases differ and are 
as described in the figure caption. 

\begin{figure}
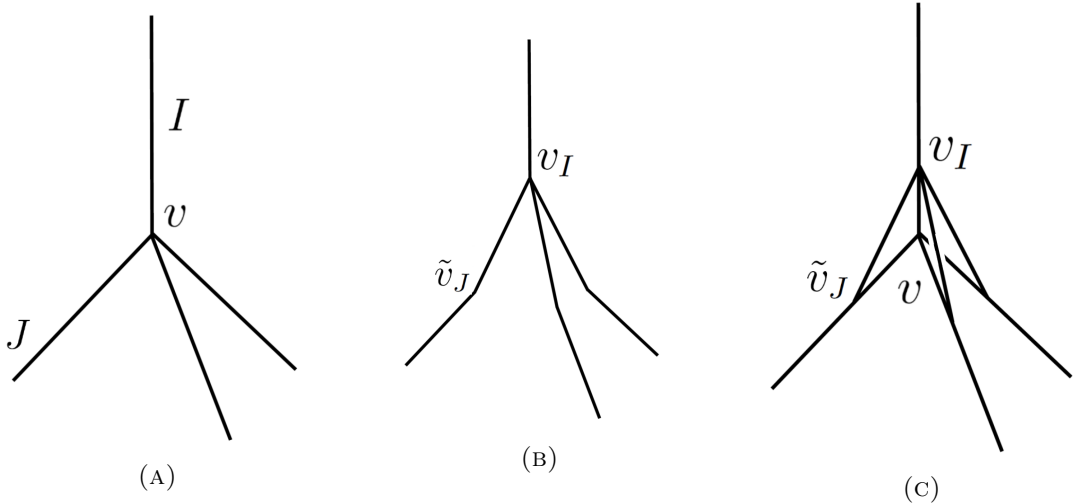

  \begin{subfigure}[h]{0.3\textwidth}
    \includegraphics[width=\textwidth]{F1-1.pdf}
    \caption{}
 \label{grpma}
  \end{subfigure}
  \begin{subfigure}[h]{0.3\textwidth}
    \includegraphics[width=\textwidth]{F1-2.pdf}
    \caption{}
   \label{grpmb}
  \end{subfigure}
\begin{subfigure}[h]{0.3\textwidth}
    \includegraphics[width=\textwidth]{F1-3.pdf}
    \caption{}
   \label{grpmc}
  \end{subfigure}
  \caption{ Fig \ref{grpma} shows an undeformed GR vertex $v$ of a chargenet $c$  with its $I$th and $J$th edges as labelled. 
  The vertex is deformed along its $I$th edge in Fig \ref{grpmb} wherein the displaced
vertex $v_I$ and the $C^0$ kink, ${\tilde v}_J$ on the $J$th edge are labelled. Fig \ref{grpmc} shows the result of an $H_{\pm}$ type deformation 
obtained by multiplying the chargenet holonomies obtained by 
coloring  the edges of Fig \ref{grpmb}  by 
flipped images of charges  on their counterparts in $c$ ,  Fig \ref{grpma}  by negative of these flipped charges and Fig \ref{grpmb} by the charges on $c$. 
If the flip is positive, the deformation is generated by $H_{+}$ and if negative, by $H_-$. As result the charges on this deformed chargenet on the
deformed edges are the {\em sum} of the $\pm$-flipped and unflipped charges and the segments from $v$ to the kinks ${\tilde v}_J$
carry the negative of the $\pm$-flipped images of their charges in $c$. 
}%
\label{fighpm}%
\end{figure}

Next, the discussion of section \ref{sec2.8} can be applied to
equations (\ref{hpfinal}) and (\ref{hmfinal}) to conclude the following. The Ket Set appropriate to these equations
contains all possible upward and downward deformed children generated by the action of ${\hat H}_{\pm}(N)$
on any parent in the Ket Set as well all possible parents of any child in the Ket Set.
The state obtained by summing over all elements of this Ket Set is killed by the actions (\ref{hpfinal}), (\ref{hmfinal}).

Since the requirement that  ${\hat H}_{\pm}(N)$ annihilated states of interest is imposed in addition to the demand that such 
states be anomaly free physical states with respect to the Hamiltonian and diffeomorphism (and Gauss Law) constraints,
the Ket Set of interest satisfies the closure properties described in the previous paragraph and also satisfies property (a)
as articulated in section \ref{sec1}. We shall use the properties described in the previous paragraph together with property (a3)
(i.e. the closure of the Ket Set with respect to diffeomorphisms) to show that the minimal Ket Set containing the simple 2 vertex state
of section \ref{sec3.1a} does encode propagation of a perturbation created by the action of ${\hat H}_+(N)$ at one of its vertices.
In what follows, for notational convenience we rename this simple 2 vertex state (called $p$ hitherto) as $c$.

Our argumentation is primarily diagrammatical and described through Figure \ref{fsr} as follows:\\
\noindent (1)We start at the left with the simple 2 vertex charge net $c$ with $N$ valent vertices  $A,B$ connected through $N$ edges in (A).
The outgoing charge on the $K$th edge emanating from vertex $A$ is denoted by $q^k_K$.  
\\

\noindent (2) 
This parent chargenet is deformed by the action of ${\hat H}_+(N)$ at the vertex
$A$ to give the child $c_{(i,I_A,+,\delta)}$ shown  in (B). 
Since the vertex A is GR, the deformation is of the type depicted
in Figure \ref{grpmc}.  As in that figure, the index $J$ will be used for edges which are different from the $I$th one.
The charges on the child may be inferred from (\ref{holprod+}). 
Denoting the $k$th component of the outgoing charge label from a vertex $v$ to a vertex ${\bar v}$ by $q^k_{v{\bar v}}$ 
it is straightforward to infer that:
\ba
q^k_{A{\tilde v}_J}= -(q_{flip})^k_J  & \;\;\;\;\;\;\;q^k_{{\tilde v}_J B}= q^k_J \;\;\;\;\;\;\; q^k_{v_I{\tilde v}_J}= q^k_J + (q_{flip})^k_J \nonumber \\
q^k_{Av_I}= -(q_{flip})^k_I  & q^k_{{v}_I B}= q^k_I 
\label{b12}
\ea
Here by $q_{flip}^k$ we mean the  positive $i$-flip (\ref{defchrgeflip}).
\footnote{The vertex $v_I$ is CGR (see section \ref{sec2.7}). Equations (\ref{b12})  imply that the net outgoing charges at this vertex are 
$q_K^k+ (q_{flip})_K^k, K=1,..,N$. We assume that the charges $q^k_K$ are such that the CGR vertex is non-degenerate. For the definition of 
non-degeneracy of a CGR vertex, see \cite{p3}.}
\\

\noindent (3) The charge net of (B) is acted upon by a seminanalytic diffeomorphism so as to `drag' the deformation from the 
vicinity of vertex A to the vicinity of vertex B.
\footnote{\label{fndiff0}We assume that the state $c_{(i,I,+1,\delta)}$ is such that it can be transformed via an appropriate diffeomorphism
 to the state depicted in (C). We shall comment further on this in section \ref{sec5a}.}
We slightly abuse notation and denote the images of $v_I, {\tilde v}_J$ by this diffeomorphism by the same symbols $v_I,{\tilde v}_J$.
\\

\noindent (4) The charge net of (C ) is deformed by the action of an appropriate electric diffeomorphism at $v_I$ to yield the charge net of (D).
This transforms the conical deformation in (C)  which is  downward with respect to the $I$th line from $A$ to $B$  to one which is
upward conical with respect to this line in (D). As a result, the deformation is now {\em downward} conical with respect to the (oppositely oriented)
line from $B$ to $A$. 
\footnote{For a downward deformation of a CGR vertex see Fig \ref{cgrk=ib}. An upward deformation may be visualised
by turning figures \ref{cgrk=ia}, \ref{cgrk=ib} upside down; see \cite{p3} and figures therein for details.
}
\\

\noindent (5) The charge net $c^{\prime}$ of (F) has the same graph as that of the charge net $c$ but its charges from $B$ to $A$ are different 
from those of $c$. Denoting these charges by $q^{\prime k}_I$, these charges are related to those on $c$ by
\be
q^{\prime k}_I = (q_{flip})^k_I .
\label{e}
\ee
Thus the outgoing charges from $B$ in $c^{\prime}$  are just the positive $i$-flipped images of the {\em incoming} charges at $B$ in $c$.
This state will play the role of a `possible parent'.
\\

\noindent (6) The charge net $c^{\prime}$ of (F) is deformed by the action of ${\hat H}_-(N)$ at the vertex B. The deformed child
$c^{\prime}_{(i,I_B,-,\delta)}$ is depicted in (E). Once again we have abused notation and re-used the symbols $v_I, {\tilde v}_J$.
The charges on this state can be  inferred from (\ref{holprod-}) and (\ref{e}). Using the fact that a negative $i$-flip is the inverse of a positive $i$-flip,
these charges turn out to be identical to their counterparts in (B):
\ba
q^k_{A{\tilde v}_J}= -(q_{flip})^k_J  & q^k_{{\tilde v}_J B}= q^k_J \;\;\;\;\;\;\; q^k_{v_I{\tilde v}_J}= q^k_J + (q_{flip})^k_J \\
q^k_{Av_I}= -(q_{flip})^k_I  & q^k_{{v}_I B}= q^k_I. 
\label{b}
\ea

\noindent (7) The chargenet $c^{\prime}_{(i,I_B,-1,\delta)}$ of (E) is deformed by the action of an appropriate electric diffeomorphism
to give exactly the chargenet of (D).
\footnote{The positions of the points ${\tilde v}_J$ in (E) and (D) should be identical. For reasons of visual clarity, these figures do not reflect this fact.}
\\

\begin{figure}
  \begin{subfigure}[h]{0.3\textwidth}
    \includegraphics[width=\textwidth]{G1.pdf}
    \caption{}
 \label{fsr1}
  \end{subfigure}
  \begin{subfigure}[h]{0.3\textwidth}
    \includegraphics[width=\textwidth]{G2.pdf}
    \caption{}
   \label{fsr2}
  \end{subfigure}
\begin{subfigure}[h]{0.27\textwidth}
    \includegraphics[width=\textwidth]{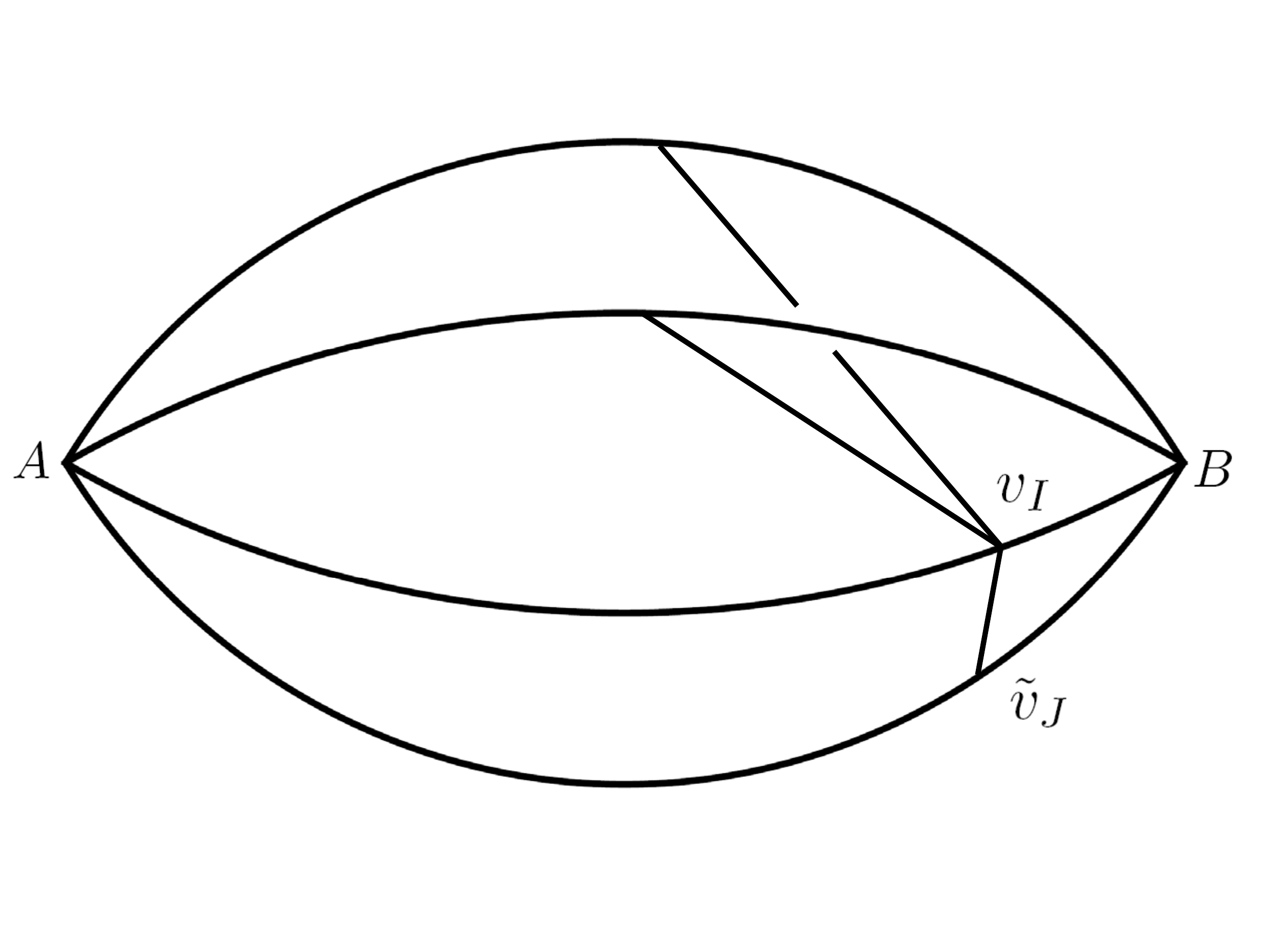}
    \caption{}
   \label{fsr3}
  \end{subfigure}
\begin{subfigure}[h]{0.3\textwidth}
    \includegraphics[width=\textwidth]{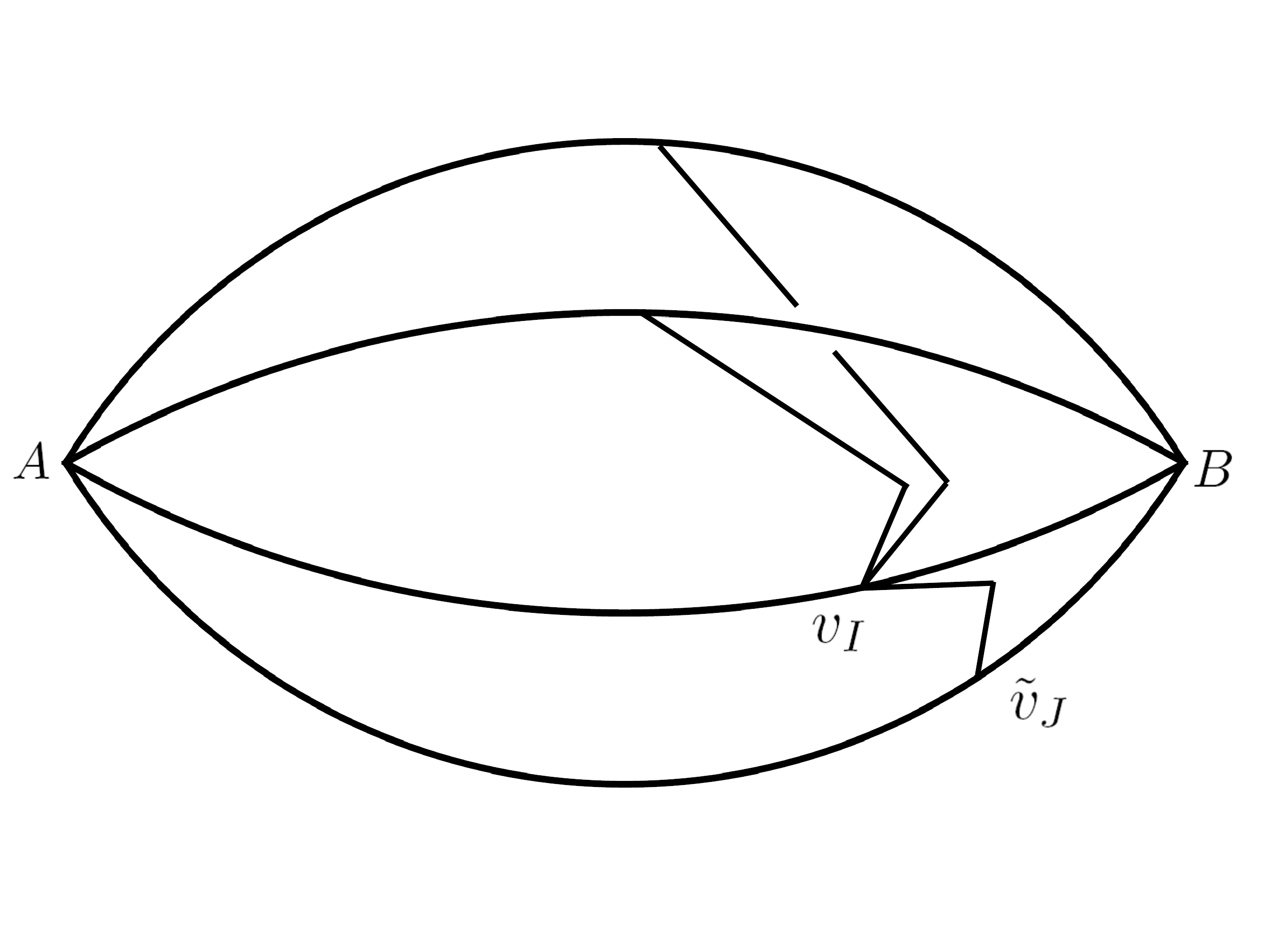}
    \caption{}
   \label{fsr4}
  \end{subfigure}  
\begin{subfigure}[h]{0.3\textwidth}
    \hspace*{7mm}
    \includegraphics[width=\textwidth]{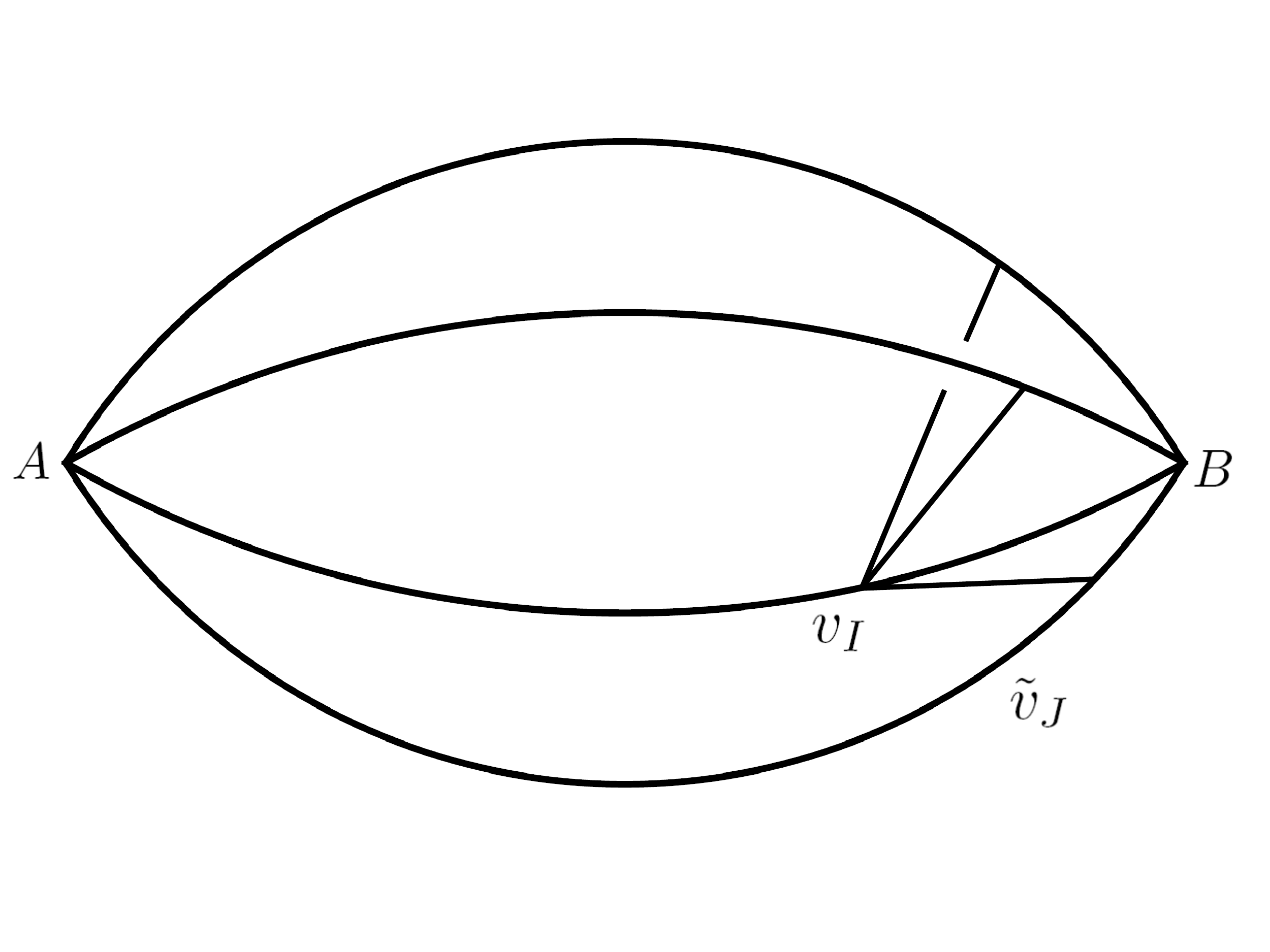}
    \caption{}
   \label{fsr5}
  \end{subfigure}  
\begin{subfigure}[h]{0.3\textwidth}
    \hspace*{12mm}
    \includegraphics[width=\textwidth]{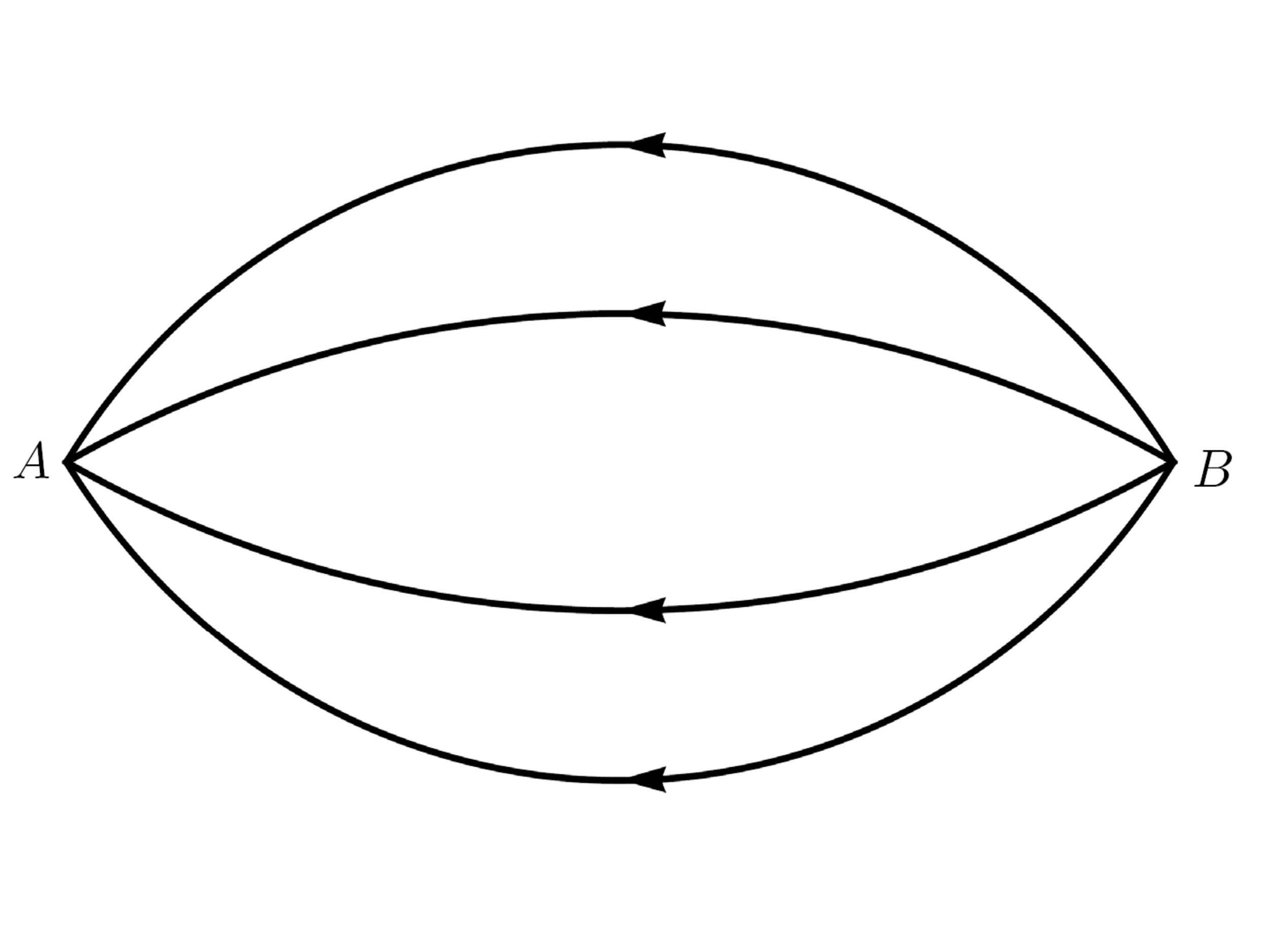}
    \caption{}
   \label{fsr6}
  \end{subfigure}  
\caption{ The figures show the sequence of  Ket Set elements (A) to (F) which encode propagation from vertex A to vertex B as described in the main text. 
The Ket Set is the  minimal
one containing (A). It underlies  a physical state  subject to the additional physical conditions of section 
\ref{sec3.2}   }%
\label{fsr}%
\end{figure}


The minimal Ket Set containing the chargenet $c$ of (A) must contain  the charge nets depicted in (B)- (F). Steps 
(1)- (7) imply that the chargenet of (D) has 2 possible ancestors, one depicted in (A)  and one in (F). The 
sequence of elements (A)-(B)-(C)-(D)-(E)-(F) is then one which encodes the `emmission' of a  conical perturbation at the vertex A of $c$
depicted  in (B) and its propagation and final `absorption' by vertex B to yield the chargenet $c^{\prime}$ in (F).
Thus the imposition of  appropriate additional physical conditions on anomaly free states can engender propagation.
Unfortunately, as we now argue, this propagation seems to be, at best, only  `1 dimensional' and even this `best case' requires very special
states.

\subsection{\label{sec3.2} No propagation between generic vertices of multivertex states}

The $N\rightarrow N$ deformations of \cite{p3} used hitherto create $N$ valent (CGR or GR) vertices from $N$ valent parental ones.
Consider a pair of GR vertices $v_1,v_2$  in a multivertex graph of different valences $N_1,N_2$. Any child vertex created from a deformation
of $v_1$ has a valence $N_1$ and any child vertex created from a deformation of $v_2$ has valence $N_2\neq N_1$. Hence the set of children
obtained through multiple deformations of $v_1$ and $v_2$ split into two disjoint classes, 
namely those with an $N_1$ valent child vertex and those with an $N_2$ valent child vertex. The former are unambiguously associated with $v_1$
and their creation can be visualised through a lineage associated with $v_1$. Similarly any lineage for the latter is associated with $v_2$.
Thus no possible parent of any child in the latter lineage can be part of the former lineage. This implies the impossibility of 
propagation between two such vertices.

Next, consider a pair of GR $N$ valent vertices $v_1, v_2$ in a graph which are connected through $M<N$ edges leaving $N-M$ edges `free' to connect with 
other parts of the graph. For generic graph connectivity,  once again children obtained through deformations of $v_1, v_2$ fall into two
disjoint sets by virtue of their connectivity with these 2 sets of free edges
and there is no propagation.

We digress here to note that the vertex deformations defined in \cite{p3} can be naturally extended to the case of a linear `multiply CGR' vertex. We 
define the multiply CGR property 
as follows. An $N$ valent linear $M$-fold multiply CGR vertex $v$ is one with the following vertex  structure:
\\
(a) There exists an open neighbourhood $U$ of $v$ and a coordinate patch thereon such that the edges at $v$ are coordinate straight lines in $U$.
\\
(b) There are $M$ pairs of edges at $v$ such that the union of each such pair forms a coordinate straight line in $U$ with $v$ splitting this line into
this pair of edges. There are $N-M$ edges which are not of this type.
\\
(c) Consider the set of outgoing edge tangents  to the remaining $N-M$ edges, together  with one edge tangent from each of the $M$ collinear pairs.
Then any triple of edge tangents from this set is linearly independent.

Deformations of such vertices can again be made, similar to the CGR case by transforming them to GR vertices
through interventions \cite{p3}. Such deformations then create child vertices whose valence (in the generalised sense described above) is the same as that 
of the parent vertex. 
\footnote{As mentioned in section \ref{sec2.7}, whether these deformations can be shown to be anomaly free in the sense of \cite{p3}
is an open question.}

The arguments in the first two paragraphs of this section indicate that long range propagation can at best be possible for special graphs.
Since the bottlenecks to propagation arise from  free edge connectivity and varying vertex valence, one `best case' scenario where propagation
could conceivably occur between multiple vertices is as follows:\\
(i) These vertices are of the same valence, say $N$.\\
(ii) These vertices are connected to each other by $N$ edges.\\
This implies that these vertices must be $N$-fold CGR with  no two edges connecting a pair of such vertices being collinear at either of the 
two vertices so connected. This leads to a graph connectivity depicted in Figure \ref{1dgraph} which is intuitively `1 dimensional'.

\begin{figure}[h]
\hspace{5cm}
\includegraphics[width=0.3\textwidth]{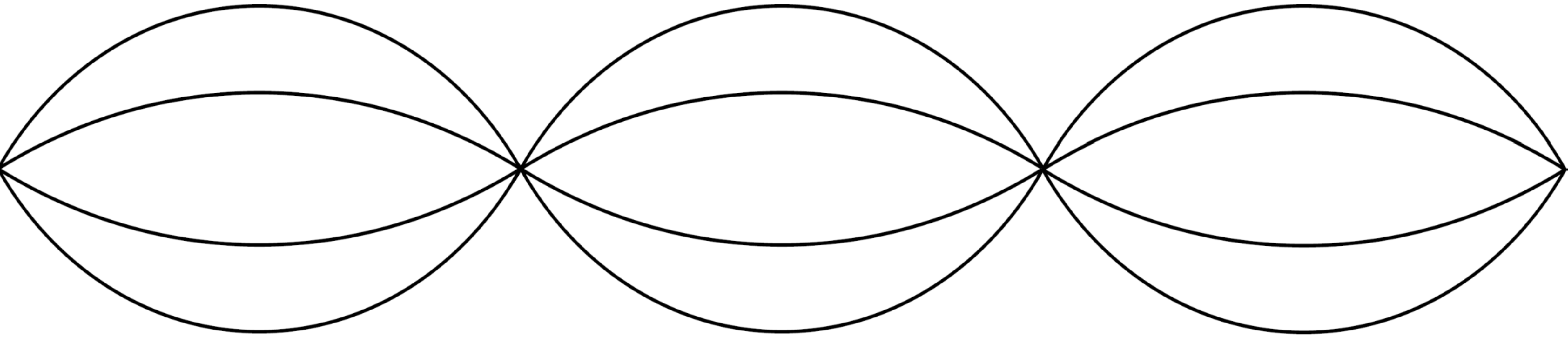}
  %
\caption{ Fig \ref{1dgraph}  shows a `best case' graph structure for the purposes of putative propagation based on $N\rightarrow N$ deformations.  
}%
\label{1dgraph}%
\end{figure}

We are unable to construct other examples of graph connectivity which could, conceivably, display propagation. 
It would be good to construct a proof that no such examples exist.
In any case, the arguments above indicate that propagation in the context of $N\rightarrow N$  deformations can
occur, at best in graphs with very special connectivity.
Hence, notwithstanding the propagation in the simple 2 vertex graph described in section \ref{sec3.1b} above, 
we seek a modification of the $N\rightarrow N$ deformation of \cite{p3} so as to engender vigorous, `3d', long range propagation for generic
graphs. As we shall see in the next section, the  $N\rightarrow 4$ modification described in section \ref{sec1} has this property.

\section{\label{sec4} Vigorous Propagation from  $N\rightarrow 4$ deformations }

In section \ref{sec4.1}  we define a modified $N\rightarrow 4$ implementation of the singular diffeomorphism encountered
in section \ref{sec2.3}. In section \ref{sec4.2} we show that this 
modification enables communication between vertices of different valence as well 
between vertices which have free edges, thus overcoming these bottle necks to propagation in the $N\rightarrow N$ case.
An immediate consequence is that of 3d long range propagation between vertices of a chargenet based on a graph which is dual to 
a {\em triangulation} of the Cauchy slice.
 In such a graph a vertex is connected to a nearest vertex by a single edge leaving 3 edges free (which in turn are connected to other nearest vertices).
 Whereas the $N\rightarrow N$ deformations do not engender propagation between vertices of such a graph, this sort of 
graph structure is not a barrier to propagation for 
the  $N\rightarrow 4$ 
deformation. We discuss this explicitly in section \ref{sec4.3}.
We note here that such graphs underlie spin nets which have a ready semiclassical interpretation   in the $SU(2)$ case \cite{twistedgeo}.
Our argumentation in sections \ref{sec4.2} and \ref{sec4.3} is largely pictorial and similar in character to that of section \ref{sec3.1b}.

\subsection{\label{sec4.1} The $N\rightarrow 4$ deformation}

The argumentation of section \ref{sec2.3} applies unchanged in the case of the $N\rightarrow 4$ defromations described here. Hence the action of the
constraint operators of interest is still built out of singular diffeomorphisms and, in the case of the Hamiltonian constraint, charge flips;
all that changes is the implementation
of the singular diffeomorphisms. 

We first define a downward conical $N\rightarrow 4$ deformation of an $N$ valent GR vertex $v$ of a charge net $c$.
This $N\rightarrow 4$ deformation replaces the  $N\rightarrow N$
deformation of Fig \ref{grb}. As in the $N\rightarrow N$  case, this $N\rightarrow 4$ deformation  corresponds to that generated by an electric
diffeomorphism action.  The deformed charge nets generated by the Hamiltonian constraint can be obtained by combining this deformation with charge flips exactly as in the
$N\rightarrow N$ case as described in the figure caption accompanying Figure \ref{gr} with the  $N\rightarrow N$ deformed charge net of Fig \ref{grb} replaced by its $N\rightarrow 4$
counterpart which we now  construct and which is displayed in Fig \ref{n4gr4}.

To construct the $N\rightarrow 4$ downward conical deformation of an $N$ valent linear GR vertex $v$ (depicted in Figure \ref{n4gr1}) 
by the singular diffeomorphism 
$\varphi(q_{I_{v}}^{i}\vec{{\hat{e}}}_{I_v}%
,\delta)$ (with  $q_{I_v}^i$ assumed to be positive as is appropriate for downward conicality), we first fix 
 3  edges $e_{J^i_v\neq I_v}, i=1,2,3$. We deform these 3 edges  exactly as for the 
$N\rightarrow N$ downward conical deformation  with $N=4$. This part of the deformation is depicted in Figure \ref{n4gr2}.
The remaining edges are pulled {\em exactly} along the $I_v$th edge as depicted in Fig \ref{n4gr3}.  The $N\rightarrow 4$ deformation
combines both these deformations and is depicted in Figure \ref{n4gr4}. Dropping the $v$ subscripts to the edge indices in what follows, if  
the outward-going edge charges at the (undeformed) vertex $v$ are $q^i_{K}, K=1,..,N$,
then in the (obvious) notation used in (\ref{b12}), the charges on the deformed charge net of Figure \ref{n4gr4} can be readily inferred from 
Figures \ref{n4gr2}, \ref{n4gr3} to be:
\ba
q^k_{v_I{\tilde v}_{J^i}} &=& q^k_{J^i}, i=1,2,3 \nonumber\\
q^k_{v_I v} &=& \sum_{J\neq I,J^1, J^2,J^3} q^k_J  = - q^k_{vv_I },    \label{n4q}
\ea
with the charges on the remaining parts of the graph being exactly those on these parts of the graph in the undeformed parent state $c$ of Fig \ref{n4gr1}.

As mentioned above, the deformed vertex structure  of Figure \ref{n4gr4} is created from the undeformed one of Figure \ref{n4gr1} by an action of 
the electric diffeomorphism constraint. The deformed vertex structure created by the Hamiltonian constraint can then be constructed exactly 
as for the $N\rightarrow N$ case by combining the $N\rightarrow 4$ deformation of Figure \ref{n4gr4} with appropriate charge flips as depicted
in Fig \ref{n4gr5} and described in the accompanying figure caption. If the charge $q^i_I$ is negative, 
the vertex $v$ is displaced along the extension of the $I$th edge and the conical deformation of the 3 chosen edges is then upward conical.
We do not discuss upward conical deformations in detail as we do not need them here; the details are straightforward and we leave their working 
out to the interested reader.

\begin{figure}
  \begin{subfigure}[h]{0.3\textwidth}
    \includegraphics[width=\textwidth]{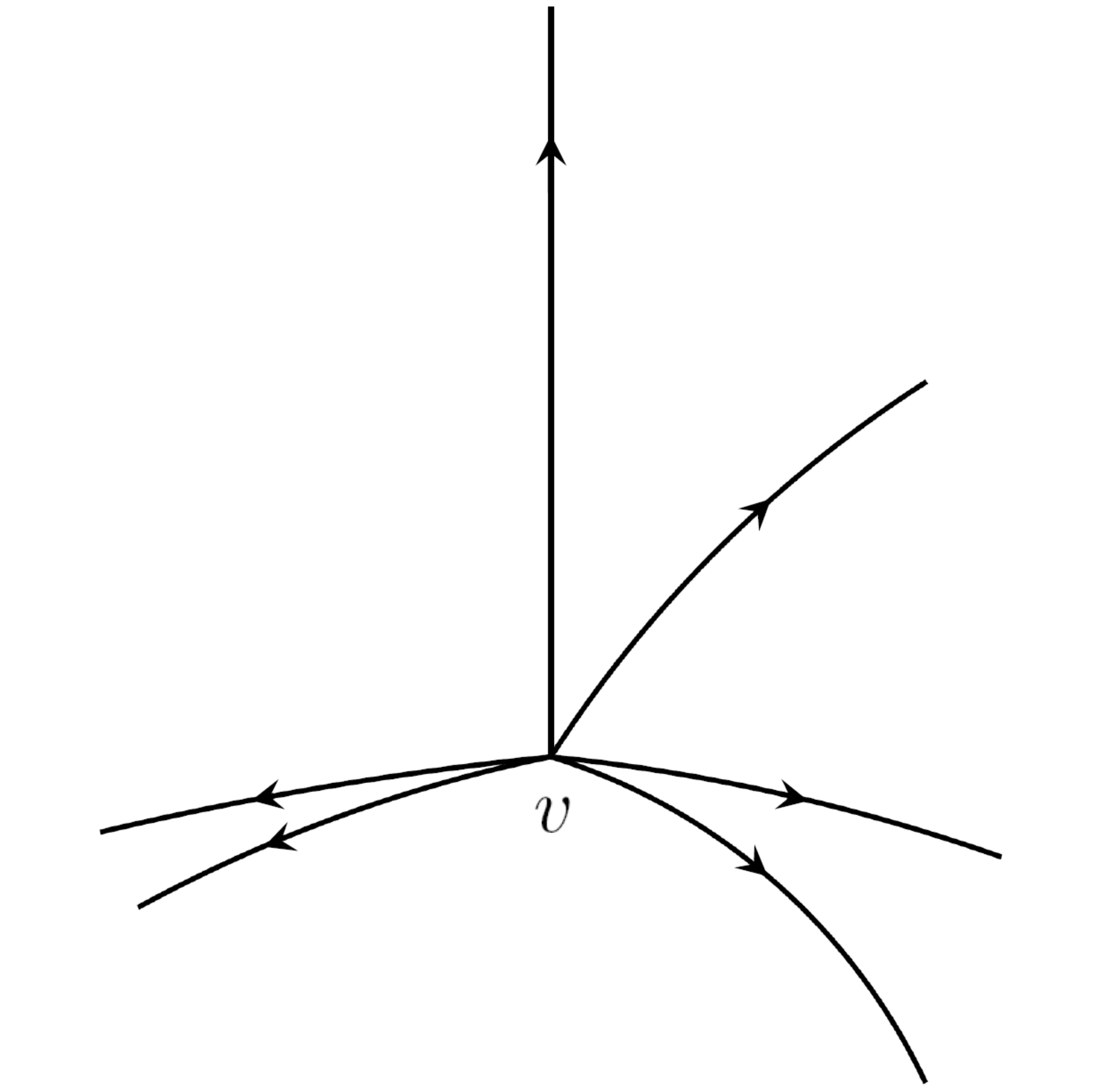}
    \caption{}
 \label{n4gr1}
  \end{subfigure}
  \begin{subfigure}[h]{0.3\textwidth}
    \includegraphics[width=\textwidth]{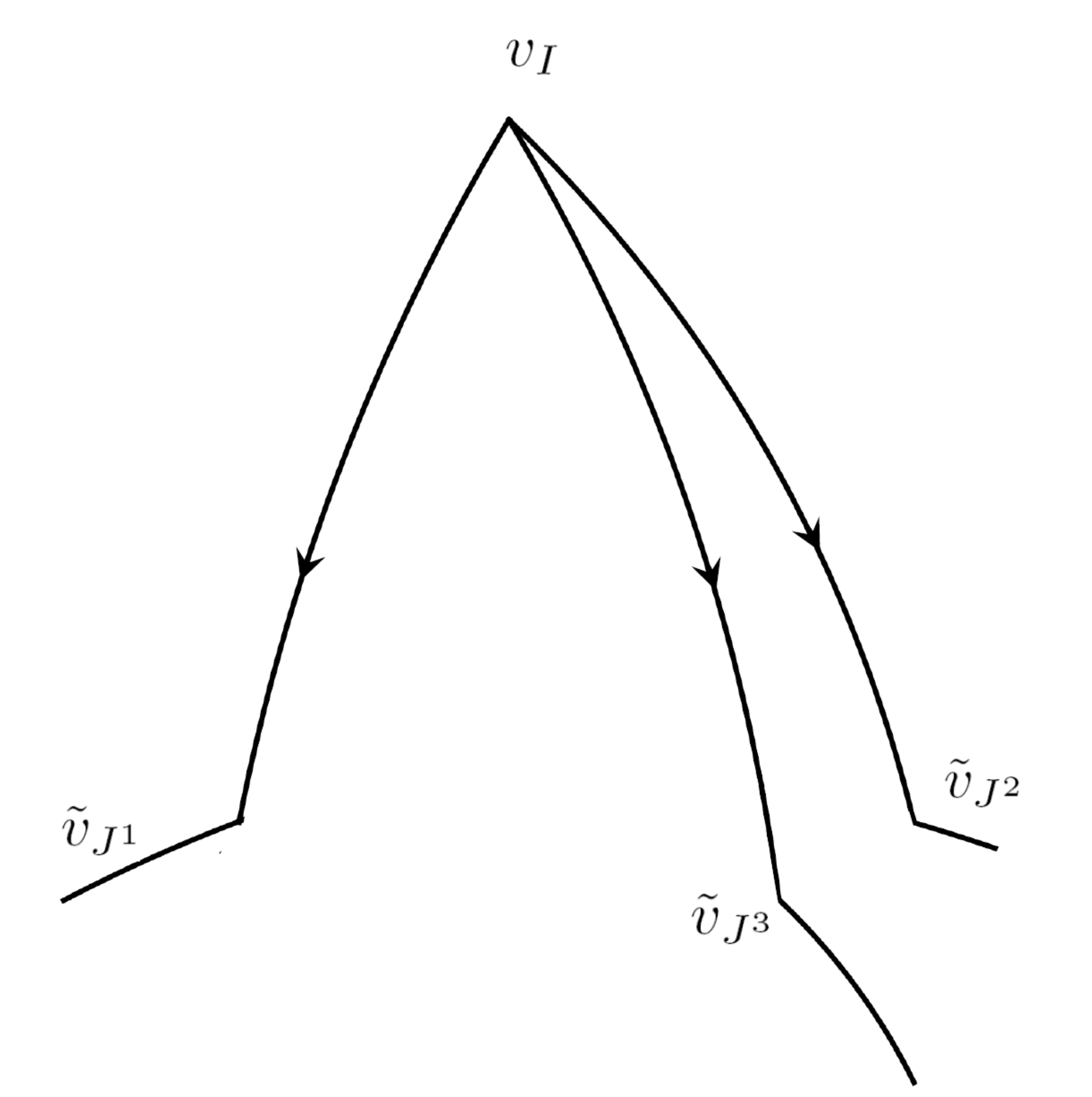}
    \caption{}
   \label{n4gr2}
  \end{subfigure}
\begin{subfigure}[h]{0.27\textwidth}
    \vspace*{-14mm}
    \includegraphics[width=\textwidth]{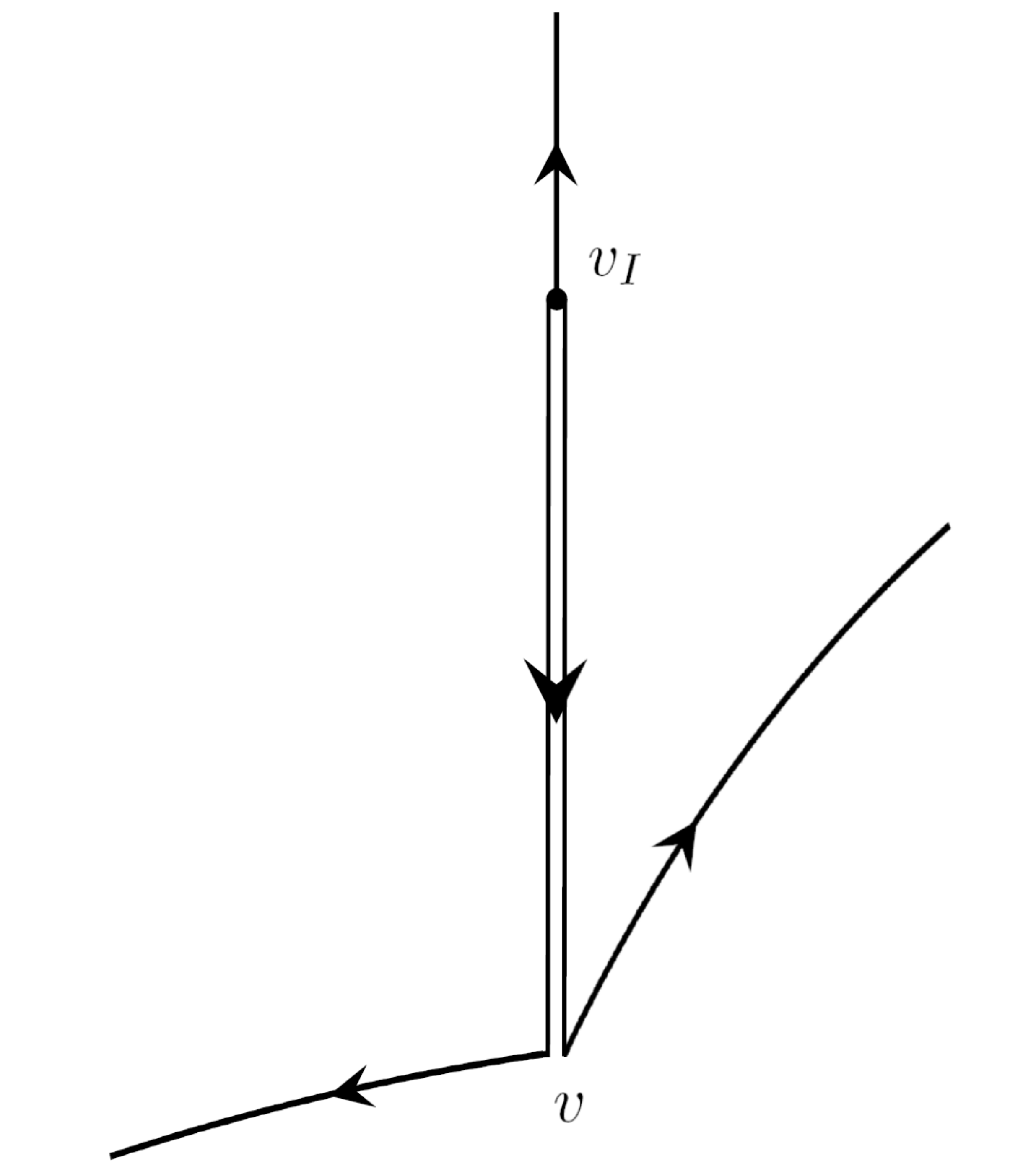}
    \caption{}
   \label{n4gr3}
  \end{subfigure}
\begin{subfigure}[h]{0.3\textwidth}
    \includegraphics[width=\textwidth]{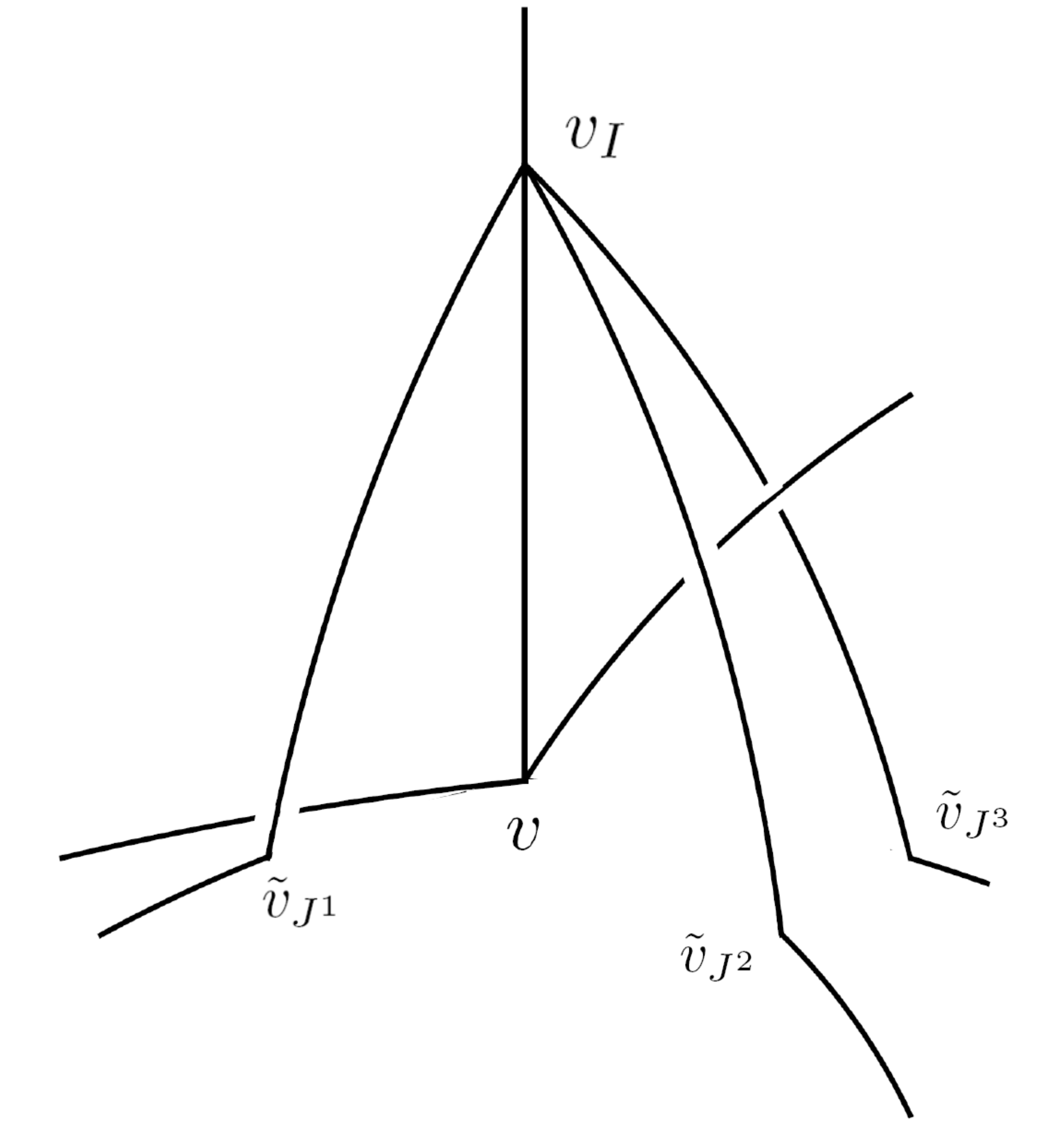}
    \caption{}
   \label{n4gr4}
  \end{subfigure}  
\begin{subfigure}[h]{0.3\textwidth}
    \hspace*{5mm}
    \includegraphics[width=\textwidth]{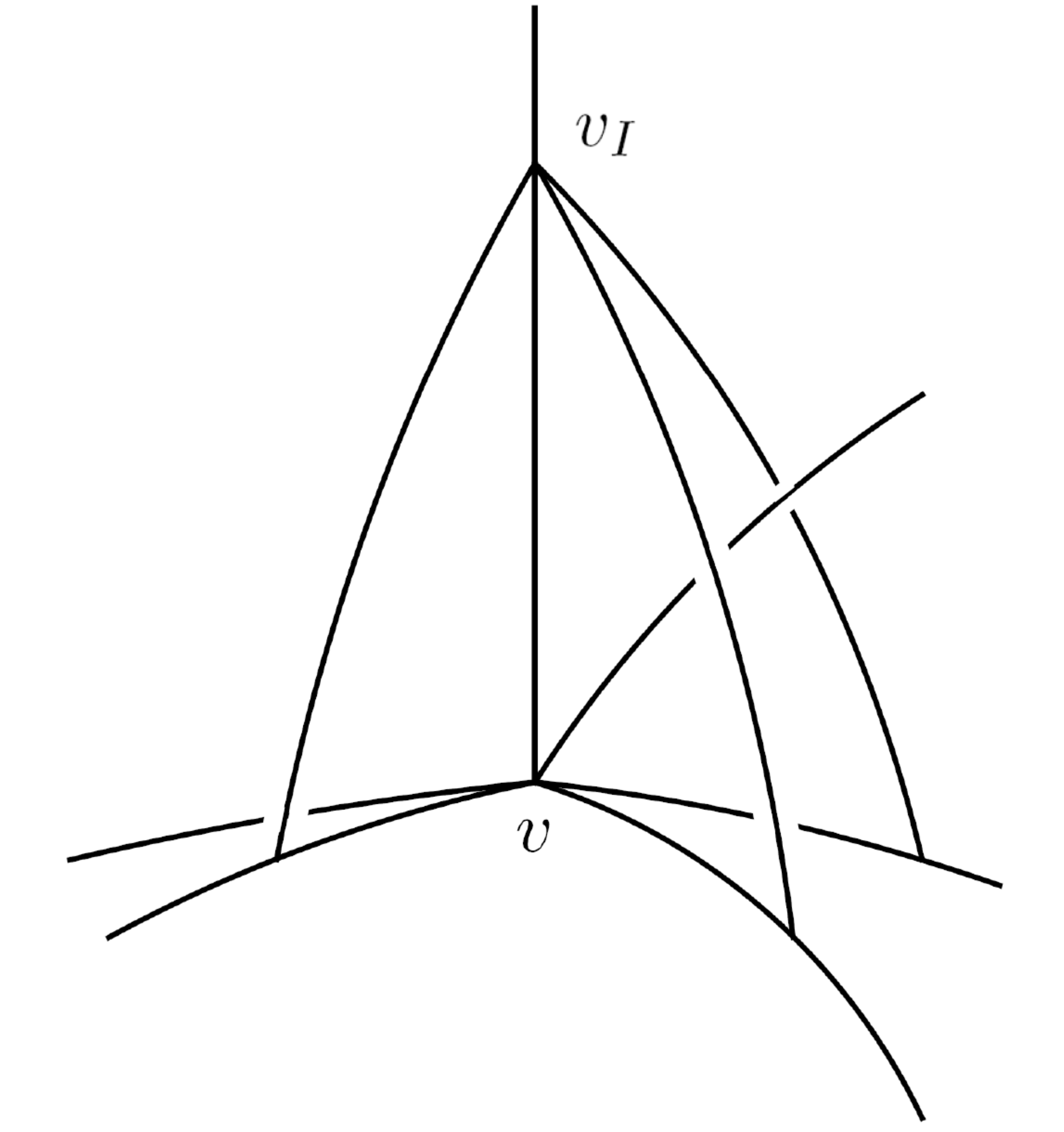}
    \caption{}
   \label{n4gr5}
  \end{subfigure}  
\caption{ Fig \ref{n4gr1} shows an undeformed GR vertex $v$ of a chargenet $c$. 
Fig \ref{n4gr2} shows the conical deformation of the 3 preferred edges $e_{J^i}, i=1,2,3$ along the $I$th edge, the 
displaced
vertex $v_I$ and the 3 $C^0$ kinks, ${\tilde v}_{J^i}, i=1,2,3$  are as labelled. Fig \ref{n4gr3} shows the remaining edges being pulled exactly along the 
$I$th edge.  The deformations of Figs \ref{n4gr2}, \ref{n4gr3} combine to yield the $N\rightarrow 4$ deformation shown in Fig \ref{n4gr4}.
Fig \ref{n4gr5} shows the result of a Hamiltonian type deformation 
obtained by multiplying the chargenet holonomies obtained by 
coloring  the edges of Fig \ref{n4gr2}, \ref{n4gr3}  by the 
flipped images of charges  on their counterparts in $c$,  of  Fig \ref{n4gr1}  by negative of these flipped charges and Fig \ref{n4gr1} by the charges on $c$. 
If the edges of Fig \ref{n4gr2}, \ref{n4gr3}  are colored by the charges on 
their counterparts in $c$ then they combine to yield the holonomy of Fig \ref{n4gr4}, this being the result of an electric diffemorphism deformation.
}%
\label{n4gr}%
\end{figure}


Due to the choice of 3 preferred edges in this deformation, the resulting charge net
is now denoted by $c_{(i, I_v,{\vec J}_v \beta_v, \delta )}$ where $\beta = +1,-1, 0$ for positive , negative and no flips and the particular choice
of edge triple is indicated by ${\vec J}_v\equiv (J^1_{v}, J^2_{v}, J^3_v)$.
The action of the constraint is then obtained by summing over all possible triples of such edges so that the 
Hamiltonian constraint action is:
\be
\hat{H}[N]_{\delta}c(A) =\frac{\hbar}{2\mathrm{i}}\frac{3}{4\pi}\sum_v \beta_v N(x(v))\nu_{v}^{-2/3}\sum_{I_{v}} \frac{1}
{\binom{N-1}{3}}
%
\sum_{{\vec J}_v}\sum_{i}
\frac{c_{(i,I_v,{\vec J}_v,\beta_v,\delta)}- c}{\delta}
\label{hamfinaln4}
\ee
and the electric diffeomorphism constraint action is:
\be
\hat{D}_{\delta}[\vec{N}_{i}]c   =\frac{\hbar}{\mathrm{i}}\frac{3}{4\pi}%
\sum_v N(x(v))\nu_{v}^{-2/3}\sum_{I_{v}}\frac{1}{\binom{N-1}{3}}
\sum_{{\vec J}_v}
\frac{1}{\delta
}(c^{}_{(I_{v},i,0, {\vec J}_v,\delta)}-c).
\label{dnfinaln4}
\ee
In equation (\ref{hamfinaln4}), $\beta_v =\pm 1$ depending on whether a positive or negative flip is chosen for the deformations at $v$.
In both the above equations we have implicitly  chosen the appropriate downward/upward conical deformation dictated by the sign of the 
edge charge $q^i_{I_v}$. However these deformations can be chosen to be either upward or downward provided, as discussed in section \ref{sec2.4},
we insert minus signs at appropriate places in  these equations.
The main implication of all this is that the set of children obtained from the action of constraint deformations 
are generated through  positive and negative charge flips as well as upward and downward conical deformations.

In what follows we shall also require the deformation generated by an electric diffeomorphism constraint on a 4 valent CGR vertex
along its collinear edges. Since $N=4$ this deformation coincides with the $N\rightarrow N$ deformation with $N=4$ depicted 
in Fig \ref{cgrk=ib}. While Fig \ref{cgrk=ib} depicts a downward conical deformation, an  upward conical deformation 
can be visualised by viewing Figures \ref{cgrk=ia}, \ref{cgrk=ib}  upside down; for details see \cite{p3} and figures therein.
The charges on the deformed edges for such deformations  are exactly the same as those
on their undeformed counterparts. 
\footnote{As mentioned in section \ref{sec2.7}, 
the derivation of these deformations and charge labellings as well as the deformations
along other edges which contribute to the action of the constraint at this vertex proceed through the use of interventions \cite{p3} which convert the 
parental CGR vertex to a GR one.  
The interested  reader may consult \cite{p3} for details with regard to the intervention procedure.}

We note here that the  charges on the deformed child in the case of a parental GR vertex   can be quickly inferred as follows without going through the 
holonomy multiplication of Figure \ref{n4gr}.
The charges on the deformed edges $e_{v_I {\tilde v}_J^i}, i=1,2,3,$ are  exactly the same as for the $N\rightarrow N$ edges with $N=4$. Thus in the case of 
Hamiltonian constraint deformation these charges are obtained through positive or negative $i$-flips of the charges on their undeformed counterparts
in $c$ whereas for an electric diffeomorphism deformation these charges are identicial to those on their undeformed counterparts in $c$.
The remaining charges maybe inferred from $U(1)^3$ gauge invariance together with the fact that the deformation is confined to a $\delta$ size vicinity
of the parental vertex.

\subsection{\label{sec4.2}Propagation between vertices with different valence and with free edges}

Consider two linear GR vertices $A,B$ of a charge network $c$ with valences $N_A,N_B$.
connected by $M<N$ edges leaving $N_A-M$ and 
$N_B-M$ edges free at $v_A$ and $v_B$.  
We now show that the $N\rightarrow 4$ constraint action engenders propagation between the vertices $A,B$.
Our argumentation is primarily diagrammatical and described through Figure \ref{figprop1} as follows:\\

\noindent (1)We start at the left with the `unperturbed' charge network structure described above depicted in Fig \ref{figprop1} (A).
We shall be interested in a Hamiltonian constraint generated deformation along the $I$th edge emanating from $A$ and connecting to
$B$. In order to keep the figure uncrowded, 
it explicitly depicts only this  single edge between $A, B$ and only a  few more edges at these vertices.
The reader may think of $M-1$ of the edges emanating from  $A$ and $M-1$ of those from $B$ 
as being connected so as to yield $M-1$ more edges connecting $A, B$. The connectivity of the remaining free edges  does not affect the arguementation.

We denote the outgoing  edge charges at $A$ by $q^k_J, J=1,..,N$.
In the $N\rightarrow 4$ deformation at $A$ along $e_I$, a choice of 3 edges $e_{J^i}, i=1,2,3$ has to be made. As we shall see below, propagation generically ensues
irrespective of which choice we make. 
\\

\noindent (2) 
The parent chargenet $c$ is deformed in a downward conical manner along the  $I$th edge at $A$ by the action of ${\hat H}(N)$ at the vertex
$A$ to give the child $c_{(i,I, {\vec J},\beta, \delta)}$ shown  in (B), where we have used the notation of (\ref{hamfinaln4}) and dropped the 
`vertex' suffix  to avoid notational clutter. Here $\beta$ can be $+1$ or $-1$. In either case, we refer to the relevant flipped image of the charge $q^i$ (\ref{defchrgeflip}),
(\ref{-defchrgeflip}) as $q_{flip}^i$. The deformation of the GR vertex $A$ is exactly that of Figure \ref{n4gr5} with charges in the vicinity of vertex $A$ 
obtained exactly as described in the figure caption accompanying Figure \ref{n4gr4}. These charges in obvious notation are:
\ba
q^k_{v_I{\tilde v}_{J^i}} &=&  (q_{flip})^k_{J^i}, i=1,2,3  \label{n4hq1}\\
q^k_{A {\tilde v}_{J^i}}  &=&  q^k_{J^i}-(q_{flip})^k_{J^i}  \label{n4hq2}                       \\
q^k_{ Av_I} & =&    
q_I^k-(q_{flip})^k_I-\sum_{J\neq I,J^1, J^2,J^3} (q_{flip})^k_J= q_I^k + \sum_{i=1}^3 (q_{flip})^k_{J^i}
=-q^k_{v_IA },  \label{n4hq3}
\ea
with the charges on the remaining part of the graph remaining unchanged and where we have used gauge invariance of at $A$ in the unperturbed charge net $c$
to go from the first equality to the second in (\ref{n4hq3}).
\\

\noindent (3) The charge net of (B) is acted upon by a seminanalytic diffeomorphism so as to `drag' the deformation from the 
vicinity of vertex A to the vicinity of vertex B.
\footnote{\label{fndiff1}We assume that the state $c_{(i,I,{\vec J},\beta,  \delta)}$ is such that it can be transformed via an appropriate diffeomorphism
 to the state depicted in (C). We shall comment further on this in section \ref{sec5a}.}
We slightly abuse notation and denote the images of $v_I, {\tilde v}_J$ by this diffeomorphism by the same symbols $v_I,{\tilde v}_J$.
\\

\noindent (4) The charge net of (C ) is deformed by the action of an appropriate electric diffeomorphism at the CGR vertex $v_I$ as described
in Figure \ref{cgrk=ib} to yield the charge net of (D).
This transforms the conical deformation in (C)  which is  downward with respect to the $I$th line from $A$ to $B$  to one which is
upward conical with respect to this line in (D). As a result, the deformation is now {\em downward} conical with respect to the (oppositely oriented)
line from $B$ to $A$. 
\\

\noindent (5) The charge net $c^{\prime}$ of (E) is based on a graph which is obtained by adding 3 edges to the graph underlying the unperturbed state $c$.
These edges emanate from the vertex $B$ and terminate at the 3 kinks ${\tilde v}_{J^i}$ of  (C).
The charges on this charge net in the vicinity of vertices $A,B$ are as follows.
\ba
q^k_{B{\tilde v}_{J^i}} &=&  (q_{flip})^k_{J^i}, i=1,2,3  \label{n41q1}\\
q^k_{A {\tilde v}_{J^i}}  &=&  q^k_{J^i}-(q_{flip})^k_{J^i}  \label{n41q2}                       \\
q^k_{AB} & =& q_I^k + \sum_{i=1}^3 (q_{flip})^k_{J^i}=- q^k_{BA },
\label{n41q3}
\ea
with the charges on the remaining parts of the graph being the same as in $c$.
\\

\noindent (6) The charge net $c^{\prime}$ of (E) is deformed in a downward conical manner by the action of the electric diffeomorphism  
${\hat D}({\vec N}_i)$ at 
its $N_B+3$ valent vertex B along the edge from $B$ to $A$ 
with the chosen 3 edges being exactly the edges from $B$ to each of ${\tilde v}_{J^i}, i=1,2,3$ to give exactly the state in (D).
\\

\begin{figure}
  \begin{subfigure}[h]{0.3\textwidth}
    \includegraphics[width=\textwidth]{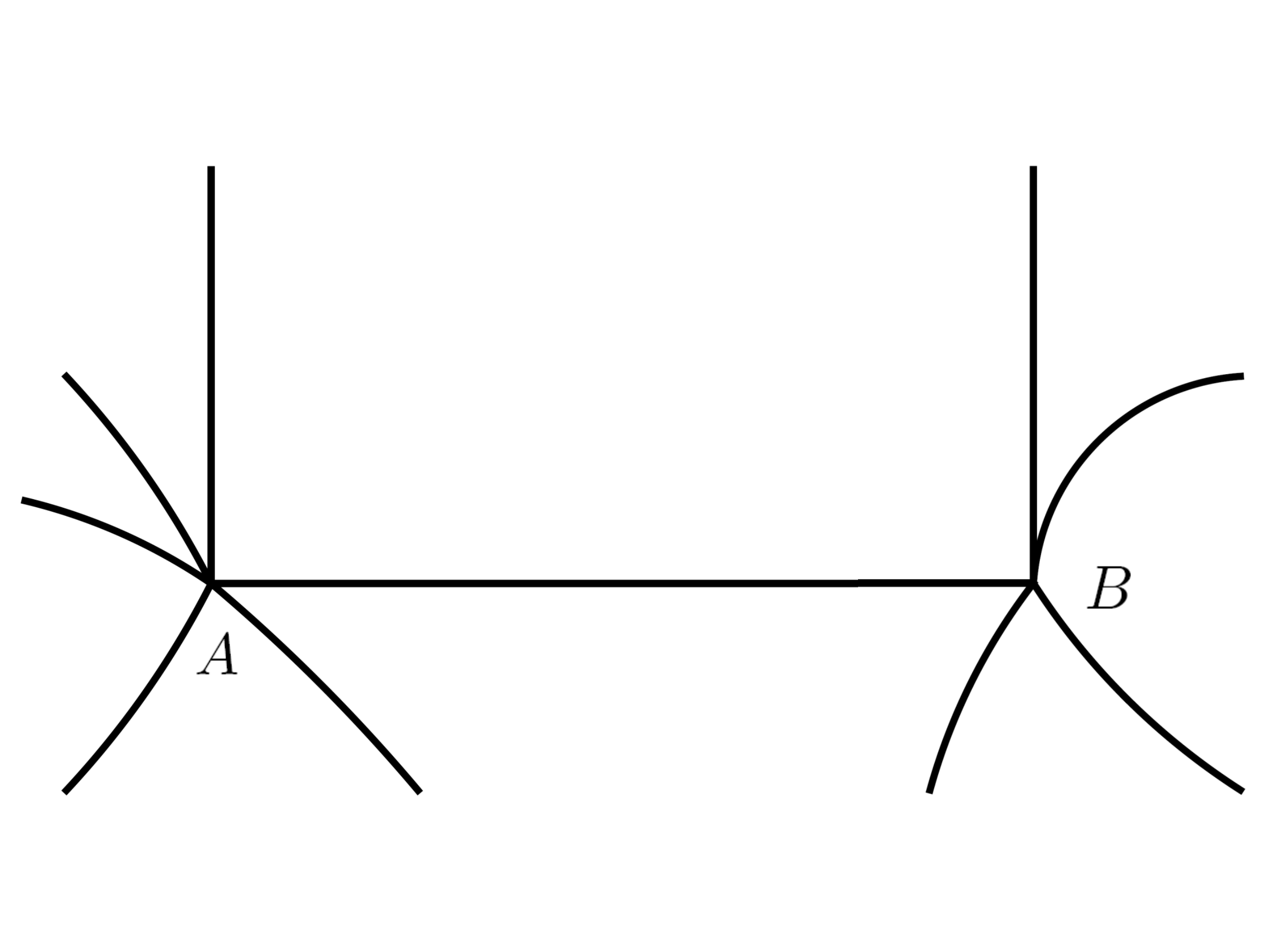}
    \caption{}
 \label{fp1}
  \end{subfigure}
  \begin{subfigure}[h]{0.3\textwidth}
    \includegraphics[width=\textwidth]{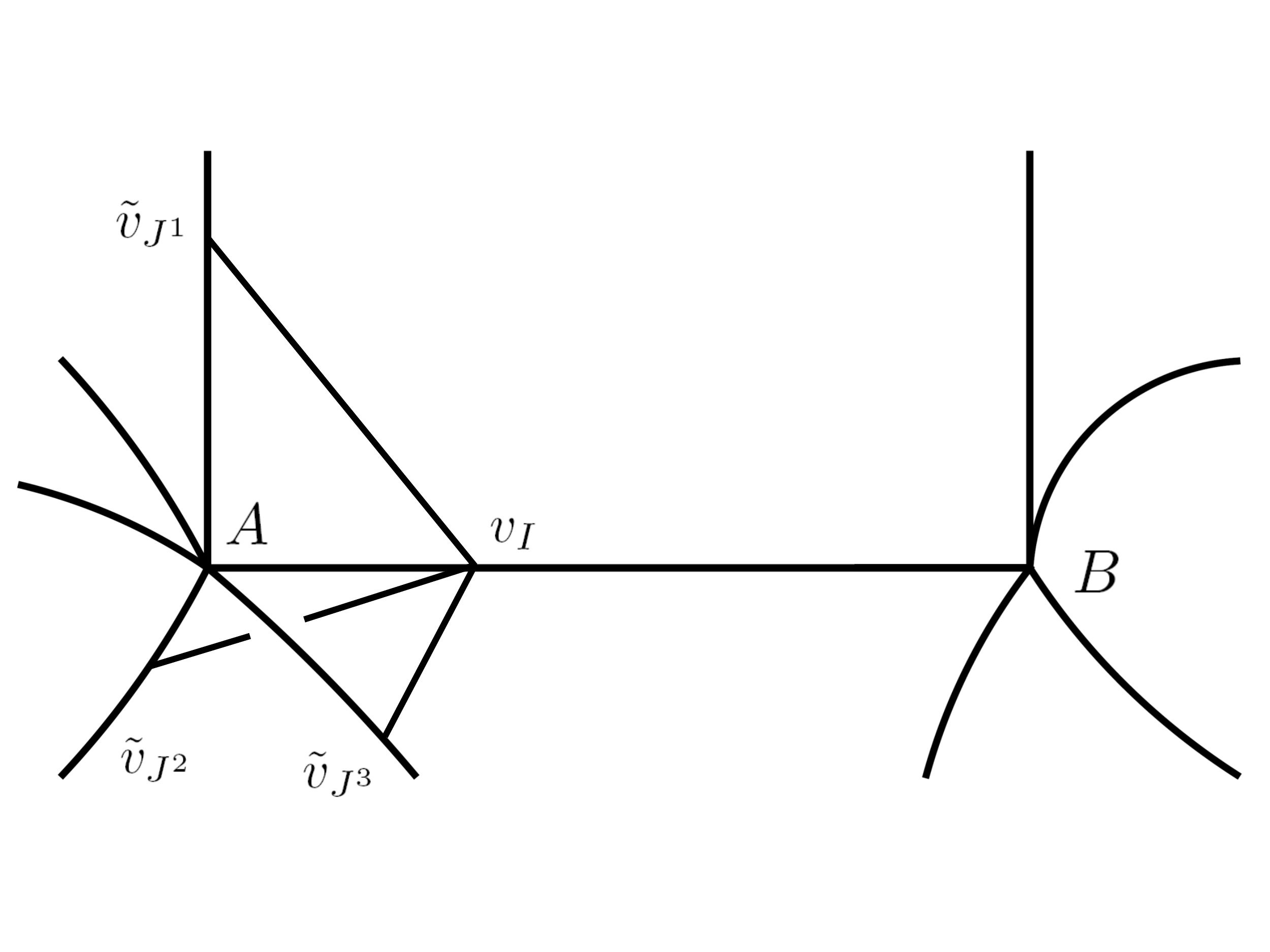}
    \caption{}
   \label{fp2}
  \end{subfigure}
\begin{subfigure}[h]{0.27\textwidth}
    \includegraphics[width=\textwidth]{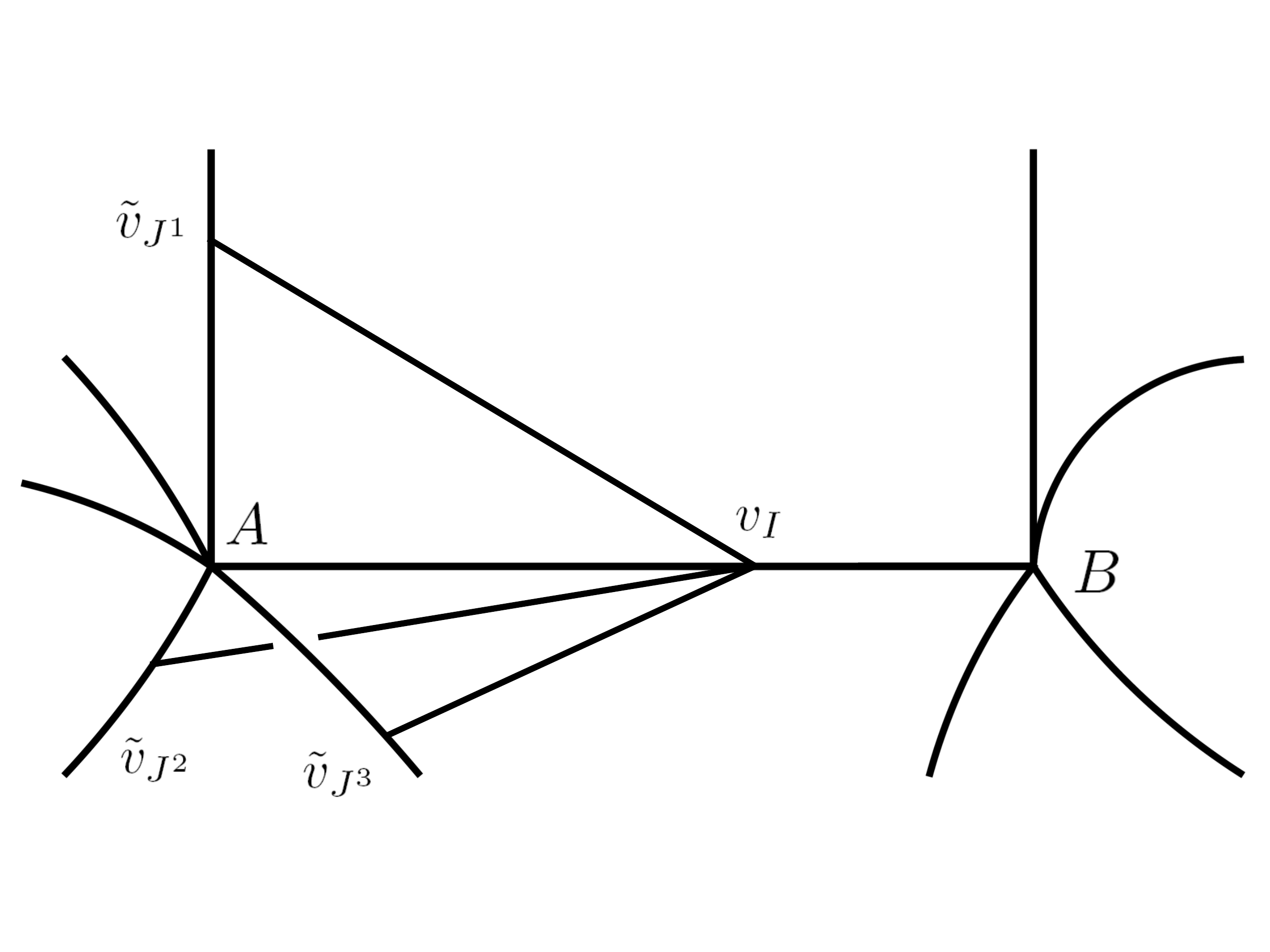}
    \caption{}
   \label{fp3}
  \end{subfigure}
\begin{subfigure}[h]{0.3\textwidth}
    \includegraphics[width=\textwidth]{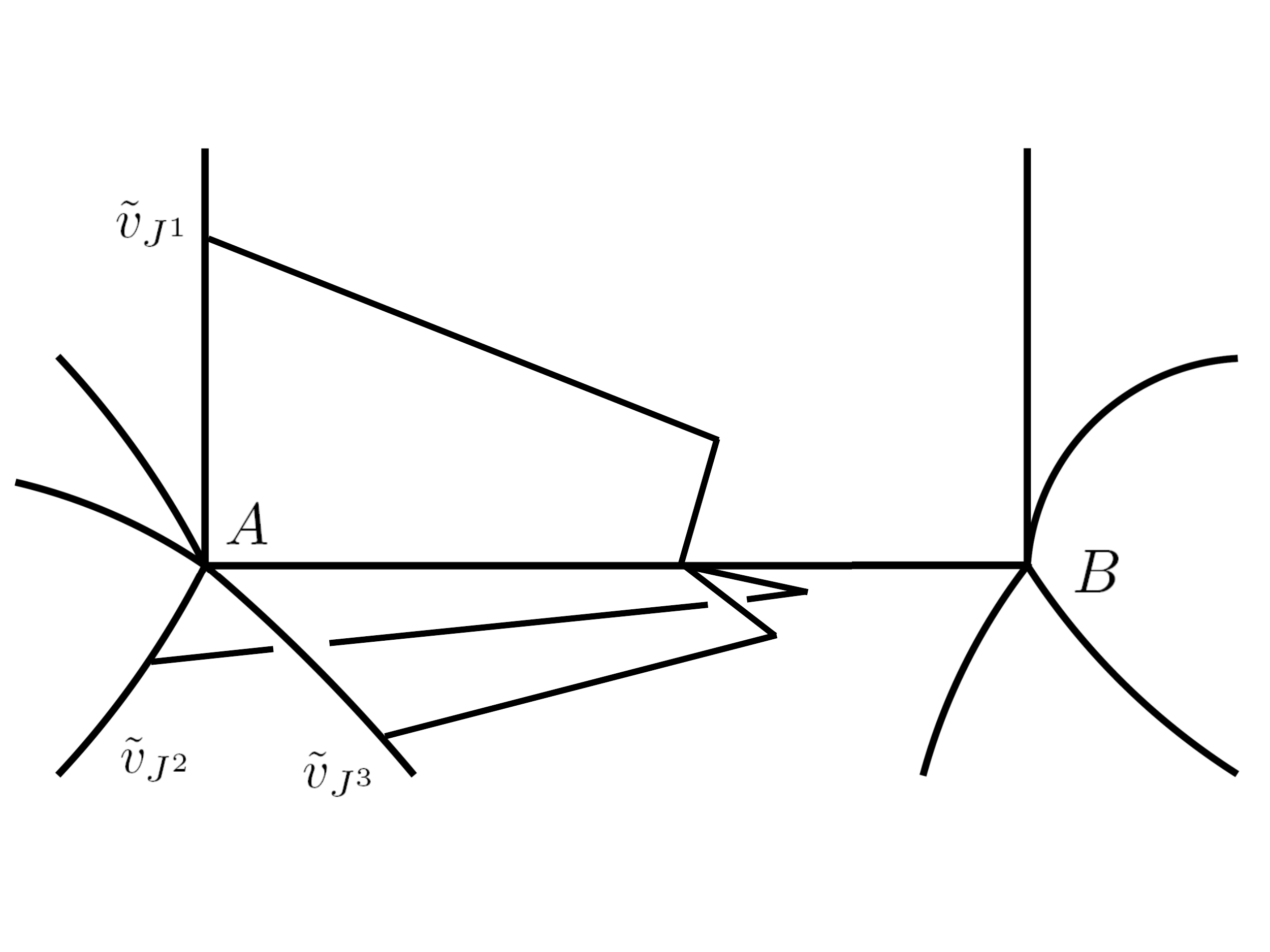}
    \caption{}
   \label{fp4}
  \end{subfigure}  
\begin{subfigure}[h]{0.3\textwidth}
    \hspace*{7mm}
    \includegraphics[width=\textwidth]{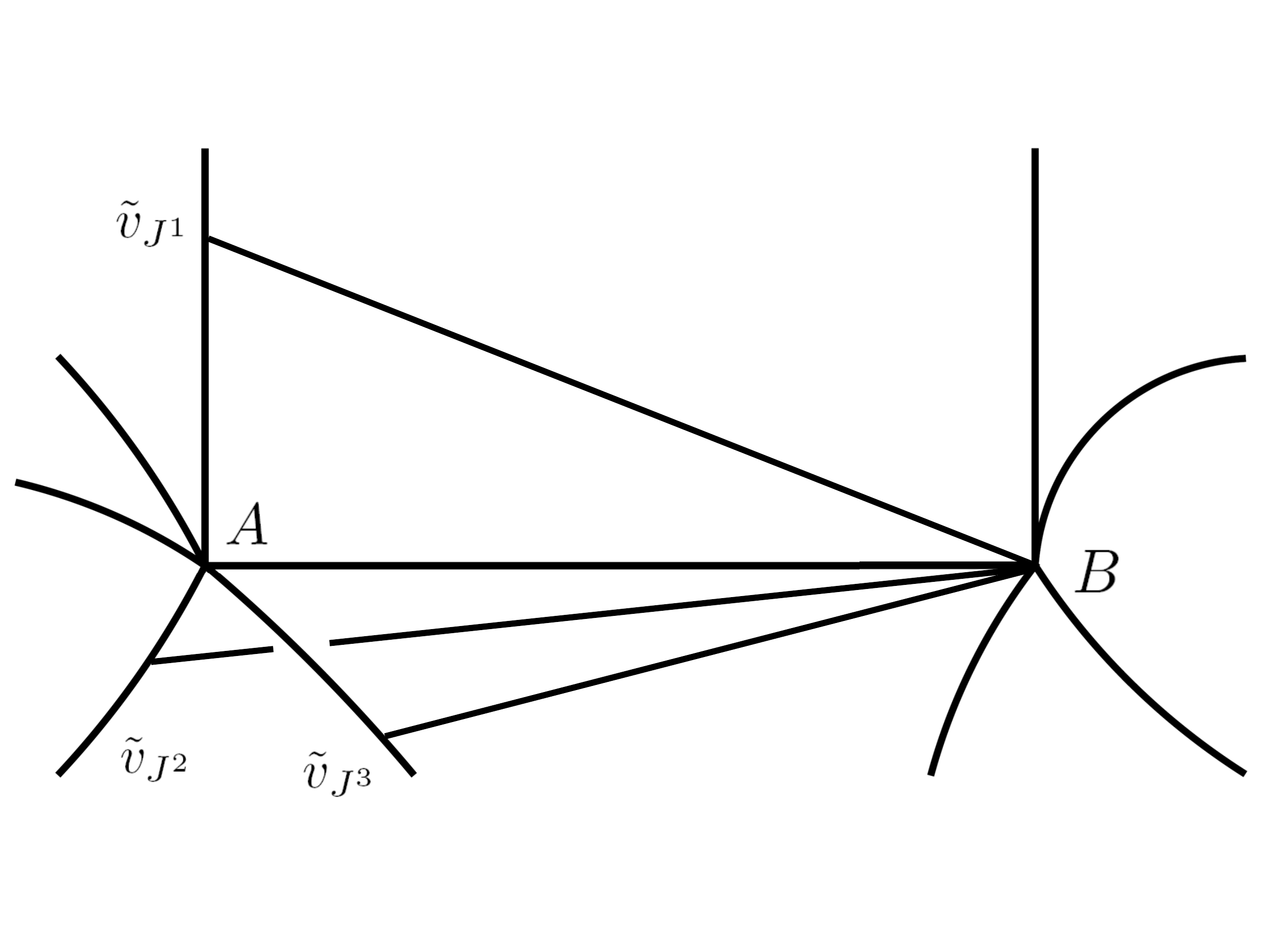}
    \caption{}
   \label{fp5}
  \end{subfigure}  
\caption{ The figures show the sequence of  Ket Set elements (A) to (E) which encode propagation from vertex A to vertex B of different valence 
as described in the main text. 
The Ket Set is the  minimal
one containing (A) appropriate to the $N\rightarrow 4$ deformation. 
}%
\label{figprop1}%
\end{figure}

The minimal Ket Set containing the chargenet $c$ of (A) must contain  the charge nets depicted in (B)- (E). Steps
(1)- (6) imply that the chargenet of (D) has 2 possible ancestors, one depicted in (A)  and one in (E). The 
sequence of elements (A)-(B)-(C)-(D)-(E) is then one which encodes the `emmission' of a  conical perturbation at the vertex A of $c$
depicted  in (B) and its propagation and final `absorption' by vertex B to yield the chargenet $c^{\prime}$.
The result of this `absorption' is an additional connectivity in the graph which additionally entangles the vertices $A$ and $B$.
Further, the valence of vertex $B$ as a result of this `absorption' has increased by 3.
It is in this sense that the $N\rightarrow 4$ action generates propagation between vertices of different valence as well as  in the presence of free edges.

Here we have implicitly assumed that the vertices $v$ in Figure \ref{figprop1} (A),  $v_I$ in Figure  \ref{figprop1}(B)  and  $B$ in Figure \ref{figprop1} (E)
are nondegenerate. The first two assumptions are simply assumptions on the charge labellings of vertex $A$ in $c$. The third is an assumption on 
the labellings of the vertex $B$ of the charge net in (E). As mentioned above the vertex structure at $B$ in (E) is obtained by adding  3 extra edges 
to the original vertex $B$ in $c$. These are positioned in the vicinity of $B$ so as to render $B$ GR in $c^{\prime}$. It seems reasonable to us that 
exploiting the available freedom in positioning these 3 edges relative to the original edges at $B$ would enable us to choose an edge configuration such 
that $B$ is non-degenerate in $c^{\prime}$. 

\subsection{\label{sec4.3}`3d' Propagation}

Let the `unperturbed' charge network $c$ in the previous section be based on a graph dual to a triangulation of $\Sigma$ by tetrahedra. Every vertex of this graph is then 
(linear) GR and 4 valent. Each vertex is connected to 4 other vertices each such connection being through a single edge.
In the language of the previous section, each vertex then has 3 free edges.
Figure \ref{fig3dprop}(A) shows the graph structure of $c$ in the vicinity of  3 of its vertices $A,B,C$.

Repeating the considerations of the previous section, we `perturb' $c$  at its vertex $A$ through the action of the  Hamiltonian constraint 
to yield $c_{(i,I,{\vec J},\beta,\delta)}$ shown in Figure \ref{fig3dprop} (B) and then 
`evolve' this perturbation at $A$ in  $c$ to $B$ yielding  the chargenet $c^{\prime}$ of Figure \ref{fig3dprop} (C)   in which the 
vertex $B$ is now 7 valent. We shall rename $c^{\prime}$ as $c_{AB}$ in what follows so as to remind us that the `perturbation' has traversed the 
path $A-B$ in $c$  to yield $c^{\prime}= c_{AB}$. 
Here we show how to further evolve this perturbation beyond the vertex $B$ through the 
exclusive use of electric and semianalytic diffeomorphisms, once again through a primarily diagrammatic argument (see Fig \ref{fig3dprop}). \\
%
In what follows we use the (obvious)  notation wherein an edge connecting the point $P$ to the point $Q$ is denoted by 
$e_{PQ}$ and the charge thereon by $q^k_{PQ}$. \\

\noindent (1) We  act by an electric diffeomorphism at $B$ of $c_{AB}$ which deforms the vertex
structure at $B$ along the edge $e_{BC}$ as depicted in Fig \ref{fig3dprop}(D). The 3 additional edges chosen for this deformation are exactly the edges $e_{B{\tilde v}_{J^i}}$.
This results in a reduction of valence of $B$ back to 4 together with creation of the displaced vertex which we call $v^{\prime}$ 
The vertex $v^{\prime}$ is now directly connected to the edges $e_{J^i}$ emanating from $A$ at the points ${\tilde v}_{J^i}$. 
Each of the segments  $l_{v^{\prime}{\tilde v}_{J^i}}$ which connect  $v^{\prime}$ to ${\tilde v}_{J_i}$ are obtained by 
deforming the edges $e_{B{\tilde v}_{J^i}}$ and thus each of these segments has a kink at ${\tilde v}^{\prime}_{J_i}$ as shown in (D).
The charges on these segments are exactly the same as those on the edges $e_{B{\tilde v}_{J^i}}$. From (\ref{n41q1}), we have (in obvious notation)
\be
q^k_{{v^{\prime}{\tilde v}_{J^i}}}= q^k_{B{\tilde v}_{J^i}} = (q_{flip})^k_{J^i}, i=1,2,3 
\label{n42q1}
\ee
From gauge invariance and the fact that the deformation only affects the immediate vicinity of $B$ we have from
(\ref{n42q1}) that:
\be
q^k_{v^{\prime} B} = - \sum_{i=1}^3(q_{flip})^k_{J^i} -q^k_{v^{\prime}C}, \;\;\;\;\;\;\; q^k_{v^{\prime}C}= q^k_{BC}
\ee
where we have denoted the charge on $e_{BC}$ in $c$ (and hence also in $c_{AB}$) by $q^k_{BC}$.
The charges on the remaining parts of the graph  are exactly those on these parts of the graph  in $c_{AB}$
\\

\noindent (3) We act by a semianalytic diffeomorphism $\phi$ so as to move the vertex $v^{\prime}$ to the vicinity of $C$ as
shown in (E).
\footnote{\label{fndiff2}An assumption similar to that in Footnote \ref{fndiff1} applies.}
Abusing notation slightly we continue to refer to the moved vertex as $v^{\prime}$.
\\

\noindent (4) Similar to (4) of section \ref{sec4.2} we  act by an electric diffeomorphism at $v^{\prime}$ so as to change the sign of the 
conicality of the vertex structure to obtain the chargenet depicted in (F).
\\

\noindent (5) Similar to (5) of section \ref{sec4.2}, the charge net depicted in (F) can be obtained by the action
of an electric diffeomorphism at the 7- valent vertex $C$ of the chargenet $c_{ABC}$ depicted in (G), with charge labels as follow:
\ba
{}^{(G)}q^k_{C {\tilde v}^{\prime}_{J^i}}  & =  &(q_{flip})^k_{J^i},\;i=1,2,3, \label{cabc1}
\\
{}^{(G)}q^k_{C B  } & =& - \sum_{i=1}^3(q_{flip})^k_{J^i} + q^k_{CB}.  \label{cabc2}
\ea
Here the left superscript $(G)$ indicates charges on the chargenet $c_{ABC}$ depicted in Fig \ref{fig3dprop} (G), $q^k_{CB}$ is the charge on the edge $e_{CB}$
of the unperturbed charge net $c$ of Fig \ref{fig3dprop}(A) and, as in (\ref{n42q1}),   $(q_{flip})^k_{J^i}, i=1,2,3$ denote the flipped images of the charges
$q^k_{J^i}, i=1,2,3$ in $c$  on the three  chosen edges $e_{J^i}, i=1,2,3$ for the  $N\rightarrow 4$ deformation.
The charges on the remaining parts of the graph in Fig \ref{fig3dprop}(G) are unchanged relative to their values on these parts of the graph in Fig \ref{fig3dprop} (F).
\\

\begin{figure}
  \begin{subfigure}[h]{0.3\textwidth}
    \includegraphics[width=\textwidth]{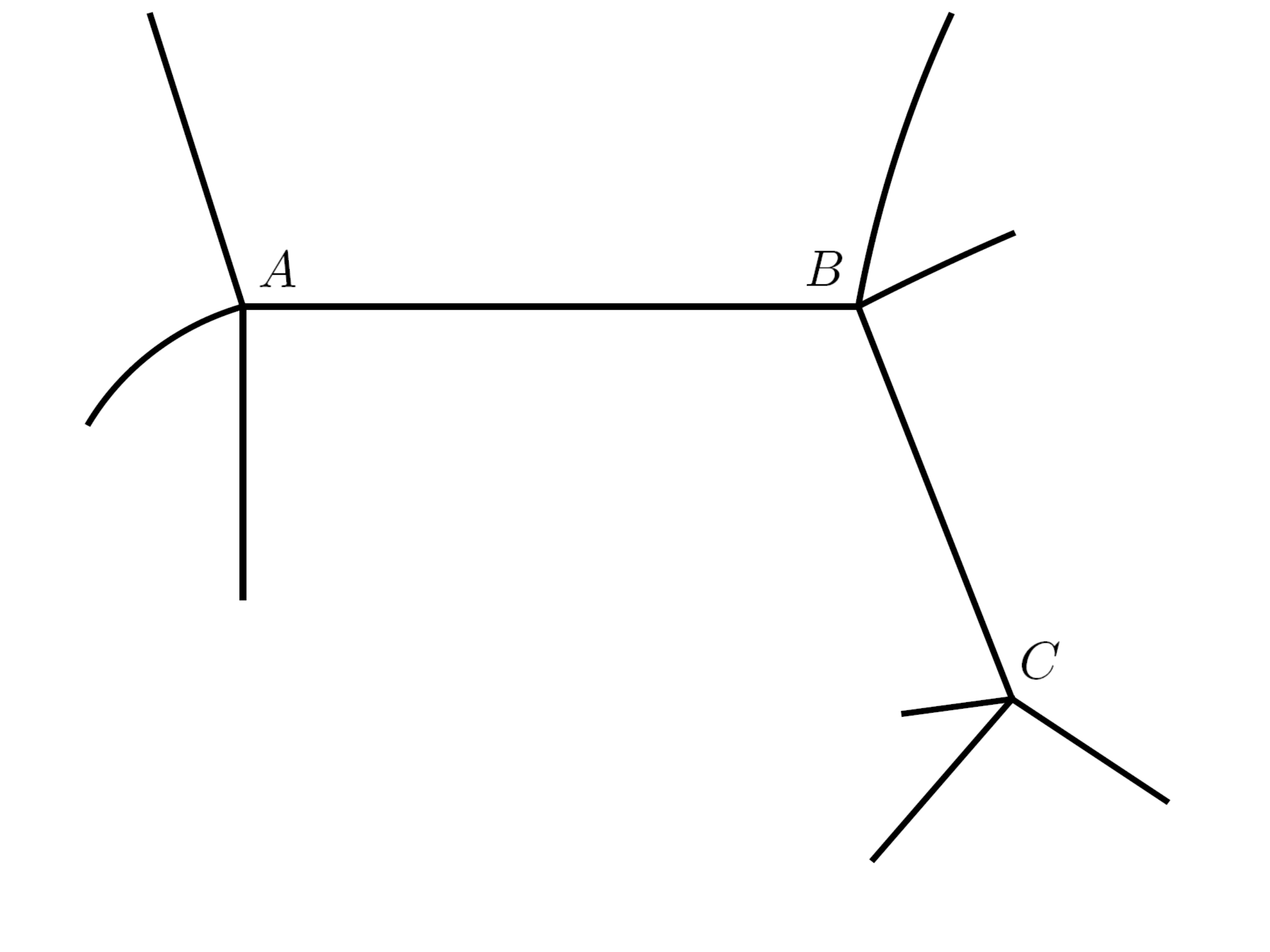}
    \caption{}
 \label{f3d1}
  \end{subfigure}
  \begin{subfigure}[h]{0.3\textwidth}
    \includegraphics[width=\textwidth]{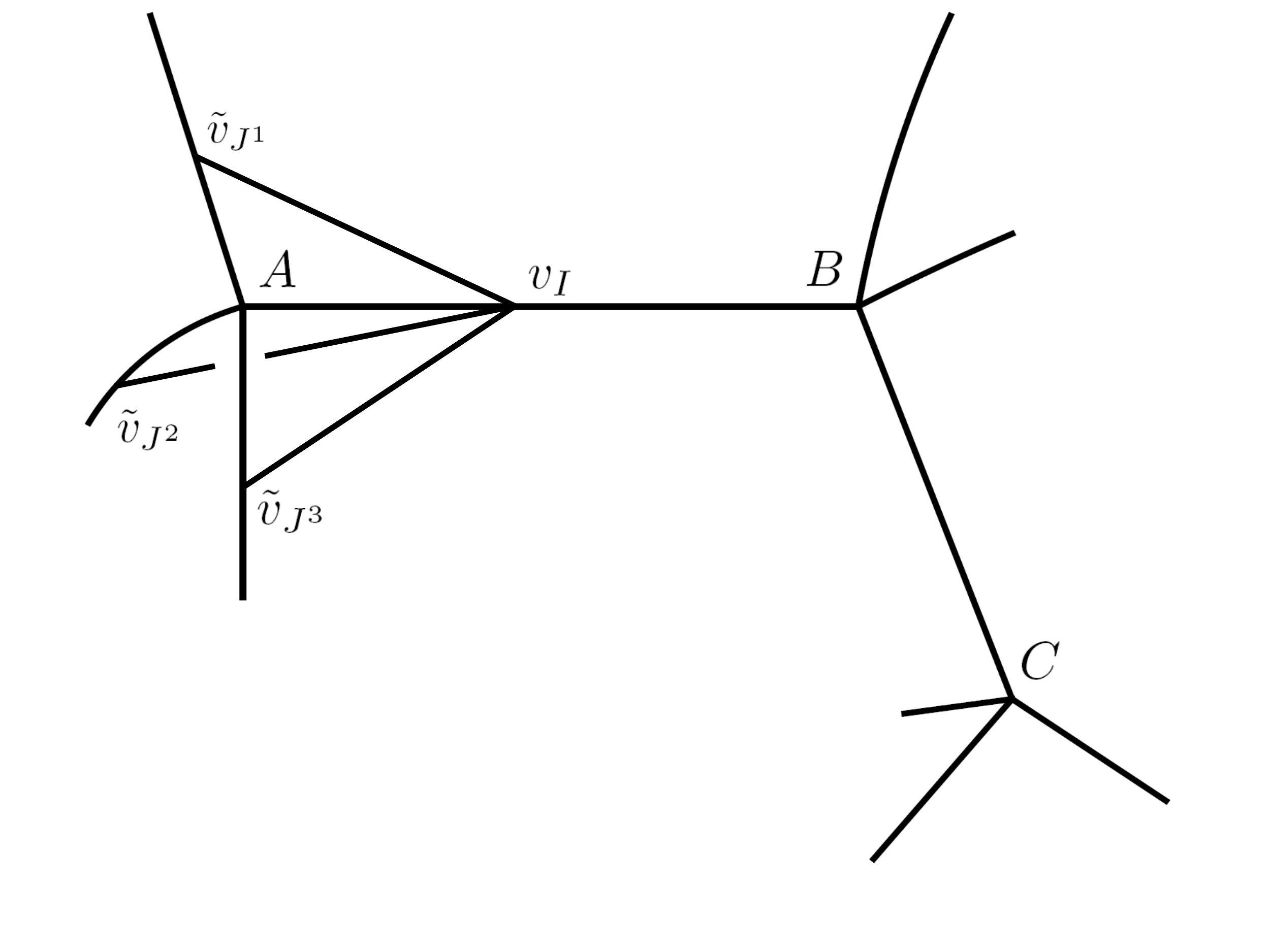}
    \caption{}
   \label{f3d2}
  \end{subfigure}
\begin{subfigure}[h]{0.27\textwidth}
    \includegraphics[width=\textwidth]{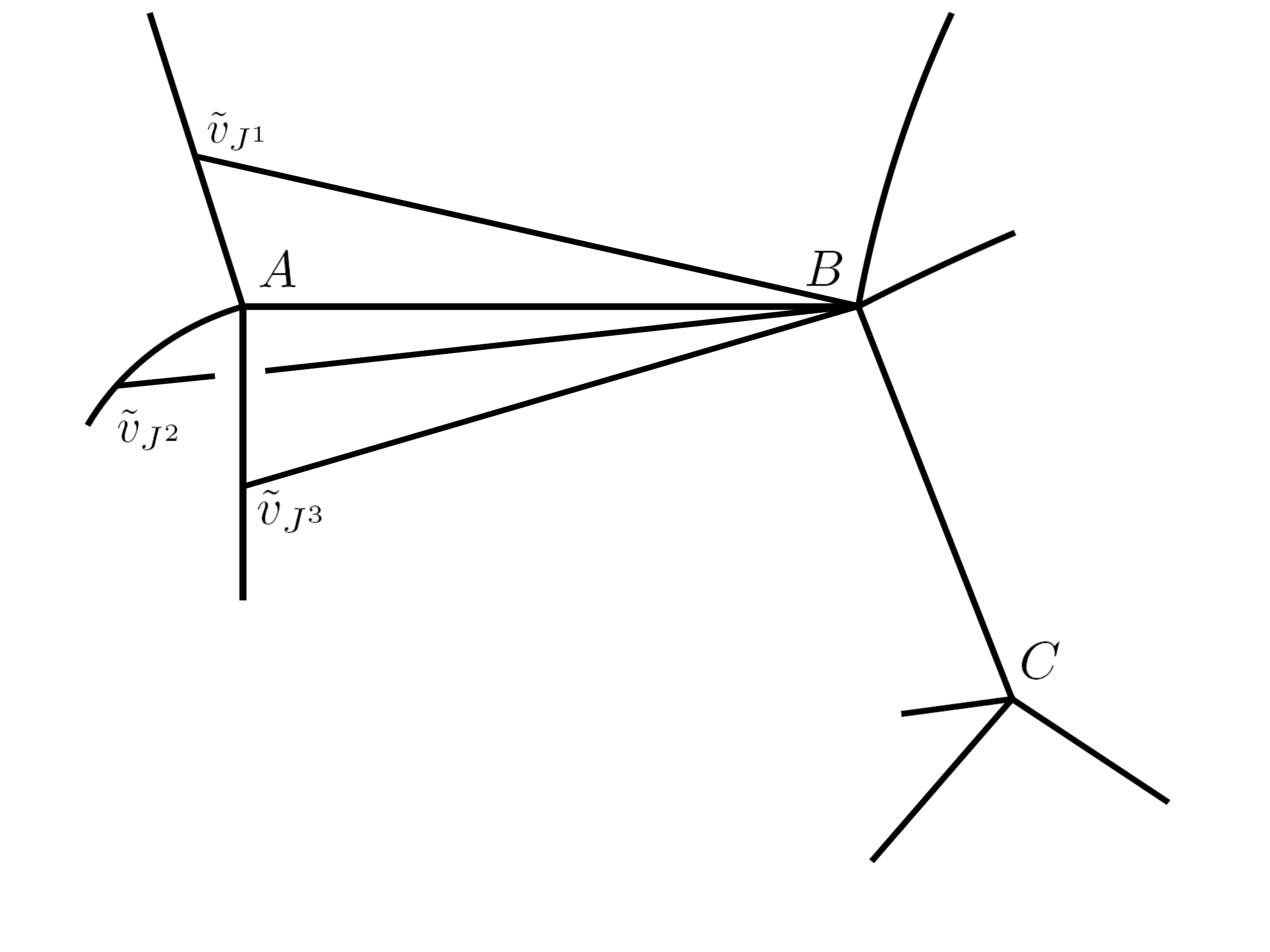}
    \caption{}
   \label{f3d3}
  \end{subfigure}
\begin{subfigure}[h]{0.3\textwidth}
    \includegraphics[width=\textwidth]{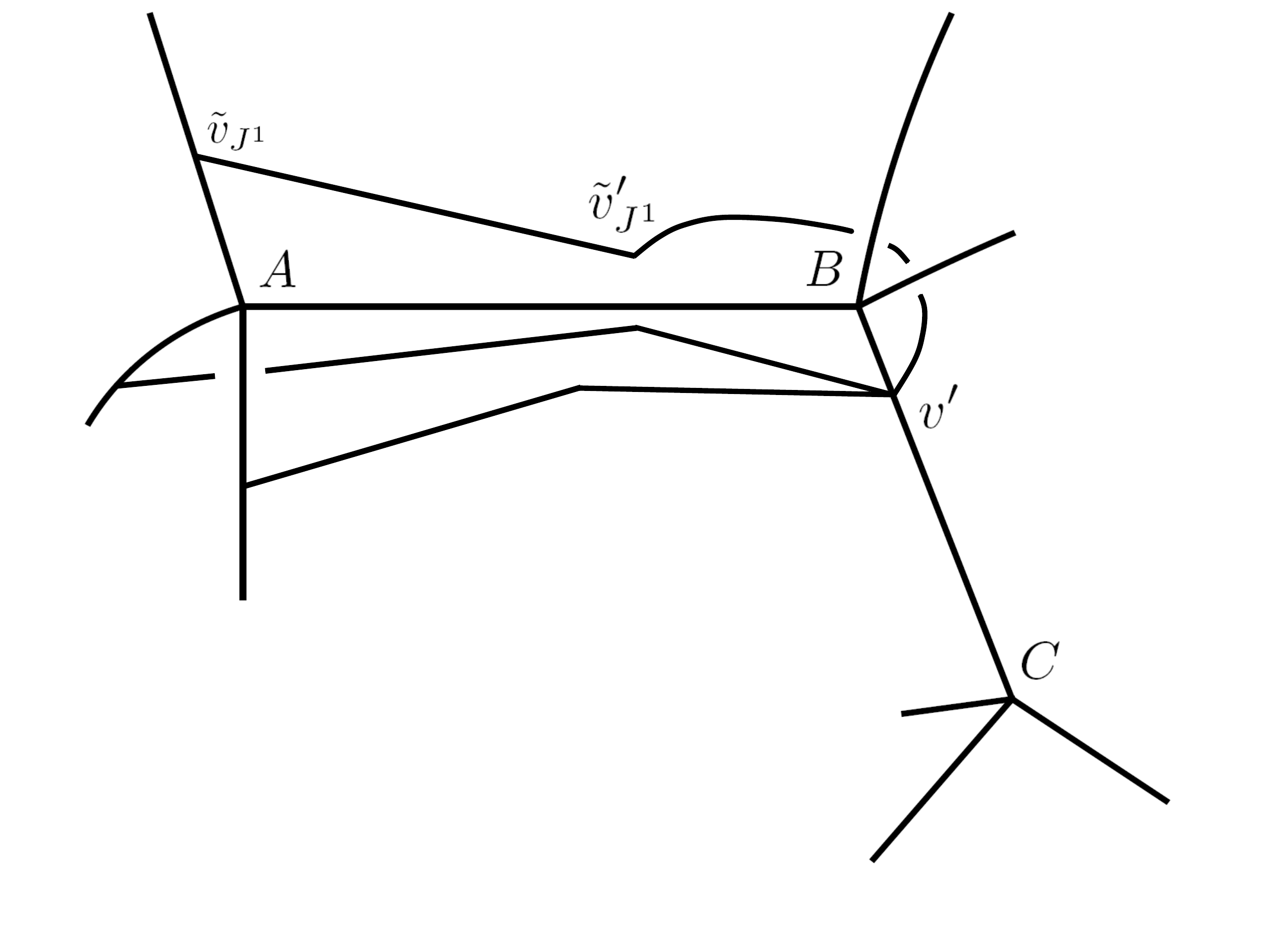}
    \caption{}
   \label{f3d4}
  \end{subfigure}  
\begin{subfigure}[h]{0.3\textwidth}
    \includegraphics[width=\textwidth]{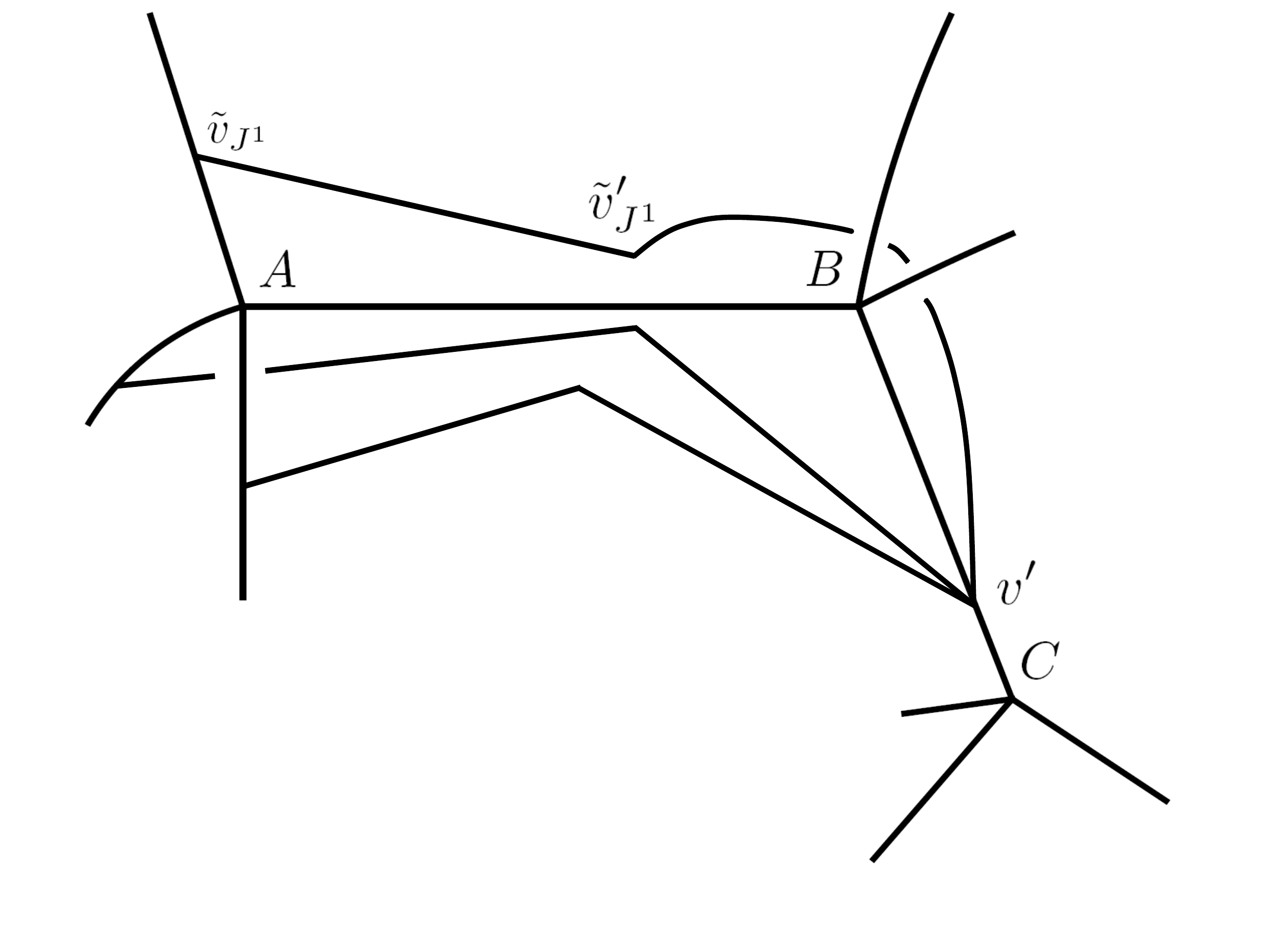}
    \caption{}
   \label{f3d5}
  \end{subfigure}  
\begin{subfigure}[h]{0.3\textwidth}
    \includegraphics[width=\textwidth]{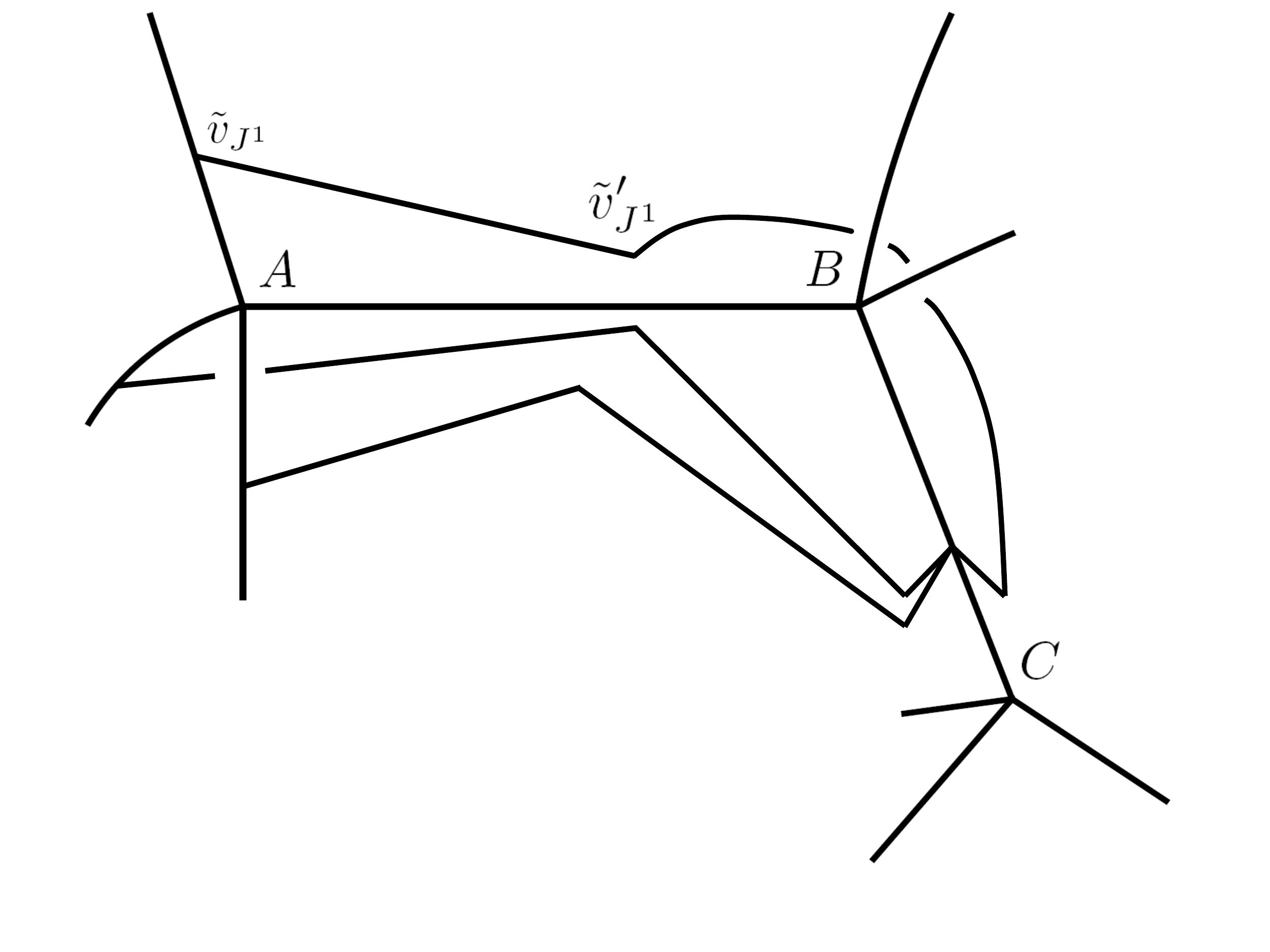}
    \caption{}
   \label{f3d6}
  \end{subfigure}
\begin{subfigure}[h]{0.3\textwidth}
    \includegraphics[width=\textwidth]{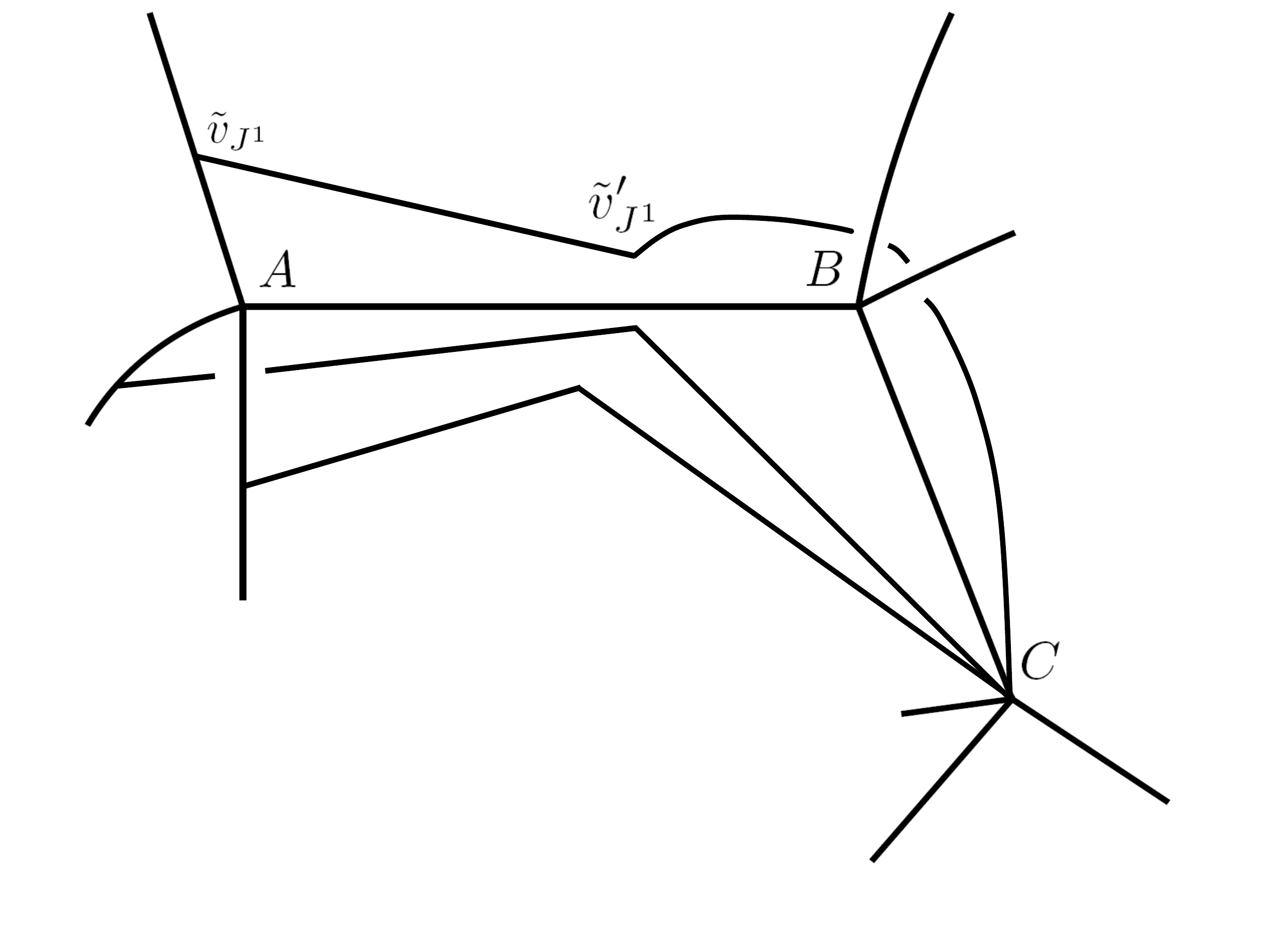}
    \caption{}
   \label{f3d7}
  \end{subfigure}  
\caption{ The figures show the sequence of  Ket Set elements (A) to (G) which encode propagation from vertex A to vertex C as described in the main text. 
The `unperturbed' chargenet in (A) is based on a graph which is dual to a triangulation of the Cauchy slice. 
The Ket Set is the  minimal
one which contains  (A) and is appropriate  to the $N\rightarrow 4$ deformation.   }%
\label{fig3dprop}%
\end{figure}

Once again, the charge nets in (A)-(G) must all be in the Ket Set and the sequence (A)-(G) describes the propagation of  the perturbation
from vertex $B$ to vertex $C$ of $c_{AB}$. Here similar to  the case of $c^{\prime}$ in section \ref{sec4.2}, we have assumed that 
the 7-valent vertex $C$ of $c_{ABC}$ is non-degenerate (the non-degeneracy of the vertex $v^{\prime}$ follows from that of the assumed degeneracy of 
$v_I$ in  section \ref{sec4.2}).

To summarise: the sequence
\be
c\rightarrow c_{(i, I, {\vec J}, \beta, \delta)} \rightarrow c_{AB}\rightarrow c_{ABC}
\ee
represents propagation of a perturbation from the vertex $A$ to the vertex $C$ which is two links away from $A$. The path of propagation is $A-B-C$.
Clearly, with assumptions of non-degeneracy similar to (5) above for the 7-valent vertices encountered in the course of propagation,
together with assumptions on the existence of appropriate semianalytic diffeomorphisms similar to that in Footnote \ref{fndiff1},
we may propagate this perturbation along a path joining $A$ to a vertex  $A_{final}$ as far away as we desire. This shows that the $N\rightarrow 4$
action generically engenders 3d long range propagation.

\section{\label{sec5.0} Propagation and Anomaly Free Action}

In section \ref{sec5} we discuss the new challenges to be confronted in rendering the constraint actions discussed in this paper consistent with 
the requirement of anomaly free commutators as articulated in \cite{p3}. 
One outcome of this discussion 
is the suggested  enlargement of the Ket Sets considered hitherto by replacing the role of semianalytic diffeomorphisms
in property (a3), section \ref{sec1} by a larger set of transformations which we call linear vertex preserving homeomorphisms or lvh transformations.
In section \ref{sec5a}  we 
describe certain technical issues  which we overlooked in our treatment of propagation hitherto and suggest that these issues may be alleviated 
as by the use of  these transformations. In section \ref{sec5b} we discuss the issue of `fake' propagation.

\subsection{\label{sec5} New challenges for a proof of anomaly free action consistent with propagation}

The physical states considered hitherto are based on Ket Sets subject to property (a). The considerations of section \ref{sec2.8} imply that these 
physical states provide a trivial anomaly free representation space for the Hamiltonian (and spatial diffeomorphism) constraints.
Several new challenges must be confronted relative to \cite{p3}  in order to construct off shell deformations of these states which support
a nontrivial anomaly free implementation of constraint commutators. We describe the ones we are able to anticipate below.\\

\noindent (1) New issues for the $N\rightarrow N$ action:\\

\noindent (1a) Multiple vertices:
\footnote{This is a slightly more detailed discussion of (2), section \ref{sec2.7}}
The work in \cite{p3} constructs the desired off shell states based on Ket Sets with elements each of which has only a single vertex where the Hamiltonian
constraint acts non-trivially through $N\rightarrow N$ deformations. The first challenge is to generalise this construction to the case of multivertex Ket Sets.
Due to the independence of the action of the Hamiltonian and electric diffeomorphism constraints at independent vertices, we feel that it should be possible to
generalise the considerations of \cite{p3} to the multivertex case with $N\rightarrow N$ deformations. The candidate Ket Set subject to property (a) would consist
of  multivertex kets. We would then seek to construct appropriate off shell states as linear combinations of (the bra correspondents of)  these kets
with coefficients which were products, over vertices, of the single vertex coefficients  of \cite{p3}. The new feature of the multiple action of constraints
would be  the appearance of contributions from different vertices. The new challenge would be developing an efficient book keeping of these contributions
and defining appropriate `$Q$-factors' \cite{p3} to ensure the existence of anomaly free continuum limit commutators.\\

\noindent (1b) Coordinate patch specification: 
In addition to (1a), an appropriate diffeomorphism covariant specification of coordinate patches for each vertex of each multivertex ket 
in the Ket Set is necessary in order both to 
define the action of the (higher density) constraints as well as to evaluate the off-shell state coefficients mentioned above.
Such a specification exists for the single vertex states of \cite{p3}. Its detailed construction rests on the choice of certain fiducial 
coordinate patches  with respect to which certain
preferred classes of kets in the Ket Set known as primaries, have linear vertices (see \cite{p3} for details). 
It then turns out that the  physical states constructed in \cite{p3} have an imprint of the choice of these fiducial coordinate patches;
this is unsatisfactory from the point of view of the `background independent' philosophy of LQG.
Work in progress suggests that this 
imprint can be removed by enlarging the single vertex Ket Sets of \cite{p3} so that the semianalytic diffeomorphisms of property (a3) in section \ref{sec1} are 
replaced by a larger set of transformations, which we tentatively identify as homeomorphisms which preserve the linearity of linear vertices and which 
we refer to as `{\em linear-vertex preserving homeomorphisms}' or {\em `lvh' transformations}.

The coordinate patches at such vertices are then related by transformations which have a much more local character than diffeomorphisms.
Since many of the considerations of \cite{p3} rest on local Jacobian transformations, it seems that the constructions of \cite{p3} can be
generalised to the context of these larger Ket Sets. Assuming such a generalization is successful, we anticipate that a further generailization of this  coordinate
patch specification  to the multivertex case  should not face any significant obstacles.
\\

\noindent (1c)Upward direction specification:
\footnote{This issue and the related exposition involves an assumed  familiarity with  fine technical issues in \cite{p3}} 
As mentioned in section \ref{sec2.7}, the constructions of \cite{p3} involve a unique specification
of upward or downward conical deformations at a vertex deriving from the signs of the charges at the vertex and the unique identification  of an upward direction at the 
vertex from the graph structure
in the vicinity of the vertex. The graph structure involved in the unique identification of an upward direction is that of the placement and type of kinks 
(i.e. $C^0,C^1,C^2$
\cite{p3}) in the vicinity of the vertex. We believe that the  creation and placement of the $C^1,C^2$ kinks is an unnecesary feature of the constructions of \cite{p3}
and that the specification of an upward direction at the parental vertex being acted upon can be made freely. What seems to be important in defining the action of 
the products of regulated constraint operators on a parent state is that the upward directions defining each operator in this product be suitably correlated
with the free choice of upward direction associated with the action of the first operator in the product on a parental vertex. 
In this way, the choice of an upward direction is a further regulator choice in defining constraint operator products. Anomaly free commutators
then refer to the equality of 2 operator products  related through a replacement of Hamiltonian constraint commutators by electric diffeomorphism commutators
in accordance with the identity (\ref{key}), wherein  the 2 operator products are assigned the same choice of upward direction at the parental vertex.
The end result is that the Ket Set would contain both upward and downward  deformations of parents  independent of the 
sign of the parental charges. While our intuition is that it should be possible to demonstrate all this, it is of course essential 
to explicitly construct such a demonstration.
\\

\noindent (1d) States subject to additional physical requirements:
As seen in section \ref{sec3.1} the $N\rightarrow N$ constraint action does not engender propagation. This is slightly ameliorated by the 
imposition of further physical requirements in section \ref{sec3.2}.  While we feel that the considerations of \cite{p3} should generalise
to multivertex Ket Sets {\em not subject to these requirements}, for Ket Sets subject to this requirement still new challenges arise related to the 
fact that the action of ${\hat H}_{\pm}(N)$ of section \ref{sec3.2} do not generically preserve the number of non-degenerate vertices.
\footnote{Note however that the action of the Hamiltonian and diffeomorphism constraints do preserve the number of non-degenerate vertices modulo the
`eternal non-degeneracy' assumption of \cite{p3}. This is in contrast to the $N\rightarrow 4$ constraint action; see the next point (2) for details.}
However
since even with the additional requirements, generic Ket Sets of  section \ref{sec3.2} do not engender vigorous propagation we shall not discuss this case 
further. Instead we proceed to a discussion of challenges in the context of the $N\rightarrow 4$ action.
\\

\noindent (2) The  $N\rightarrow 4$ action:  Coordinate patch specification remains an issue  and the discussion in (1b)
above applies equally well to the $N\rightarrow 4$ action. 
New issues arise  from  the fact  that, unlike the $N\rightarrow N$ action,  
the $N\rightarrow 4$ action does not preserve the number of non-degenerate vertices.
For example, if we have a single vertex  charge net with a non-degenerate vertex of valence $N$, a Hamiltonian constraint $N\rightarrow 4$ action on this
vertex yields a state with an $N-3$
valent vertex and a 4 valent vertex, which for $N \geq 7$ are both generically non-degenerate. Thus unlike the $N\rightarrow N$ case, it is not meaningful 
to talk of `fixed vertex sectors'.
Moreover the second action of a constraint  acquires possible contributions not only from the 4 valent child vertex created by the first action but also
from the $N-3$ valent vertex at the location of the original parental vertex.
Since the off shell states constructed in \cite{p3} were geared to the preservation of the number (and valence) of the non-degenerate vertices 
by the $N\rightarrow N$ action, accomodating this new feature is a challenge. 
%
%

We feel that one avenue for possible progress in overcoming these new challenges may be to seek a specification of the 
$N\rightarrow 4$ action through the use of interventions (similar but not identical to 
those of \cite{p3}) which convert $N$ valent vertices to 4 valent ones 
\footnote{\label{fnupn4} Once this is done, we would like the upward direction specification to be free as discussed
in section (1c).}
and which is such that this specification is still consistent with 
the propagation described in section \ref{sec4}.
Such a specification may also constrain the space of `possible parents' whose existence 
derives from the merger of a 4-valent vertex with an $M$ valent one to yield an $M+3$ valent vertex (see sections \ref{sec4.2}, \ref{sec4.3})
by constraining the placement of the three new edges relative to those of the  $M$ existing edges. 

\subsection{\label{sec5a} Issues related to the movement and absorption  of child vertices}
Our demonstrations of propagation hitherto overlooked issues concerned with the specification of coordinate patches
at vertices where the Hamiltonian and electric diffeomorphism constraints act. More in detail,
propagation ensues due to our demonstration of non-unique ancestry of the same child. Our arguments purport to show that  
the same child can be generated, upto diffeomorphisms,  by constraint actions
in the vicinity of one ancestral vertex and equally well by constraint actions in the vicinity of a neighbouring ancestral vertex.
In our arguments we have made two assumptions, one explicit and one implicit. The first is an assumption that semianalytic diffeomorphisms
can be used to `move' child vertices in the vicinity of one ancestral vertex to  the vicinity of a neighbouring ancestral vertex, and is 
explicitly mentioned in Footnotes \ref{fndiff0}, \ref{fndiff1} and \ref{fndiff2}.
The second, implicit, assumption relates to the `absorption'  of the moved vertex by the second ancestral vertex resulting in a distinct `possible' ancestor, or, equivalently,  
the creation of this moved vertex by 
constraint actions at this second ancestral vertex  in this distinct  ancestor. These actions create deformations which are conical with respect to
the coordinate patch associated with this second ancestral vertex and this patch is in general different from the one associated with the first ancestral vertex 
so it is not clear if the moved vertex can be  conical with respect to the patch associated with this second ancestral vertex.
If it is not, then the candidate `possible ancestor' cannot be an ancestor of such a child and propagation does not ensue. Hence the implicit assumption
is that the child  vertex can be moved from the vicinity of one ancestral vertex to a neighbouring ancestral vertex in such a way that it does represent 
a  deformation of this neighbouring vertex in the candidate `possible ancestor' which is conical with respect to the ancestral coordinate patch at this neighbouring vertex
in this possible ancestor.

We believe that both  these assumptions  are unnecessary if we admit the `lvh' transformations described in (1b) of section \ref{sec5}. 
In other words, these lvh transformations can be used to move
a child vertex from the vicinity of one ancestral vertex to the vicinity of a neighbouring ancestral vertex in such a way that the moved child vertex 
can indeed be created by the candidate possible ancestor used in our arguments for propagation in sections \ref{sec3.1b}, \ref{sec4.2} and \ref{sec4.3}.
Showing this requires developments along the lines of (1b), section \ref{sec5} and constitutes a problem for future work.

\subsection{\label{sec5b} The issue of `fake' propagation.}
As discussed in section \ref{sec1}, propagation is said to be encoded by a physical state if in the set of 
its kinematic summands there exists a `propagation' sequence of states starting with a parent $p$ and describing the propagation of its  perturbations from one parental vertex
to another followed by absorption of these perturbations to yield another possible parent $p^{\prime}$. In order that this notion of propagation be strongly tied to the 
properties of the quantum dynamics (i.e. to the properties of  the constraint operators), it is essential that the coefficients in the sum over kinematic summands
are {\em uniquely} determined (upto an overall constant factor) by the requirement that the physical state in question is annhilated by the constraints.
This rules out {\em `fake'} propagation wherein the physical state in question is itself an arbitrary linear combination of physical states such that each 
physical state in this combination contains only a subset of elements of the desired propagation sequence as summands. Such `fake' propagation arises only because
of the `artificial' choices of linear combination of individual physical states and is not strongly tied to the dynamics.
Thus it is important to establish that the examples of propagating physical states we construct in section \ref{sec4} are ones whose summand coefficients
{\em are} uniquely determined.  

In this regard, 
we note that the physical states which encode propagation in section \ref{sec4} are `minimal' physical states containing a parent $p$ of interest.
By a `minimal' physical state  containing  a charge net $p$ we mean  one which is  built out of a sum (with unit coefficients) over all elements of the minimal Ket Set containing $p$.
In turn, by the minimal Ket Set containing $p$ we mean 
the smallest  Ket Set subject to property (a) which contains  $p$.
Section \ref{sec4} then constructs examples of a  minimal Ket Set containing a specific  parent $p$ of interest such that this Ket Set 
also contains propagation 
sequences of states which start with $p$. The question is then if the specific $N\rightarrow 4$ implementation of the constraints uniquely fixes the coefficients
in a linear superposition of elements of a minimal Ket Set to be unity upto an overall factor.
The answer to this is almost certainly in the affirmative provided we define the $N\rightarrow 4$ action as suggested in 
(2), section \ref{sec5} (see especially Footnote \ref{fnupn4}). A proof of this is desireable but should be straightforward to construct given that both upward and downward conical 
deformations are generated independent of the sign of  charge labels, if need be by the use of interventions (see  section \ref{sec2.7} and \cite{p3}) together with 
judicious sign insertions (see the discussion around equation (\ref{-liee})). The idea is to then construct  legitimate expressions for the constraints 
in which the `(deformed child  $-$ parent)' contributions to (\ref{prop2}) can occur with either positive or negative signs independently of each other so that
the condition that a physical state $\Psi$ is annihilated by all these expressions reduces to (\ref{prop3}). 

While a proof along the lines sketched above is appropriate with regard to minimality in the context of property (a) of section \ref{sec1}, there is a further fine technical
point which arises due to the desired enlargement of these minimal Ket Sets through the replacement of semianalytic diffeomorphisms in the articulation of 
property (a3) by the larger set of `lvh' transformations advocated in (1b) and (2) of section \ref{sec5}.
We shall refer to properties (a1), (a2) together with this replacement of (a3) as the `{\em lvh modification of property (a)}'.
We note that the  enlargement of the Ket Set  through the requirement of consistency with the lvh modified property (a) 
is {\em not} a direct result of the implementation of the Hamiltonian, electric diffeomorphism and semianalytic diffeomorphism constraints
but is motivated by the general requirement of background independence as discussed in (1b). More in detail, recall that this requirement indicates 
that physical states should not depend on ad-hoc regulating coordinate choices.
Since we strongly believe that any trace of ad-hoc regulating structures in our physical state space would lead to unphysical 
consequences, we are only interested in propagation for those physical states which do not bear any trace of such regulating structures.
In the context of our discussion of `fake' propagation we then adopt a working definition for the engendering of propagation as follows.

The {\em minimal lvh extended Ket Set containing $p$} is defined to be the 
smallest Ket Set containing $p$ which is consistent with the lvh modification of property (a). A physical state
 $\Psi_{lvh}$ will be said to  encode  propagation from a parent chargenet state  $p$ if:\\
\noindent (i) $\Psi_{lvh}$ is constructed as the sum over elements, with unit coefficients, of elements of the Ket Set $S_{ket,lvh}$.\\
\noindent (ii) $S_{ket,lvh}$ is the minimal lvh extended Ket Set containing $p$.\\
\noindent (iii) There is a propagation sequence in $S_{ket,lvh}$ starting from $p$ with elements of this sequence being related 
by the action of Hamiltonian constraint deformations, electric diffeomorphism diffeomorphism deformations and lvh transformations.

Clearly this definition requires clarity on the exact nature of the lvh transformations alluded to in (1b), section \ref{sec5}.
As indicated in section \ref{sec5}, this clarification along with other issues discussed in section \ref{sec5} constitute 
open problems worthy of further study.

\section{\label{sec5.1} Vertex Mergers and the Issue of Coarse Graining}

While, in section \ref{sec4.3},  we illustrated {propagation} in the context of a  parent charge net based on a graph which is dual to a triangulation 
of the Cauchy slice, 
it is instructive (and fun!) to explore other consequences of the $N\rightarrow 4$ action for states based on graphs with this,  as well as different,
vertex structures. In this section we focus on the phenomenon of {\em vertex merger}.

For  parent graphs, in which a pair of parental vertices are connected by a single line which extends beyond these vertices
(rendering the vertices `CGR'), it turns out that this action  can effectively {\em merge} parental vertices,
the chargenets with these merged vertices corresponding  to `possible parents' 
of appropriately deformed  children of the parent chargenet.
To see this imagine that in Fig \ref{f3d1} the edge $e_{AB}$ extends beyond $A$ and connects to some other vertex $D$ in the graph. 
Applying an downward conical electric diffeomorphism deformation with respect to $e_{BC}$  to the (now CGR) parental vertex $A$, we effectively `move'
$A$ towards $B$. We bring the deformed vertex in the vicinity of $B$ (see section \ref{sec5a} for pertinent discussion), 
reverse the conicality of the
deformed vertex structure by an appropriate electric diffeomorphism and then get $B$ to `absorb' this conical deformation through the use of an electric diffeomorphism
on a possible parent in which $B$ is now 7 valent and $A$ has disappeared!
For special graphs with appropriately connected  (multiply)  CGR  parental vertices (see section \ref{sec3.2} 
for the definition 
of a multiply  CGR vertex),  a repeated application of appropariate electric and semianalytic diffeomorphisms  can cause these parental vertices to merge to 
form a single high valent vertex reminiscent of classical `collapse'. 
Since we have not explored 
the {\em classical dynamics} of $U(1)^3$ theory adequately, we do not know if this dynamics admits such collapsing solutions.
In case (as we are inclined to believe)  it does not, we would view the (physical state based on the minimal Ket containing the) parental state  above   
as not of physical interest. Of course the identification of vertex merging with collapse  is only intuitive because, similar to the notion of propagation, 
it is based on the behaviour of charge net summands in the sum which defines a physical state rather than on the behavior of physical expectation values
of Dirac observables. On the other hand, one may  speculate that if  the $N\rightarrow 4$ action can be incorporated into the action of the constraints for gravity,
such an incorporation could possibly carry the seeds of both graviton propagation and gravitational collapse.

From Figures \ref{figprop1}, \ref{fig3dprop}, the simple modification from the $N\rightarrow N$ to the $N\rightarrow 4$ action not 
only creates and merges vertices it also creates new connections between
structures in the vicinity of vertices thus `entangling' them. Thus this modification has the desirable property of a simple local rule
which leads to rich non-local structure. Due to this property, the $U(1)^3$ model offers an ideal testing ground for {\em coarse graining} proposals in the 
context of the $N\rightarrow 4$ action. Specifically we have in mind a $U(1)^3$ implementation of  the  proposals of Livine and Charles \cite{etera}
for the $SU(2)$ case. Since coarse graining and the recovery of effectively smooth fields is a key foundational issue in LQG, the $U(1)^3$ model
hereby acquires additional significance as a toy model for LQG.

Reverting to the example of a charge net based on a graph which is dual to a triangulation, it is possible to construct  an evolution sequence from  the unperturbed state $c$ of 
Figure \ref{fig3dprop} which leads to  a single vertex of high valence surrounded by parental vertices which are rendered degenerate. 
A sketch of this construction is as follows.
We start from the parental chargenet $c$ of Fig \ref{f3d1} and generate the chargenet $c_{AB}$ of Fig \ref{f3d2}. In doing so we have increased
the valence of vertex $B$ to 7 with a concomittant loss of non-degeneracy
of vertex $A$ cause by the vanishing of the $i$th component of all the outgoing charges there (see the discussion in section \ref{sec2}). We then 
deform $c_{AB}$ as follows:\\
\noindent (a) We perturb  some nearest nondegenerate
parental vertex $D$ connected to $A$ by the action of a Hamiltonian constraint. Since the connection must be through one of the 3 edges $e_{J^i}$, the perturbation
will encounter the  trivalent kink ${\tilde v}_{J^i}$ for some fixed $i$ (more precisely this kink is a CGR bivalent vertex) before  getting to $A$.\\
\noindent (b)
We may then, through the exclusive use of semianalytic diffeomorphisms and electric diffeomorphisms together with appropriate assumptions on non-degeneracy
of vertices of possible parents:\\
\noindent (b1) get ${\tilde v}_{J^i}$ to `absorb' the conical perturbation rendering it 5 valent,\\
\noindent (b2) emit it towards $A$ rendering  ${\tilde v}_{J^i}$ bivalent CGR as before, \\
\noindent (b3)  get the vertex $A$ to `absorb' this perturbation   so as to render $A$ 7 valent, and, \\
\noindent (b4)  get $A$ to emit this perturbation and $B$ to absorb it rendering $A$ degenerate and 4 valent,
and $B$, 10 valent and connected to $D$, in addition to its prior connectivity with $A$.   \\

Thus at the end of (a)- (b4) we have again increased the valence of $B$ by 3 at the cost of rendering
a non-degenerate vertex, in this case vertex $D$, degenerate. 
We may then, with appropriate non-degeneracy assumptions, 
repeat this procedure for any path connecting a non-degenerate vertex to $B$, thereby increasing the valence of $B$ by 3 each time and simulataneously rendering the 
non-degenerate vertex, degenerate. The end result is the vertex $B$ of high valence with all vertices in a region surrounding $B$ rendered degenerate.
Note that this process of vertex merger is  distinct from the 
case alluded to above wherein the original parental vertices merge through the exclusive application of  electric and semianalytic diffeomorphisms.  
Indeed, since no parental  vertex of a graph dual to a triangulation (see 
Fig \ref{f3d1}) is CGR, that mechanism of vertex merger fails. Instead, as sketched above, we need to employ the Hamiltonian constraint deformation of step (a)  as well.

It is not clear if the state obtained from repeated applications of the steps (a)-(b4), with many degenerate parental vertices and a single non-degenerate one at the `microscopic' level,
represents a   classically singular configuration precisely because any such interpretation is interwined with the issue of coarse graining of such a state.
This discussion points to  a clear need for an unambiguous interpretation of physical states and their kinematic summands. 
If, following such a putative interpretation, generic physical states display properties at variance with generic classical
solutions, one may need to further modify the dynamics perhaps by a suitable mixture of $N\rightarrow 4$ and $N\rightarrow N$ deformations. For example, 
if we want to avoid the state obtained through repeated applications of (a)-(b4), 
we may create an obstruction to the accumulation of valence by $B$ beyond some fixed valence $N_{max}$ as follows.
%
We define a dynamics which generates $N\rightarrow 4$ deformations for all nondegenerate vertices of valence $N< N_{max}$   and which generates $N\rightarrow N$ deformations
for vertices of valence $N\geq N_{max}$. Clearly, if through  repeated  applications of (a)- (b4), we increase the valence of $B$ to $N$ such that $N_{max} >N \geq N_{max}-3$, 
a further such application
will be obstructed at step (b4) because any possible parental vertex  of valence  $N\geq N_{max}$ can only yield
a child vertex of valence $N$ rather than one of valence 4.
Thus, for such a dynamics,  evolution sequences for generic parental graphs will not generate vertices of valence greater than or equal to $N_{max}$ by vertex mergers.
Note however that the evolution sequence describing long range 3d propagation in section \ref{sec4.3}  remains consistent with such a dynamics  provided $N_{max} >7$, a valence of 7 
being the maximum valence encountered in this evolution sequence.
Also note that our discussion of vertex mergers above is subject to the same technical caveats  discussed in section \ref{sec5a} in connection with propagation.

\section{\label{sec6} Discussion}

Early pioneering works \cite{early} on the quantization of the Hamiltonian constraint for gravity in the late 80's and early 90's together with 
the development of a rigorous quantum kinematics (see for e.g. \cite{kinrigor} as well as  \cite{aajurekreview, alm2t,ttbook} and references therein) and some ideas from other
researchers \cite{aalewand} lead to the 
detailed framework for the construction of a Hamiltonian
constraint operator in Thiemann's seminal work\cite{qsd1}. This framework organises this construction as the continuum limit
of quantum correspondents of classical approximants to the Hamiltonian constraint. 
The resulting operator carries an imprint of the  choice of these approximants and is hence infinitely non-unique. 
Current work seeks to subject the resulting operator to physically and mathematically well motivated requirements so as to reduce this non-uniqueness.
On the other hand, since these requirements are extremely non-trivial, the  mathematical tools and techniques needed to confront them 
also have to be constantly upgraded. 
For example, the work  \cite{hannojurek} seeks to construct this operator such that it is well defined on a {\em Hilbert} space rather than 
a representation space with no natural inner product. This leads to the consideration of new Hilbert spaces which lie between the 
kinematic Hilbert space of LQG and the linear representation space known as the habitat \cite{habitat1}.
The works \cite{mediff,p1,p2,p3} seek constructions which are consistent with non-trivial anomaly 
free commutators and lead to new tools such as electric field dependent holonomy approximants \cite{mediff}, the use of electric diffeomorphism deformations arising from the 
discovery of a new classical identity \cite{p1},
diffeomorphism covariant choices of coordinate patches \cite{p2},  interventions \cite{p3}  and a new mechanism for diffeomorphism covariance \cite{p3}.  

In this work we confront the Hamiltonian constraint of the $U(1)^3$ model by the requirement of propagation. 
The new mathematical elements which allow us to bypass the obstructions to propagation (modulo the caveats discussed in section \ref{sec5a}) pointed out by Smolin \cite{leeprop}
are the  structural property of the constraint discussed in section \ref{sec2.7} (and uncovered in \cite{proppft}) together with  
the $N\rightarrow 4$ action.   As with all new additions to our toolkit, it is necessary to accumulate intuition as to what 
they do and, if necessary improve them further  or discard them. It is in this general context that the developments presented in this
paper should be viewed.

Thus, while on the one hand the $N\rightarrow 4$ action engenders vigorous
propagation thereby showing the basic LQG framework for the construction of the Hamiltonian constraint is powerful
enough to bypass the `no propagation' folkore in the field (see also, however, the discussion at the end of this section), 
it also leads to the phenomenon of vertex merger discussed in section \ref{sec5.1}. 
More generally, the $N\rightarrow 4$ action seperates vertices, merges vertices and increases graphical connectivity leading to rich non-local
structure. The work in this paper studies aspects of this structure  in the context of a few examples of parental graph structures.
It is necessary  to study a larger diversity of such examples so as to explore the full power of the $N\rightarrow 4$ action, and if necessary 
subject it to further improvement.

In this regard, it is an open issue as to  
whether the $N\rightarrow 4$  or some other choice of improved constraint actions is physically appropriate for the $U(1)^3$ model.
In view of the discussion in sections \ref{sec5.1} and \ref{sec1}, 
any resolution of this issue involves (a) an understanding of coarse graining of kinematic states and a consequent interpretation of physical states, (b) an understanding of the 
classical solution space of the model, and,  (c) an analysis of constraint commutator actions on a suitable off shell state space so as to check if  the chosen
actions are anomaly free in the sense of \cite{p3}.
In relation to (c), focussing exclusively on the $N\rightarrow 4$ action, we have discussed the challenges inherent in a putative demonstration of non-trivial anomaly free action
in section \ref{sec5}. If these challenges can be overcome, 
 we believe that a  mixture of $N\rightarrow 4$ and $N\rightarrow N$ actions  of the type discussed towards the end of section \ref{sec5.1} 
should then also not present significant obstacles
to such a demonstration.  
With regard to 
issues (a) and (b),  as indicated in section \ref{sec5.1}, 
we believe that the $U(1)^3$ model provides   a valuable toy model to test proposals for coarse graining in LQG such as those in 
\cite{etera}.

The next step beyond the $U(1)^3$ model  is that of the construction of a satisfactory quantum dynamics for full blown LQG. 
By `satisfactory' we mean, at the very least, `anomaly free' and `consistent with propagation'.
By `anomaly free', we mean a constraint action  which admits the construction of a space of off shell states which support non-trivial anomaly free commutators in the sense of \cite{p3}.
In this regard, we believe that it is important that the constraint action be such that 
a second constraint action acts non-trivially on the deformed structure created by a first such action; this property (which holds for the $N\rightarrow 4$ deformation) 
is crucial for the putative emergence of the desired `$M \partial_a N- N\partial_a M$ dependence of the commutator on the lapses $M,N$ which smear  
the two constraints. By `consistent with propagation' we mean the existence of  sequences of kinematic state summands  which describe
propagation. While the seminal constructions of the Hamiltonian constraint by Thiemann \cite{qsd1,qsd2} do not have the property that second constraint actions
act on deformations generated by the first, it turns out that {\em contrary to common belief} \cite{leeprop}, preliminary calculations \cite{ttmv} 
seem to suggest that propagation {\em is} partially realised by appropriate physical states in the kernel of Thiemann's  Hamiltonian constraint.
More in detail, as mentioned in Footnote 1, while there is no flaw in the analysis of Reference \cite{leeprop}, that work did not
analyse the full kernel of the constraint.  Work in progress \cite{ttmv} hints that the  part of the kernel of the Lorentzian Hamiltonian constraint not analysed by Smolin contains
states which do encode propagation. We shall report on this elsewhere. Our hope is that the framework of \cite{qsd1,qsd2} together with the structures developed
in \cite{p1,p2,p3} and in this work will prove useful for the putative construction of an anomaly free quantum dynamics, consistent with propagation, 
for Euclidean, and finally, for Lorentzian gravity.

\section*{ Acknowledgements:} I am very grateful  to Fernando Barbero for his  comments on a draft version of this manuscript
and for his kind help with the figures. I thank Thomas Thiemann for his comments on a draft version of this manuscript and for
a series of intense discussions which provide the basis for the work  \cite{ttmv}.


\begin{thebibliography}{999}





\bibitem{aajurekreview}
A. Ashtekar and J. Lewandowski, {\em Classical and Quantum Gravity}, 21(15):R53, 2004.

\bibitem{ttbook}
T. Thiemann, {\em Modern Canonical Quantum General Relativity}, 
Cambridge Monographs on Mathematical Physics. Cambridge University Press, 2007.

\bibitem{mebook} {\em Loop Quantum Gravity: The First 30 Years} in {\em 100 Years of General Relativity: Volume 4}, 
Edited by Abhay Ashtekar and Jorge Pullin, World Scientific (2017)



\bibitem{leeprop} L. Smolin, e-Print: gr-qc/9609034

\bibitem{ttmv} T. Thiemann and M. Varadarajan, Note in Preparation.

\bibitem{proppft} M.  Varadarajan 
{\sl Class.Quant.Grav.} {\bf 34} 015012 (2017)


\bibitem{alm2t} A. Ashtekar, J.Lewandowski, D. Marolf, J. Mour${\tilde {\rm a}}$o and T. Thiemann, {\sl J.Math.Phys.}
 {\bf 36} 6456 (1995).

\bibitem{habitat1} J. Lewandowski and D. Marolf, {\sl Int.J.Mod.Phys}
 {\bf D7} 299 (1998).

\bibitem{mediff} A. Laddha and M. Varadarajan, {\sl Class.Quant.Grav.}
 {\bf 28}   195010 (2011).

\bibitem{habitat2} R. Gambini, J. Lewandowski, D. Marolf, and J. Pullin, {\sl Int.J.Modern Physics} {\bf D7} 97 (1998).


\bibitem{p1} C. Tomlin and M. Varadarajan, {\sl Phys.Rev.} {\bf D87} 044039  (2013).


\bibitem{p2} M. Varadarajan,  {\sl Phys.Rev.} {\bf D87} 044040 (2013).

\bibitem{jureklin} J. Lewandowski and C-Y. Lin
{\sl Phys.Rev.}{\bf D95} 064032     (2017). 

\bibitem{p3} M. Varadarajan, {\sl Phys.Rev.} {\bf D97} 106007 (2018).


\bibitem{leeu13}  L. Smolin, {\sl Class.Quant.Grav.} {\bf 9} 883 (1992)

\bibitem{aanewpersp} A. Ashtekar, {\em Lectures on Non-perturbative Canonical Gravity} (Notes prepared in collaboration with 
R.S. Tate), World Scientific Singapore  (1991)


\bibitem{twistedgeo} L.Freidel and  S. Speziale, 
{\sl Phys.Rev.} {\bf D82}  084040 (2010).



\bibitem{etera} C. Charles and  E. Livine, 
{\sl Gen.Rel.Grav.} {\bf  48}   113 (2016) .

\bibitem{qsd1}T. Thiemann, {\sl Class.Quant.Grav.}{\bf  15} 839  (1998).

\bibitem{qsd2} T. Thiemann, {\sl Class.Quant.Grav.}{\bf  15} 875 (1998). 

\bibitem{early}
C. Rovelli and L. Smolin,{\sl Phys. Rev. Lett.}{\bf 61}  1155 (1988);
T. Jacobson and L. Smolin, {\sl Nucl.Phys.}{\bf B299} 295 (1988);
R. Gambini, {\sl Phys.Lett.}{\bf  B255} 180 (1991);  M. Blencowe, {\sl Nucl.Phys.}{\bf B341} 213 (1990);
B. Bruegmann and J. Pullin, {\sl Nucl.Phys.}{\bf B390} 399 (1993).

\bibitem{kinrigor} A. Ashtekar and C. Isham, {\sl Class.Quant.Grav.}{\bf  9} 1433 (1992).



\bibitem{aalewand}  A. Ashtekar and J. Lewandowski, Regularization of the
Hamiltonian constraint (unpublished), cited and partially
invoked in Ref \cite{qsd1}.


\bibitem{hannojurek} J. Lewandowski and H. Sahlmann, {\sl Phys.Rev.}{\bf  D91} 044022 (2015).





\end{thebibliography}
\end{document}